\documentclass[a4paper,12pt]{article}
\usepackage[left=3cm,right=3cm,top=2cm,bottom=2cm,includeheadfoot]{geometry}

\usepackage{float} % to put figures H

\usepackage{natbib}
\usepackage{ragged2e}
\usepackage{xcolor}
\usepackage{amsmath}
\usepackage{caption}
\usepackage{amsfonts}
\usepackage{booktabs}
\usepackage{geometry}
\usepackage{graphicx}
\usepackage{multirow}
\usepackage{tabularx}
\usepackage{rotating}
\usepackage{dsfont}                   %% for indicator variable
\usepackage{mathpazo}
\usepackage{pgfplots}                %% for plots 
\pgfplotsset{compat=newest}     %% for plots 
\usepackage{amssymb}

\usepackage[normalem]{ulem}
\usepackage[obeyspaces]{url}
\usepackage[utf8x]{inputenc}
\usepackage{rotating,subfigure,lscape,pdflscape} 
\usepackage{setspace} % per interlinea 
\onehalfspace               % interlinea 1.5 
\def\sym#1{\ifmmode^{#1}\else\(^{#1}\)\fi} % for asterisks in tables
\PassOptionsToPackage{allbordercolors=white, colorlinks=true, allcolors=blue, hyperfootnotes=false}{hyperref} % blue ref

\usepackage{geometry}
\graphicspath{{Figures/}}  % path for figures

\begin{document}

\title{Risk Sharing and the Adoption of the Euro\thanks{We acknowledge support by the ADEMU project, `A Dynamic Economic and Monetary Union', funded by the European Union's Horizon 2020 Programme under grant agreement No. 649396. We are grateful to the editor Sebnem Kalemli-Ozcan, two anonymous referees as well as to Cinzia Alcidi, Giancarlo Corsetti, Juan Dolado, Mathias Dolls, Dalila Figueiredo, Matteo Gatti, Mathias Hoffmann, Mathias Klein, Ramon Marimon and all seminar participants at the Joint Research Centre, at the `Risk-Sharing Mechanisms for the European Union' workshop held at the European University Institute, and at the FIRSTRUN Workshop on Fiscal Policy Coordination in the EMU held at LUISS.}}

\author{Alessandro Ferrari\thanks{University of Zurich. Email: \protect\url{alessandro.ferrari@econ.uzh.ch}.} \and Anna Rogantini Picco\thanks{Sveriges Riksbank.  Email: \protect\url{Anna.Rogantini.Picco@riksbank.se} }\footnotemark[2]} 

\maketitle

\begin{abstract}

This paper empirically evaluates whether  adopting a common currency has changed the level of consumption smoothing of euro area member states. 
We construct a counterfactual dataset of macroeconomic variables through the synthetic control method. We then use the output variance decomposition of Asdrubali, Sorensen and Yosha (1996) on both the actual and the synthetic data to study if there has been a change in risk sharing and through which channels. We find that the euro adoption has reduced risk sharing and consumption smoothing. We further show that this reduction is mainly driven by the periphery countries of the euro area who have experienced a decrease in risk sharing through private credit. 

\bigskip 

\end{abstract}
 \thispagestyle{empty}  % to eliminate page numbering 
\clearpage

\maketitle

\newpage
\section{Introduction}
In 1999, eleven countries adopted a newly created currency, the euro. Sharing the same currency eliminated the exchange rate risk and guaranteed a high degree of price stability, which favoured both trade and financial integration across euro area members.\footnote{See \cite{Kalemli_etal_2010} for a discussion on the impact of the euro on financial integration and \cite{saia} for a focus on trade.} 
As explored by the vast literature on currency areas, deeper trade and financial linkages help countries increase risk sharing. 
At the same time, euro adopters had to give up on using monetary policy as a stabilisation tool against idiosyncratic shocks. 
Higher integration and the loss of monetary policy independence affect the shock absorption capacity of countries in opposite directions. 
Thus, it is not theoretically clear whether and to what extent the adoption of the euro should have affected the degree of consumption smoothing in euro area member states. 
\newline
\newline This paper aims to empirically address this question by evaluating how the adoption of a common currency has changed the degree of consumption smoothing for euro area member states. To do so, it is not enough to compare risk sharing capacity before and after the euro. Rather, what is needed is an appropriate counterfactual for a scenario in which euro area members had not adopted the common currency. A major obstacle when evaluating the effect of a policy intervention like the adoption of the euro is the absence of a natural counterfactual. Our first contribution is to build a synthetic counterfactual dataset of macro variables for the scenario of no adoption of the euro. For this purpose, we rely on the synthetic control method (SCM) of \cite{abadie_economic_2003}. Equipped with such a synthetic dataset, we are able to assess how the adoption of the euro has changed the level of consumption smoothing in euro area countries. 
We want to systematically identify different channels through which risk may be shared across and within countries. We therefore carry out an output variance decomposition \`a la \cite{asdrubali_channels_1996}. This method allows us to isolate four possible channels through which shocks are absorbed:
private cross-border investments, government taxes and transfers, private savings, and public savings. The first two channels are purely international as they involve cross border absorption of shocks, while the other two channels refer to consumption smoothing that happens also domestically. 
\newline
\newline We present our results on the risk sharing channels decomposition in a difference-in-differences framework. In this context, our treatment is the adoption of the euro, our treated group are the euro area member states,\footnote{To be precise, we include the first eleven countries that adopted the euro:  Austria, Belgium, Finland, France, Germany, Greece, Ireland, Italy, Netherlands, Portugal, and Spain.} and our control group are their synthetic counterfactuals for the scenario of no adoption of the euro. 
Our main result is that with the adoption of the euro consumption smoothing has decreased.
 We find a heterogeneous effect of the adoption of the euro between core and periphery countries.\footnote{We follow \cite{baldwin2015eurozone} and define {\sl core} countries: Austria, Belgium, Finland, France, Germany, and Netherlands; {\sl periphery} countries: Greece, Ireland, Italy, Portugal, and Spain. Also \cite{cimadomo_etal_2020} adopt a similar classification.} While the euro has not significantly affected consumption smoothing in core countries, it has substantially and significantly reduced it in periphery countries, mainly due to a lower absorption through private savings. We further show that the decrease in consumption smoothing is not attributable to the Great Recession and the period thereafter. In fact, we find that the decrease in consumption smoothing is even sharper once we exclude the period from the Great Recession. 
 We provide a potential explanation for our result on the lower consumption smoothing by showing that it may be attributable to the higher GDP growth and volatility prompted by the adoption of the common currency.
 %Further, our empirical results on lower consumption smoothing are compatible with the theoretical contributions in \cite{aguiar_gopinath} and \cite{benigno_fornaro}. According to the former, a higher consumption smoothing can be the consequence of a change in the trend growth of output, which may have been prompted by the euro adoption. Following the latter, a more robust consumption response in the periphery countries may be the result of a stronger interest rate convergence in periphery countries.
%Furthermore, the lower consumption smoothing effect is particularly strong in countries who had the largest decline in their government bonds spreads (euro dividend). We interpret these results as indicative that the adoption of the common currency has generated a positive income shock and a trend increase, which, in turn, has led to  consumption frontloading. In summary, we find evidence of a positive effect of the euro on both contemporaneous and future income. When agents increase their contemporaneous consumption this shows up as a higher comovement of income and consumption, and, therefore, lower consumption smoothing. This mechanism is consistent with the theoretical contributions in \cite{aguiar_gopinath} and \cite{benigno_fornaro}.  
\newline
\newline Adopting the SCM provides us with an alternative methodology to the two main approaches that can be found in the literature analyzing the effect of the euro on risk sharing -- the before/after analysis and the difference-in-differences. We believe the SCM to be a valuable method as it allows us to control for any common trend between the before and after treatment period by using a selected control group. While 
a difference-in-differences setting would also allow us to control for pre-treatment common trends, the SCM further helps us building an appropriate counterfactual -- one that best mimics our treated group before treatment -- even in absence of many control units.
Given the specificity of our setting where we are limited in the available number of potential control units, we believe the SCM to be especially suited to tackle our research question.
%this common trend would not be possible with a before/after analysis. By adding an unselected control group, a difference-in-differences analysis is able to capture common trends between the treated and the control group.
%However, due to the specificity of our setting where we are limited in the available number of potential control units, the SCM is especially suited. In particular, it allows us to build a control group that best mimics our treated group before treatment even with few control units, a limitation which could instead potentially invalidate a difference-in-differences analysis.
To exemplify the value of our approach,  we show that the SCM allows us to decompose the results found by previous literature between what is common to treatment and control group and what can actually be attributed to the introduction of the euro. An example of this is the finding by \cite{furceri_euro_2015}, \cite{balli_etal_2012} and \cite{cimadomo_etal_2020} that the introduction of the common currency has increased risk sharing through international capital markets. Through our approach we clarify that such increase is common to both our treatment and control group, while we find no significant effect of the common currency on capital market risk sharing for the full sample of countries. 
\newline
\newline 
As it is crucial for the validity of our results to have a robust counterfactual, we carry out several checks. We show that our results hold when using different strategies to build our synthetic counterfactuals. Our baseline results are confirmed if we generate the synthetic variables by matching
on additional observables such as the share of rents from natural resources, by matching on first differenced data rather than on variables in levels, by changing the matching window, or by keeping fixed weights for all macro variables of each synthetic country. Results are also robust to changing the assumed year of adoption of the euro, to dropping recessionary periods or to excluding European Union (EU) countries, which are not part of the euro area, from the group of countries used to generate synthetic variables. We use placebo tests to confirm that our main results are not driven by noise or common trends. Finally, we show that our main results remain when correcting from a potential bias arising from estimating regressions with variables that may contain some measurement error.
\newline
\newline 
Our paper is related to two main strands of the literature. First, by building our counterfactual dataset via the SCM, we are related to the research stream that employs the SCM to generate counterfactual scenarios in absence of natural ones. 
The SCM is introduced by \cite{abadie_economic_2003} to test for the impact of the outbreak of terrorism in the Basque Country in the late 60s, and further employed by \cite{abadie_synthetic_2010} to estimate the effect of a large-scale tobacco control programme implemented by California in 1988. \cite{billmeier_assessing_2012} use it to investigate the impact of economic liberalization on real GDP per capita in a worldwide sample of countries. Closer to our focus, \cite{campos_economic_2014} use the SCM to evaluate the benefits of being part of the European Union, while \cite{saia} employs it to estimate counterfactual trade flows between the UK and Europe under the counterfactual scenario that the UK had joined the euro. \cite{born_etal_2019a} use the SCM to study the impact of the Brexit vote on the UK output, while \cite{born_etal_2019b}  to study the effect of Trump's election on the US growth and job creation. \cite{Terzi_2020} relies on a propensity score matching model to build a counterfactual for the per capita GDP in the euro area countries and study the impact of macroeconomic adjustment in the euro area. The closest to our paper are \cite{puzzello_gomis_2018} and \cite{duque_pessoa_2020}, who use the SCM to evaluate the impact of the euro on real per capita income. These two papers focus on the effects of the euro on the general economic performance, which is measured through real per capita income. Our paper goes a step further, and evaluates how the adoption of the euro directly affects the degree of consumption smoothing of member states and through which channels this has occurred.  
\newline
\newline Our paper is also related to the stream of the literature that studies cross-regional risk sharing --see \cite{baxter_business_1995} among the earliest contributions. To identify different risk sharing channels, we follow the output variance decomposition introduced by \cite{asdrubali_channels_1996} to examine risk sharing in the US. 
This methodology is also used  by \cite{balli_etal_2012} to study risk sharing channels including capital gains in the euro area, the EU and the other OECD countries, and \cite{furceri_euro_2015} to analyse and compare risk sharing across euro area countries and the US states.
The analysis of \cite{furceri_euro_2015} is updated with more recent data by \cite{van_beers_cross-country_2014}, who assess the functioning of insurance mechanisms in the euro area, and  \cite{kalemliozcan_debt_2014} who consider separately countries hit by the sovereign debt crisis in 2010. \cite{Poncela_etal_2016} use the \cite{asdrubali_channels_1996}'s decomposition to study risk sharing channels across OECD members. Our paper innovates over the previous literature by studying how the adoption of the euro has changed the risk sharing channels of euro area member states. We do so by adopting a new approach as we build a counterfactual for the euro area countries. Some existing papers like \cite{balli_etal_2012},  \cite{furceri_euro_2015}, \cite{Poncela_etal_2016}, and \cite{cimadomo_etal_2020} aim to assess the effect of the euro on risk sharing channels by comparing the channels before and after the adoption of the common currency. While this exercise compares risk sharing channels between two different time periods, it does not allow them to properly assess how the adoption of the euro has changed risk sharing after 1999. Only a counterfactual for euro area countries under the scenario of no adoption of the euro provides an accurate comparison against which to measure how risk sharing has changed. This is precisely the exercise that our paper carries out. Also related to our paper, \cite{Hoffmann_etal_2019} study how the inception of the euro area has affected risk sharing through banking and capital market integration, focusing especially on the Great Recession. Finally, \cite{cimadomo_etal_2020} study how financial integration and official financial assistance have contributed to consumption smoothing in the euro area. 

The rest of the paper is organised as follows. Section \ref{sec: method} describes the SCM, the output variance decomposition and the data. Section \ref{Results} presents our main results concerning both the building of the synthetic dataset and the estimation of how risk sharing channels have changed. Section \ref{sec: robustness} shows our robustness checks. Section \ref{sec: conclusion} concludes.

\section{Methodology}
\label{sec: method}
In this section we first lay out the methodology we use to create counterfactuals in the scenario in which countries had not joined the euro. Next, we discuss the \cite{asdrubali_channels_1996} output variance decomposition and how we estimate it on our treated and counterfactual data.
\subsection{The Synthetic Control Method}
\label{SCM}
This paper aims to assess whether the adoption of the euro has had any effects on risk sharing among member states. To address this question, we need to estimate risk sharing between euro area member states under the alternative scenario in which the currency area had not been established. 
As a real counterfactual for this scenario does not exist, we use the SCM  by \cite{abadie_economic_2003} to generate a synthetic counterpart. The SCM is a data driven procedure that allows us to estimate the effect of a policy intervention in the absence of a natural counterfactual.

Our first step is to generate the synthetic counterpart of the following macroeconomic variables in per capita terms:  gross domestic product (GDP), household final consumption (C), government consumption (G), national income (NI), and disposable national income (DNI). We will need these variables to isolate risk sharing channels as discussed in section \ref{GDP Decomposition}.
To generate the synthetic counterpart of our macroeconomic variables, we proceed as in \cite{abadie_economic_2003}. The idea of the SCM is to create a synthetic counterpart for the macro variables of the countries that are subject to a policy intervention (the euro area countries) by using a convex combination of macro variables of countries that are not subject to the policy intervention (some non euro area countries). More formally, let $N$ be the number of countries in the potential counterfactual pool (some non euro area countries), and let $W=(w_i)_{i=1}^N$ an $N \times1$ vector of country weights such that $\sum_iw_i=1$ for $i=1,...,N$. Let $X_1$ be the $K\times 1$ vector of our variables of interest for {euro area} member states before the introduction of the euro.  Similarly, let $X_0$ be the $K \times N$ matrix  values of the same $K$ variables of interest for all $N$ {non euro area} countries in our counterfactual pool before the introduction of the euro. In addition, let $V$ be a $K \times K$ diagonal matrix with non negative components representing the relevance of our variables of interest in determining the macroeconomic outcome variables.  As discussed in \cite{abadie_economic_2003}, while the choice of the matrix $V$ could be arbitrarily based on economic considerations, we compute it through a factor model.
Then, the algorithm of  \cite{abadie_economic_2003} looks for the vector $W^*$ of weights that minimises
\begin{align*}
 (X_1-X_0W)'V(X_1-X_0W)
 \end{align*}
  subject to 
  \begin{align*}
  w_i\geq0 \text{  and  } \sum_iw_i=1 \text{ for } i=1,...,N.
   \end{align*}
The vector $W^*$ determines the convex combination of macroeconomic variables for non euro area countries, which best reproduces each variable of interest for the euro area countries in the period before the introduction of the euro. Therefore, let $Y_1$ and $Y_0$ be the outcome variables for respectively  the euro area and the non euro area countries. Then, the method uses $Y^*_1=Y_0W^*$ as counterfactual for the outcome variables of euro area countries after the introduction of the euro. 
The choice of the variables in matrix $X_0$ is such that it maximises the ability of the synthetic series to reproduce the  behaviour of the series of the euro area countries in the period before the introduction of the euro. 
For example, to generate the counterfactual series of Portuguese consumption (C) for the scenario in which Portugal had not adopted the euro, the method uses the variables GDP, C, G, NI, DNI and other covariates of the non euro area countries in our sample and it chooses the vector of weights $W$ so as to minimise the distance between Portugal and the control group the macro variables which predict Portuguese C before the introduction of the euro. Given these weights we can generate Portuguese consumption as the convex combination of the consumption series of control group countries. Once we have a synthetic series of Portuguese C that mimics the actual series in the matching period before the euro, we can use that series as a counterfactual for Portuguese C in the scenario where Portugal had not joined the euro in the period after the introduction of the euro.
The matching is carried out for one euro area country at a time, so that the procedure always involves one euro area country and N non-euro area countries.
A relevant assumption for the correct use of the SCM is that the non euro area group is unaffected by the adoption of the euro. This assumption can be troublesome since, given the potential magnitude of the effect of the euro, one might think that its introduction has indirectly affected all countries in the world. This could be especially true for the countries in our non euro area group, which is composed of OECD countries with strong trade and financial linkages with our euro area sample.
This concern is legitimate if we look at the total effect of the introduction of the euro. However, this effect can be thought of as being made of two components: i) the effect of the mere existence of the euro; ii) the effect of having adopted the euro and being a member of the currency union. Under this decomposition, even though all countries in the world are potentially subject to the first effect, only euro area member states are subject to the second one. Hence, the effect that we isolate is the adoption of the euro, conditional on the existence of the euro itself.  

While the literature has discussed ways of evaluating the robustness of the SCM estimates, no analytical result is available to compute their standard deviation. Robustness checks can then be carried out in three possible ways: i)  performing permutation, by randomly resampling the donor pool of non euro area countries \citep[see][]{abadie2021}; ii) estimating a difference-in-differences regression and testing  whether the outcome is significantly different from zero (see \citealp{campos_economic_2014}); iii) running placebo studies on units in the donor pool in order to assess whether the method delivers spurious effect of the adoption of the euro. We use all three approaches to study the robustness of our results. 

The SCM provides an alternative methodology to the before/after and difference-in-differences approaches. By using a selected control group, it allows us to capture potential common trends between pre- and post-treatment periods that are independent of the treatment itself. Missing any common trend between the pre- and post-treatment periods would result in the mis-attribution of these trends to the treatment itself.
%allows us to improve on the before/after and difference-in-differences approaches by having a control group which is selected to best approximate the treated units before treatment. A before/after analysis would miss any common trend between the pre- and post-treatment periods that are independent of the treatment itself. A direct consequence of this would be the mis-attribution of these trends to the treatment effect. 
This could be solved by having a control group of untreated units, instead of a synthetically generated one. As discussed in \cite{abadie2021}, though, there are a number of advantages in using the SCM in settings like ours. First, we are limited by few potential control units for which the required data is available. Secondly, as a consequence of the few control units, we cannot apply standard matching methods that would give us a control group similar to the treated group along relevant observables. This is particularly harmful to the extent that these observables might be key determinants of the outcomes we are interested in and therefore inhibit any causal identification. The use of the SCM helps along both dimensions. In particular, it helps creating a control group that best mimics our treated units pre-treatment, much like a matching estimator would but it does so even in the presence of very few untreated units. Secondly, it prevents extrapolation since it imposes that the weights on control units are non-negative. Third, as discussed in \cite{abadie_synthetic_2010} and \cite{abadie2015}, it is very transparent since it allows us to directly report the weights on each control unit (see Tables \ref{oecd}-\ref{oecd_g} in the Online Appendix). 
 If we were to run a simple regression we would also have weights on each individual unit, which could actually be negative. Lastly, the method allows us to directly assess how suitable our synthetic control group is. In particular, it allows us to directly check whether each observation is statistically different from its synthetic counterpart and report the distribution of these differences. Finally, to complete the comparison with alternative existing methods, note that our method embeds the difference-in-differences and the before/after. In particular, SCM can produce the difference-in-differences or 1-to-1 matching by finding a vector of optimal weights composed of $N-1$ zeros and a single 1 for each treated unit. Similarly, 
 as we estimate our treatment effect of interest through a difference-in-differences on treated and synthetic data,
 if we were to find no difference between the pre and post treatment period for our synthetic data, then we would obtain the same results as in a before-after analysis.

\subsection{Identifying Different Risk Sharing Channels}
\label{GDP Decomposition}
To identify different channels of risk sharing, we follow \cite{asdrubali_channels_1996}. Starting from an output decomposition, we isolate different channels of risk sharing and study how these channels are able to absorb output shocks. 
%The idea of this analysis is to check which of the potential risk sharing channels absorb output shocks.
This is implemented by decomposing GDP into the following national account aggregates: Net National Income (NI), Disposable National Income (DNI), and Private and Government Consumption (C+G). According to this decomposition, GDP can be disaggregated as the following accounting identity:
\begin{equation}
\label{GDP dec}
\text{GDP}=\frac{\text{GDP}}{\text{NI}}\frac{\text{NI}}{\text{DNI}}\frac{\text{DNI}}{\text{DNI+G}}\frac{\text{DNI+G}}{\text{C+G}}\text{(C+G)}.
\end{equation}
Because of the differences in the national account aggregates, the ratios on the right-hand side can be interpreted as specific channels through which risk is absorbed.
The first ratio, $\frac{\text{GDP}}{\text{NI}}$, accounts for income insurance stemming from internationally diversified investment portfolios. This is because NI measures the income (net of depreciation) earned by residents of a country, whether generated on the domestic territory or abroad, while GDP refers to the income generated by production activities on the economic territory of the country. Therefore, the ratio $\frac{\text{GDP}}{\text{NI}}$ captures the {\sl private} insurance channel due to private cross-border investments or, as \cite{kalemliozcan_debt_2014} refer to, holding of claims against the output of other regions. The ratio $\frac{\text{NI}}{\text{DNI}}$, instead, can be interpreted as the {\sl public} insurance channel due to government taxes and transfers as DNI is the income that households are left with after subtracting taxes and adding transfers. Finally, the ratios $\frac{\text{DNI}}{\text{DNI+G}}$ and $\frac{\text{DNI+G}}{\text{C+G}}$ account for smoothing through respectively {\sl public} and {\sl private} saving channels.  

To measure how much of the variation in output is absorbed by each channel, we proceed as in \cite{asdrubali_channels_1996}. We take logs of equation \ref{GDP dec}, difference the series,   multiply by the change of log GDP, and take expectations to get:
\begin{align*}
\text{Var}(\Delta \log GDP_{i,t})&=\text{Cov}(\Delta \log GDP_{i,t},\Delta \log GDP_{i,t}-\Delta \log NI_{i,t})\\
&+\text{Cov}(\Delta \log GDP_{i,t},\Delta \log NI_{i,t}-\Delta \log DNI_{i,t})\\
&+\text{Cov}(\Delta \log GDP_{i,t},\Delta \log DNI_{i,t}-\Delta \log (DNI_{i,t}+G_{i,t}))\\
&+\text{Cov}(\Delta \log GDP_{i,t},\Delta \log (DNI_{i,t}+G_{i,t})-\Delta \log (C_{i,t}+G_{i,t}))\\
&+\text{Cov}(\Delta \log GDP_{i,t},\Delta \log (C_{i,t}+G_{i,t})).
\end{align*}
Dividing both sides by $\text{Var}(\Delta \log GDP_{i,t})$ we get the following identity:
\begin{align*}
1=\beta^m+\beta^g+\beta^p+\beta^s+\beta^u,
\end{align*}
where  we define
\begin{align*}
\beta^m &\equiv \frac{\text{Cov}(\Delta \log GDP_{i,t},\Delta \log GDP_{i,t}-\Delta \log NI_{i,t})}{\text{Var}(\Delta \log GDP_{i,t})}\\
\beta^g &\equiv \frac{\text{Cov}(\Delta \log GDP_{i,t},\Delta \log NI_{i,t}-\Delta \log DNI_{i,t})}{\text{Var}(\Delta \log GDP_{i,t})}\\
\beta^p &\equiv \frac{\text{Cov}(\Delta \log GDP_{i,t},\Delta \log DNI_{i,t}-\Delta \log (DNI_{i,t}+G_{i,t}))}{\text{Var}(\Delta \log GDP_{i,t})}\\
\beta^s &\equiv \frac{\text{Cov}(\Delta \log GDP_{i,t},\Delta \log (DNI_{i,t}+G_{i,t})-\Delta \log (C_{i,t}+G_{i,t}))}{\text{Var}(\Delta \log GDP_{i,t})}\\
\beta^u &\equiv \frac{\text{Cov}(\Delta \log GDP_{i,t},\Delta \log (C_{i,t}+G_{i,t}))}{\text{Var}(\Delta \log GDP_{i,t})}.
\end{align*}
All $\beta$  coefficients can be estimated through the system of equations proposed by \cite{asdrubali_channels_1996}:

\begin{subequations}

\begin{align}
&\Delta \log GDP_{i,t}-\Delta \log NI_{i,t}=\beta^m \Delta \log GDP_{i,t}+\epsilon^m_{i,t} \label{reg1}\\
&\Delta \log NI_{i,t}-\Delta \log DNI_{i,t}=\beta^g \Delta \log GDP_{i,t}+\epsilon^g_{i,t}\\
&\Delta \log DNI_{i,t}-\Delta \log (DNI_{i,t}+G_{i,t})=\beta^p \Delta \log GDP_{i,t}+\epsilon^p_{i,t}\\
&\Delta \log (DNI_{i,t}+G_{i,t})-\Delta \log (C_{i,t}+G_{i,t})=\beta^s \Delta \log GDP_{i,t}+\epsilon^s_{i,t}\\
&\Delta \log (C_{i,t}+G_{i,t})=\beta^u \Delta \log GDP_{i,t}+\epsilon^u_{i,t}, \label{reg5}
\end{align}
\end{subequations}
where each $\beta$ coefficient represents the share of output variation smoothed by a given channel. In particular, $\beta^m$ accounts for the share of GDP variation smoothed by capital markets, $\beta^g$ by fiscal transfers, $\beta^p$ by public savings, and $\beta^s$ by private savings.
What is left, $\beta^u$, is the unsmoothed part of the GDP variation.  A $\beta^u$ equal to zero means that a shock to GDP is fully absorbed through capital markets, fiscal transfers, public and private savings, thus leaving consumption unchanged. Instead, a high $\beta^u$ means that only a minor part of the shock is absorbed through risk sharing, while a significant part stays unsmoothed.

The estimation of coefficients in Equations \ref{reg1} - \ref{reg5} is carried out using ordinary least squares (OLS) with time fixed effects and clustered standard errors. The inclusion of time fixed effects is important as it allows us to take out euro area business cycle fluctuations. In this way, we make sure that the effects that we find are deviations from the euro area business cycle and not fluctuations of the euro area business cycle itself. 
\newline
 \newline We show the results of this estimation as computed in a difference-in-differences setting. We stack together our actual and synthetic samples and include the independent variable interacted with the four possible combinations of actual/synthetic and euro/no euro. In particular, the regressions that we estimate are:
\begin{align}
\label{eq:DiD}
y_{i,t}=\beta_0+ \beta_1x_{i,t}+\beta_2\text{Tr}_i x_{i,t}+\beta_3\text{Eur}_t x_{i,t}+\beta_4(\text{Tr}_i*\text{Eur}_t) x_{i,t}+\nu_t+\epsilon_{i,t},
\end{align}
where $x_i$ is $[\underline{1} \; \Delta \log \text{GDP}_{i,t}]$ and $y_i$ are the dependent variables in Equations \ref{reg1} - \ref{reg5}. $\text{Tr}_i$ is a dummy variable taking the value of 1 if the series comes from the actual dataset and 0 if it comes from the synthetic dataset. $\text{Eur}_t $ is a dummy taking the value of 1 since the adoption of the euro in 1999 and 0 otherwise. $\nu_t$ are time fixed effects. We prove in appendix \ref{proofdecomp} that this is equivalent to estimating the regression in each subsample separately and therefore still constitutes a variance decomposition.
$\beta_2$ represents the additional share of GDP variation smoothed by a given channel for our actual data before the introduction of the euro in deviations from its synthetic counterpart $\beta_1$. If our matching is successful, we should find that $\hat\beta_2$ is not significantly different from zero. $\beta_3$ then represents the change in risk sharing for the control group in the post-treatment period, in deviation from $\beta_1$.
For our euro period, i.e. for the period in which $\text{Eur}_t=1$, we should  study $\hat\beta_4$. If the euro has had an effect on the analysed risk sharing channel, $\hat\beta_4$ should be significantly different from zero. Note that we estimate Equation \ref{eq:DiD} for each of the dependant variables in \ref{reg1} - \ref{reg5}. This means that we will have five difference-in-differences estimations, which result in a variance decomposition. Since we started from an accounting identity (\ref{GDP dec}), all the estimated coefficients of the five equations sum up to one. Finally, note that if our empirical model was misspecified and the best possible approach was a before/after analysis, then we would have to estimate $\hat\beta_3=0$, suggesting that there were no changes for the control group between the pre- and post-treatment periods.

\subsection{Data}
Our data is taken from the OECD National Account Statistics. 
In particular, we use  household final consumption expenditure for C, general government consumption for G, gross domestic product computed following the output approach for GDP, net national income for NI, and net disposable income for DNI.
Our dataset has an annual frequency, spans the sample period 1990-2018, and includes 24 countries. Of these, 11 are euro area member states: Austria, Belgium, Finland, France, Germany, Greece, Ireland, Italy, Netherlands, Portugal, Spain; 13 are OECD countries not in the euro area: Australia, Canada, Denmark, Israel, Japan, Korea, Mexico, New Zealand, Norway, Sweden, Switzerland, UK, and US. As the SCM requires the data not to display any missing values, we cannot extend the sample along the time or country dimension any further.

\section{Results}
\label{Results}
We start our analysis by introducing two benchmark estimations. The point of departure is given by the simple comparison of the euro area countries risk sharing before and after the introduction of the common currency. Next, we estimate the same channels using non euro area countries as a control group. Lastly, we create synthetic national accounts for our treated countries in the absence of the euro. We discuss this methodology in the next subsection and then return to the comparison between these approaches.

\subsection{Before/After and Difference-in-Differences}
We start by simply comparing the euro area countries risk sharing decomposition before and after the adoption of the common currency.
Formally, we estimate the following regression on a sample comprised only of euro area countries:
\begin{align}
\label{eq:before_after}
y_{i,t}=\gamma_0+ \gamma_1x_{i,t}+\gamma_2\text{Eur}_t x_{i,t}+\chi_t+\varepsilon_{i,t}.
\end{align}
In this setting $\gamma_1$ represents the level of risk sharing before the introduction of the euro and $\gamma_2$ the change after the common currency. $\chi_t$ is a set of time fixed effects. The results are reported in Table \ref{before_after_oecd}. As discussed in the methodology, these estimates are not causal since they do not allow us to separate i) general trends which would have occurred in absence of the treatment from ii) the treatment effect itself. 

The before and after analysis would suggest that in the euro period, the level of consumption smoothing of EA member states remained unchanged, relative to the 1990s. This seems driven by the net effect of three different forces: i) a large increase in the contribution of capital markets to consumption smoothing; ii) a decrease in the contribution of public savings, and iii) a large decrease in smoothing via private savings. 

If we were to stop here, we would conclude that the euro did not change the level of consumption smoothing, mostly by reshuffling the channels through which fluctuations are absorbed. 

Building towards our main estimation we then include a control group and estimate a standard difference-in-differences. Our control group is composed of non euro area OECD countries. Table \ref{dd} reports the results. First, we note that our treatment and control group display different levels of risk sharing in the pre-treatment period. This, per se, does not invalidate the estimation, but suggests that the control group might not be very similar to our treated group. Nonetheless, for the purpose of benchmarking our SCM estimation, we discuss the results of the difference-in-differences. We find that before the introduction of the euro, euro area countries had significantly higher consumption smoothing, relative to the control group as summarized by the negative coefficient in the second row for the unsmoothed component. This was driven by (albeit not statistically significant) higher smoothing through capital markets and private savings. In the post treatment period the control group countries increased their consumption smoothing mostly through improvement in risk sharing through capital markets and private savings. At the same time the euro area member states saw a relative reduction in the smoothed component of fluctuations driven by a significantly lower absorption through the private savings channel.  The key takeaway is that while the control group countries were improving their consumption smoothing, the euro area member states experienced virtually no change relative to the pre-treatment period, which, in relative terms, represents a worsening of risk sharing. 

We now move to our main specification based on our synthetic national accounts.  
\subsection{Generating Synthetic National Account Aggregates} 

We start by generating the synthetic national account aggregates that will be used as counterfactual series for the period after the introduction of the euro. 
As discussed in section \ref{SCM}, the SCM algorithm produces the vector $W=(w_1,...,w_N)$ of weights that maximise the matching between the actual and the synthetic series before the adoption of the euro. Tables \ref{oecd} to \ref{oecd_g} display the optimal weights that generate the synthetic GDP, C, DNI, NI, and G. For example, the Finnish synthetic GDP is best reproduced by a vector of Mexican, Swedish, and British macro variables in the percentages of 14.3, 72.7, and 13.0 respectively. This is the convex combination of non euro area countries' GDP, which best matches the Finnish GDP before the adoption of the euro. 
As an example of our matching, Figure \ref{fig:matching_GDP} displays the actual and the synthetic series of GDP for all euro area countries in our sample.\footnote{For completeness, Figures \ref{fig:matching_C}-\ref{fig:matching_G} also report the actual and synthetic series for consumption, net national income, disposable national income, and government consumption.}
The two series are very close in the matching period 1990-1998 and diverge in the post euro period.  While Austria, Belgium, Finland, Germany and the Netherlands have benefited from the euro for all the years subsequent to its adoption, the picture is less clear-cut for other countries.  Greece has seen a greater GDP growth from the adoption of the euro up to the beginning of the Great Recession, and a lower GDP growth afterwards. Italy has never benefited from adopting the euro in terms of GDP growth. Ireland has benefited before the Great Recession and after 2015. 

 \subsection{Testing the Parallel Trend Assumption}
For the validity of our analysis, it is important that we have a high quality matching and that the actual and the synthetic series are close to each other over the matching window. Furthermore, to be able to estimate a causal effect through our difference-in-differences we need the usual parallel counterfactual trend assumption. As we highlight later when presenting our results, and as is shown in Figures \ref{oecd}-\ref{oecd_g}, we find no significant differences between our treated and synthetic data before treatment. We also provide a formal test for pre-trends. Formally, we test that the treatment and control group do not have different trends and different interactions between trends and $\Delta \log GDP$ in the pre-treatment period. Table \ref{parallel_trend} reports the p-values of the test for differences.\footnote{Formally, for each variable $y$ on the left hand side of \ref{reg1} - \ref{reg5} we estimate
 \begin{align}\label{reg_parallel}
     y_{i,t}=\beta_0+\beta_1 \text{year}_t +\beta_2 \text{year}_t \text{ Treated}_i+ \beta_3 \Delta \log GDP_{i,t}+\beta_4 \text{year}_t \Delta \log GDP_{i,t}+\beta_5 \text{year}_t \text{ Treated}_i \Delta \log GDP_{i,t}+\varepsilon_{i,t},
\end{align}
only on the pre-treatment period.
Our coefficients of interest are $\beta_2$ and $\beta_5$ as they capture the differential trends between treated and control group.  They are reported in Table \ref{parallel_trend}.} Reassuringly we observe that there is no significant difference between treatment and control. We also report these results separately for the core and periphery countries, reaching the same conclusions. Note that this is a byproduct of our matching procedure, albeit not a mechanical one, since we targeted our macro variables in levels, rather than in differences.\footnote{We provide an alternative matching which includes first-differenced variables in Section \ref{sec: robustness}.}
 
Finally, to further alleviate the concern that our matching on levels would deliver synthetic counterfactuals with different properties relative to our treated pool, Figure \ref{fig:beta_distrib} reports the level differences in risk-sharing in the pre-treatment period country by country. Formally, we estimate equation \ref{eq:DiD} separately for each country $i$ and all our variables in the decomposition and report the distribution of $\hat{\beta}_{2,i}$. The left panel of Figure \ref{fig:beta_distrib} shows the distribution of $\hat{\beta}_{2,i}$, while the right panel shows the distribution of p-values. The main takeaway is that the estimates are bunched around zero and we never reject the null hypothesis that they are zero. That is, in the pre-treatment period, we find no significant differential risk-sharing between treated and synthetic counterfactuals not only in the cross-sectional average, but also country by country.
 
\subsection{Risk Sharing Channels}\label{sec: main_results}

Table \ref{ols} - \ref{periphery } display the results of our difference-in-differences estimation as in Equation \ref{eq:DiD} for Equations \ref{reg1} - \ref{reg5}. Estimates are carried out using OLS with country clustered standard errors using actual and synthetic national account series over the full sample period 1990-2018. 
In each table, results should be read as follows. Every column corresponds to the difference-in-differences estimation as in Equation \ref{eq:DiD} of each Equation \ref{reg1} - \ref{reg5}. For example, the first column is the difference-in-differences estimation of Equation \ref{reg1} and reports coefficients $\beta_1$ to $\beta_4$ of Equation \ref{eq:DiD}.\footnote{While usually only the estimated difference-in-differences coefficient is reported ($\beta_4$ in Equation \ref{eq:DiD}), here we report also coefficients $\beta_1$ to $\beta_3$. As a matter of fact, the estimated coefficient $\beta_1$ allows us to compare our estimates to the existing results in the literature, while the estimated coefficient $\beta_2$ to test the quality of our match.} $\beta_1$ is the portion of GDP variation absorbed through capital markets before the adoption of the euro (Pre euro) and computed with the synthetic data (Synthetic).  $\beta_2$ is the portion of GDP variation absorbed through capital markets before the adoption of the euro (Pre euro) and computed with the actual data (Actual) in deviations from the portion computed with its synthetic counterpart.  $\beta_3$ is the portion of GDP variation absorbed through capital markets after the adoption of the euro (Post euro) and computed with the synthetic data (Synthetic) in deviations from the portion computed before the adoption of the euro. $\beta_4$ is the portion of GDP variation absorbed through capital markets after the adoption of the euro (Post euro) and computed with the actual data (Actual) in deviations from the portions computed before. Coefficients reported in the other columns have a similar interpretation, but refer to different channels of risk sharing. 
Notice that since we started from an accounting identity, the estimated coefficients of each table add up to one. 
\newline
\newline Table \ref{ols} shows estimates for the full sample of countries and displays three main results. First, the row of coefficients {\sl Pre euro Synthetic} reports the risk sharing channel decomposition for the pre euro period, 1990-1998. 
There is no purely international risk sharing, that is shock absorption through  capital markets or international transfers. Instead, shock absorption happens through public and private savings, which smooth respectively 13\% and 40\% of GDP shocks. The unsmoothed portion of risk is  46\%.
\newline
\newline The second result displayed by Table \ref{ols} concerns the quality of our matching in the pre euro period and shown by the row of coefficients {\sl Pre euro Actual}. These coefficients capture the portion of GDP variation absorbed through the different channels before the adoption of the euro and computed with the actual data in deviations from the portion computed with its synthetic counterpart. None of these coefficients is significant. As we computed our synthetic series by minimising the distance from the actual series in the pre euro period, the synthetic and the actual series should be very close to each other in the pre euro period. This should also imply that the regressions run by using either the actual or the synthetic series should give very similar results. This implication is tested by finding that the channels of absorption computed by using the actual data are never significantly different from those computed by using the synthetic data for the pre euro period. 
\newline
\newline Our third and main result concerns consumption smoothing after 1998. The row of coefficients {\sl Post euro Synthetic} shows the risk sharing channel decomposition for the synthetic data, that is for a counterfactual scenario in which euro area countries had not adopted the euro. As shown by the last coefficient, the unsmoothed component of the shock would have decreased by 24\% in the period after 1998. However, as reported by the row of coefficients {\sl Post euro Actual}, the unsmoothed component has been higher because of the adoption of the euro. In particular, the adoption of the euro has increased the unsmoothed component of GDP variation by 22\%. In summary, we find that the before and after result that between the pre and post euro periods the level of consumption smoothing did not significantly change is actually masking important variation. In particular, our results suggest that absent the introduction of the common currency, the smoothed fraction of output variations should have significantly increased but that the euro prevented this from happening.
\newline
\newline 
As changes in the shock absorption due to the adoption of the euro might have been heterogeneous across countries, we carry out the same difference-in-differences estimation as in Table \ref{ols} on two subsamples of countries, core and periphery. Following \cite{baldwin2015eurozone}, countries included as core are: Austria, Belgium, Finland, France, Germany, and the Netherlands. Countries included as  periphery are: Greece, Ireland, Italy, Portugal, and Spain. Tables \ref{core } and \ref{periphery } report the results for the two subsamples. Before the adoption of the euro, both subsamples of countries shared risk mainly through public and private savings, and core countries in an even greater proportion -- core countries shared 19\% and 46\% of the shocks through private and public savings versus a share of 11\% and 40\% for the periphery countries. Core countries appeared to share 8\% of the shocks through international transfers, while capital markets were only able to amplify the shocks. To the contrary, periphery countries did not absorb any portion of the risk through capital markets or international transfers before the adoption of the euro. Interestingly, the two groups had very similar levels of consumption smoothing before the common currency, with approximately 45\% of income fluctuations being transferred to consumption.
\newline
\newline The main heterogeneity between the two subsamples, though, concerns the period after 1998. As shown by the row {\sl Post euro Synthetic}, in a counterfactual world core countries would have had a substantial increase in shock absorption through capital markets, and a decrease through international transfers and public savings. However, the adoption of the euro has reduced the increase in smoothing through capital markets compared to the counterfactual world. Moreover, no change has happened to the unsmoothed component of the shock after 1998. To the contrary, the reduction in the smoothing that we found for the full sample of countries is driven by the subsample of periphery countries. In the counterfactual scenario, these countries would have increased their risk sharing through private savings. However, this increase in risk sharing through private savings has been undone by the adoption of the euro. As shown by the row {\sl Post euro Actual}, the unsmoothed component of the shock has increased by 34\%, mainly due to a decreased absorption through private savings. 
\newline
\newline  Our analysis shows the importance of having an accurate counterfactual to draw the right conclusions. Our results are in line with other studies once we make the appropriate comparison with the before/after euro estimates in the existing literature. For example, our estimates of how risk sharing channels changed with the euro adoption are consistent with those estimated by \cite{furceri_euro_2015}.\footnote{The relevant comparison is between the sum of our estimated coefficients in the post-euro rows in Table \ref{nocrisis} and the difference of the columns post-EMU and pre-EMU in Table 13 of \cite{furceri_euro_2015}. For example, we estimate an increase in the unsmoothed component of 17\%, which is comparable to their estimated increase of 18\%. Results in table \ref{nocrisis} will be discussed in greater detail in the next section. Yet, they are the relevant benchmark to compare to \cite{furceri_euro_2015} since they are estimated on the sample up to the financial crisis, which is closer to the 1999-2010 time span of \cite{furceri_euro_2015}.} Similarly, our estimates of risk sharing through capital markets are comparable to \cite{balli_etal_2012}.\footnote{In this case,  the relevant comparison is between \cite{balli_etal_2012}'s estimates of $\beta_f+\beta_k$ in Table 5 and our estimates of capital markets in Table \ref{nocrisis}.} In particular, we both find that there has been an increase in risk sharing through capital markets in the post-euro period. The larger difference between our baseline estimate of capital market channel (Table \ref{ols}) and \cite{balli_etal_2012} comes from including the post-crisis period up to 2018. Our relatively high coefficient of capital market smoothing in the full sample estimation (1990-2018) is actually in line with the results of \cite{cimadomo_etal_2020}, who find that there has been a higher degree of financial integration after the financial crisis, as reflected in particular in cross-border portfolio holdings of corporate and government bonds.

Our method allows us to break down these effects further. Namely, it is possible to observe that euro area countries and control group countries had markedly different behaviors in the post-treatment period. As Table \ref{core } shows, while the synthetic control for the core countries has seen an increase in smoothing through capital markets after the euro, the treated group has faced a lowering in capital market smoothing compared to the synthetic control. At this point it is important to recall that, given our successful match, we can interpret the results for the synthetic counterfactual as what would have happened to euro area countries absent the common currency. In this sense we can interpret the increase in the unsmoothed component for the treated as a reduction in consumption smoothing due to the introduction of the common currency. Absent the appropriate counterfactual we would not be able to draw this conclusion, as we would confound the effects of the introduction of the euro (treatment) with what would have happened in the post-treatment period with national currencies. In particular, we would only be able to attribute the effect estimated in the before/after analysis to the introduction of the euro under the strong assumption that, absent the treatment, nothing would have changed in these countries. Our estimates suggest, on the contrary, that absent the common currency these countries would have had a higher level of consumption smoothing, particularly through higher risk-sharing via capital markets. As a consequence, we conclude that the introduction of the euro prevented these developments and therefore induced a reduction in consumption smoothing.
\newline
\newline In summary, the adoption of the euro has decreased consumption smoothing for member states. This is mainly due to a lower absorption through private savings in the periphery countries.

\subsection{Exclusion of the Period from the Great Recession}
\label{sec:nocrisis}
It could be that the effect of the euro adoption on the shock absorption has changed throughout the twenty years after the adoption of the common currency. To check whether the baseline results might be different if we only considered the first ten years of the euro, instead of also including the Great Recession and the period thereafter, we rerun our estimations including only the years up to 2007. 
Table \ref{nocrisis} shows our estimates for the full sample of countries over the period 1990-2007. Our baseline results are confirmed also for this subsample.\footnote{Our estimated changes in the risk sharing channels are in line with the literature and can be compared, for example, to Table 13 in \cite{furceri_euro_2015}.}  
In fact, the coefficient for the unsmoothed component grows even further. In particular, we find that 30\% of an idiosyncratic variation to GDP remains unsmoothed against 22\% found for the full sample period. This suggests that consumption smoothing was actually higher during the Great Recession. Table \ref{nocrisis_core} and   \ref{nocrisis_periphery} show the estimated coefficients over the period 1990-2007 for the country subsamples of core and periphery and again the results of the full sample estimates are confirmed. While the coefficient for the unsmoothed component is not significant for the core countries, it is positive and significant for the periphery countries (42\%). We interpret this as evidence that the increase in the unsmoothed component for the full sample is driven by the periphery countries. In particular, the decrease in consumption smoothing is mainly due to a reduction in their shock absorption through the private savings channel (-42\%) already in the first ten years of the euro adoption.

\subsection{Discussion}

Before highlighting possible mechanisms underlying our results, we briefly discuss the potential caveats and limitations of our approach.

First, it is clear that the adoption of the common currency is not a random event and that countries self-select into treatment. As in all quasi-experimental settings, this issue is at best only partially solved by controlling for covariates that are informative towards selection. Our methodological approach does so by creating a counterfactual unit which is as close as possible to the treatment unit on observable characteristics. We are still not shielded by important unobserved characteristics determining selection into treatment. It is however important to remark that the same criticism applies a fortiori to a difference-in-differences approach in which the control group is not matched or selected. The reason is that, by making treatment and control as close as possible, our approach is likely to reduce the selection bias relative to a control group that might have self-selected out of the treatment. 

Secondly, our method might confound with treatment the effect of idiosyncratic shocks occurring to our treated units at the same time of treatment. Importantly, as we control for aggregate fluctuations via time fixed effects, these potentially confounding idiosyncratic shocks would need to hit all our treated units and none of our control units in the same period of the euro adoption. While still possible, we regard this instance as unlikely.

Thirdly, our approach is subject to potential criticism stemming from the selection of control group countries. This concern would generally apply to all quasi-experimental methods such as a difference-in-differences estimation. Nonetheless we attempt at mitigating the concern by reporting the results when we use a different donor pool. In particular we use a much larger set of countries from the World Bank. This comes at the cost of a shortened decomposition as not all the necessary national accounts series are available in this data source. Table \ref{worldbank} reports the result for the last regression of the decomposition to compare the headline result of the reduction in consumption smoothing. First, note that this analysis broadly confirms, at least qualitatively, our finding. The first observation is that the before and after analysis delivers the same result. This is unsurprising since there is no control group and so the regression is estimated on the same countries and data, albeit from a different source. Next, we note that when we compare the simple difference-in-differences results between the OECD and World Bank samples they are markedly different. Here we find that in the post period neither the control nor the treated group saw any change in the level of consumption smoothing. Lastly, when we apply the SCM to create our counterfactual we confirm our headline result: the level of consumption smoothing would have increased absent the euro but the introduction of the common currency prevented this. Quantitatively, our effect is even larger than the one estimated on OECD data. This result suggests that the SCM approach significantly reduces the potential for bias induced by our selection of the control group, relative to alternative methods. 

Lastly, we conclude with a short discussion on inference in our estimation. To evaluate the significance of our estimates we follow \cite{abadie2021} and use permutations to get a distribution of treatment effects. The idea is that we can re-assign treatment at random, perform our synthetic control to generate the counterfactuals for these units and re-estimate our coefficient of interest. Since our original treated group included 11 units we re-assign treatment to 11 random units 1500 times. We use re-assignment with replacement. For each permutation we generate the synthetic series for GDP, government consumption and private consumption, so that we can estimate the treatment effect on each unsmoothed component. For each of the 1500 re-assignments we estimate equation \ref{reg5}. As in  \cite{abadie_synthetic_2010} and \cite{abadie2021} we use the following statistic for our inference procedure. First, we build the difference between treatment and control pre-treatment in our coefficient of interest, which in the notation of equation \ref{eq:DiD} is given by $\hat\beta_2$. Next we take our difference in the post-treatment, given by $\hat\beta_4$. Following the literature we build the statistic $r\equiv |\hat\beta_4/\hat\beta_2|$, which measures the ratio between the  post and pre treatment difference between treated and control groups. This statistic accounts for the fact that a large effect in $\hat\beta_4$ can be spurious if the pre-treatment fit is poor as measured by a large $\hat\beta_2$. We then have a distribution of $r_i$  where $i$ denotes the re-assignment. The p-value of our estimated treatment is then given by the mass to the right of of $r_{EURO}$, which measures the effect in our true assignment process. Intuitively, this measures how likely it is to estimate a larger effect than our true one if we were to assign treatment randomly in our sample. Formally, the p-value is defined as 
\begin{align*}
    p\equiv\frac{1}{N} \sum_{i=1}^N I_+(r_i-r_{EURO}),
\end{align*}
where, $I_+$ is an indicator function that returns one for non-negative arguments and zero otherwise.

Figure \ref{permutation} plots the kernel density of the distribution of $\log(r_i)$. We take a log transform because some values are extremely large, which would make it impossible to see the distribution.\footnote{This is actually due to our match being successful, which in turn implies that many of our estimated $\hat\beta_2$ are very close to zero and therefore $r_i$ can be extremely large.} Clearly this does not affect the ordering and therefore our p-value calculation. As the figure clarifies our original estimate is an extreme value in this distribution, suggesting that the likelihood that this was the result of noise is small. Formally, the empirical p-value, defined as the mass to the right of our point estimate in the estimated treatment effect distribution, is 4\%. 

\subsection{Why Has the Adoption of the Euro Prompted a Lower Consumption Smoothing?}\label{mechanism}
 As shown in section \ref{sec: main_results}, the adoption of the euro has decreased consumption smoothing for member states. This effect is mainly due to lower shock absorption through private savings in the periphery countries. This section argues that the lower consumption smoothing may be attributed to the higher GDP growth and volatility, which have followed the adoption of the euro. 
Panel a) of Figure \ref{fig:growth} compares the cross-country average of actual per capita GDP (blue solid line) to its synthetic counterpart (red dashed line). The actual GDP is higher than its synthetic counterpart, showing that the adoption of the euro has prompted a boost in GDP growth. A similar message is conveyed by Panel b) of Figure \ref{fig:growth}, which shows the cross-country average growth rate of the actual GDP over its synthetic counterpart (red line), along with the central 80 percentiles (blue area). From the adoption of the euro to the beginning of the Great Recession GDP has increased significantly more than it would have without the euro. 
  \newline
\newline Our empirical results are compatible with the idea that the euro has prompted a higher GDP growth in both the trend and the cycle. %While the literature has generally focused on either one or the other, it may well be that the adoption of the euro has affected growth along both dimensions. 

We check whether there has been a change in the {\sl cyclical} growth of GDP. We  thus detrend GDP with a linear quadratic trend. Then, relying on the same difference-in-differences setup that we used in section \ref{sec: main_results}, we regress GDP growth and logged GDP variance on a set of dummies spanning the possible combinations of pre euro/post euro and euro/no euro. Table \ref{growth_did} shows that, compared to a scenario without the euro, the adoption of the common currency has increased both the growth and the variance of the cyclical portion of GDP of member states.

While we have brought empirical evidence that there has been a change in GDP growth and volatility at business cycle frequency, our results are also compatible with the narrative that big policy regime switches, such as the adoption of a common currency, may generate changes in the GDP {\sl trend}. This is the narrative put forward by models along the lines of \cite{aguiar_gopinath} and with which our empirical results are consistent.

\section{Robustness Checks}
\label{sec: robustness}
In this section we provide a battery of robustness checks for our main results. We report multiple alternative specifications of our matching strategy, definitions of treatment period, definitions of control group, and specification of the estimated regressions. 
\subsection{Panel Correlated Standard Errors}
We check the robustness of our baseline results to the way standard errors are computed, by computing panel correlated instead of clustered standard errors, following \cite{furceri_euro_2015}. 
Table \ref{pcse1} and Table \ref{pcse2} display OLS estimates with panel correlated standard errors for the full sample of countries, as well as the core and periphery subsamples, both for the full time period and for the period up to 2007.\footnote{For brevity we only report results for the cases {\sl Pre euro Synthetic} and  {\sl Post euro Actual}, but the full set of coefficients is available upon request.} Our baseline results are robust. The significance of the public and private savings channels in the pre euro period is confirmed. Moreover, the adoption of the euro has decreased the consumption smoothing and the full sample result is driven by the periphery countries, which substantially reduced their shock absorption through private savings. 

\subsection{Matching Including Natural Resources}
\label{sec:nat}
To generate our baseline synthetic control group, we have matched on log per capita GDP, consumption, government consumption, national income, and disposable national income. It could well be that other features of the economies could help generating a better synthetic control group. For example, it could be that some countries are richer in raw materials than others. To address this concern, we generate a new synthetic control group where the matching is done using also the total natural resources rents (\% of GDP). Table \ref{nat_resources} reports the results. 

\subsection{Matching on First Differences}
Our main results are based on a difference-in-differences estimation, which requires the parallel trend assumption to be satisfied. As in our setup the control variables are generated through the SCM, the fulfilment of the parallel trend assumption should be straightforward. As a matter of fact, the SCM method aims to minimise the distance between actual and synthetic series in the pre-treatment period. Nonetheless, while we produce our matching on variables that are in levels, we run the difference-in-differences estimation on first-differenced data. 
In fact, even though our matching in levels is such that the dynamics of the synthetic series are very close to the ones of the actual series, this is not enough to ensure that the first-differenced data have the same trend.

 To address this potential issue, we redo the matching by using covariates and outcomes that are first-differenced, while still maintaining among the predictors pre euro averages in levels. In other words, when using covariates $X_0$, we actually match on $\{\Delta X_{0,t}\}_{t=0}^{T^*-1}$ and $\overline{X}_0$, where the barred variables stand for pre euro period averages. The reason for this matching strategy is that we want to replicate as closely as possible the first differenced data, hence the matching on $\Delta X_0$. The drawback of this methodology is that, by replicating the first-differenced data, we may find some countries with similar year-to-year changes, but very distant fundamentals from the actual series, to be an excellent match. To shield against this possibility, we keep some predictors in levels and match with a relatively homogeneous non-euro area group, namely OECD countries.

The results of this estimation are displayed in Table \ref{firstdiff1} and Table \ref{firstdiff2}, respectively for the full time span and for the sample up to the Great Recession. 
Also under this matching strategy, our baseline results are confirmed. The adoption of the euro has decreased consumption smoothing, especially for periphery countries. This reduction is due to a lower shock absorption through private savings. 

\subsection{Changing the Matching Window}
For the validity of our analysis, we need to make sure that our matching is sound and does not crucially depend on the matching window. As we cannot extend our matching window back in the past, we can nevertheless shorten it. We therefore regenerate our synthetic series by running the matching over the period 1990-1995, in place of the baseline period 1990-1998.
\newline
\newline Table \ref{ols1995} shows the difference-in-differences estimations run with the new synthetic series. We confirm our result that the adoption of the euro has  not changed consumption smoothing of core countries, but it has decreased that of periphery countries. This decrease is mainly due to a lower shock absorption through private savings. As displayed by  Table \ref{ols1995_nocrisis}, a similar picture emerges also if we exclude the period from the Great Recession onwards.

\subsection{Accounting Identity and Country Fixed Weights for the Synthetic Variables}
\label{fixed_weights}

In our baseline analysis, we allow each country's macro variable to be generated using different weights. For example, we allow Austrian synthetic consumption and Austrian synthetic GDP to be generated by different combinations of donor countries. Such an approach buys us more flexibility in the estimation, but comes at the cost that some national accounts identities could be violated. Furthermore, an additional concern is that as a consequence we may mechanically measure a lower level of risk sharing as the macro aggregates are not constrained by the accounting relationships between them. 

To address this, first in Figure \ref{identity} we report the comparison between synthetic GDP and a version of GDP computed as the sum of synthetic components. As a benchmark, if the accounting identity were to hold at all times the two would be exactly identical. The average percentage distance between the two versions of synthetic GDP is 0.1\% with a standard deviation of 3.6\%. This result reassures us that the interpretation of the risk sharing channel decomposition is tenable given that the accounting identity is satisfied. Nonetheless, we provide two additional robustness tests in which we impose the identity and check that our results obtain.

To test the robustness of our results to this approach we estimate our counterfactual with only one set of weights per country. More specifically, we combine the optimal weights of each variable into a single set of weights. More precisely, for $i=GDP,C,G,DNI,NI$ denote $W^*_i$ the optimal weight for each individual variable in our decomposition. We create $\tilde W\equiv 1/5\sum_i W^*_i$. We weigh them equally 1/5 as this is equivalent to targeting all variables at the same time. 
We then create our synthetic variables for the decomposition using the $\tilde W$ matrix. This mechanically insures that the accounting identity is satisfied in our counterfactual data. 

With this synthetic dataset of variables with country fixed weights, we rerun our difference-in-differences regressions. Estimations are displayed in Table \ref{fw1} and \ref{fw2}. We find that before the adoption of the euro, most part of income shocks remains unsmoothed, both for core and periphery countries. Our baseline results are confirmed both for the full sample and for that up to the Great Recession.
%While we do not find any change in the unsmoothed component of the shock due to the adoption of the euro for the full sample period, we do find a further significant increase, both for  core and periphery countries, for the sample period up to the Great Recession. 

\subsection{Placebo Test}
A standard check to evaluate the robustness of an estimated treatment effect are placebo tests. In our framework, this involves matching macro variables for non euro area countries, which have never adopted the euro, as if they had adopted it. For example, we try to find the best match for a country like the US, which has never adopted the euro, as a convex combination of other countries that have never done so. If we were to find any effect of the adoption of the euro on countries that have never adopted it, then it would be possible that our euro effect picks up some spurious correlation.

After building a synthetic dataset for all OECD non euro area countries, we run the same risk sharing decomposition that we carried out for the euro area countries. The results of this estimation are displayed in Table \ref{placebo1}. All our difference-in-differences estimators are never significant, meaning that we find no effect of the adoption of the euro on our non euro area group.

\subsection{Different Year of Adoption of the Euro}
One of the identifying assumptions of the SCM is that the covariates on which the matching is carried out are not affected by the adoption of the euro. If this assumption is violated, the matrix of weights may be biased by matching on series that already incorporate the effect of the adoption of the euro. It is possible that some effect of the introduction of the euro has materialized between the announcement and the actual introduction of the physical currency. In this sense, our approach is already conservative as it uses 1999 as year of adoption of the euro. This year corresponds to the introduction of the euro as an accounting currency, while  physical euro coins and banknotes started circulating only in 2002.
\newline
\newline 
As evidence of anticipation effects has been found -- see \cite{frankel} for an application to trade --, we run our analysis again using 1998 as the year of adoption.\footnote{Anticipation effects might be at play even before 1998. However, given the length of our matching window, which only starts in 1990, moving the treatment year back in time would come at a cost of further shortening our matching span.} The results of this estimation are displayed in Table \ref{anticipation1} (full sample) and \ref{anticipation2} (up to the Great Recession). Our estimates are in line with our baseline results. We find that the adoption of the euro has increased the unsmoothed component of GDP variation, and this result is driven by periphery countries, which reduce their consumption smoothing through private saving. This effect is even stronger for the sample period up to the Great Recession.

\subsection{Excluding EU members from the Donor Pool}
An additional robustness check that we carry out is to exclude EU countries, which are not part of the euro area, from the group of countries used to generate synthetic variables. The rationale for this check is that countries geographically in Europe may have endogenously decided not to join the common currency, as the UK, or simply be indirectly affected by the existence of the euro. For this reason, we exclude them from our non euro area group.
\newline
\newline
Table \ref{noeu1} and \ref{noeu2} show our estimates of the risk sharing channel decomposition run with this different dataset of synthetic variables. We confirm our main result that the adoption of the euro has decreased the level of consumption smoothing of euro area countries. This  result is driven by periphery countries, which lowered their shock absorption through private savings.

\subsection{Excluding Recessionary Periods}
\label{sec:no_rec}
To double check whether the reduction in consumption smoothing is happening mainly during recessionary periods or also during expansions, we follow the EABCN recessionary dates and exclude the years with at least three quarters of recession -- these are 2008 and 2012.\footnote{Recessionary dates can be found at \url{https://eabcn.org/dc/recession-indicators}.} Table \ref{recessions} shows that also when excluding recessions, we find that the euro adoption has prompted a reduction in consumption smoothing. 

\subsection{Differential Trends}
We argued above that we observe two potentially distinct effects of the euro on risk sharing. First, we see an increase in the unsmoothed share of fluctuations; second, we see an increase in the growth rate of GDP for euro area member states. Our main specification includes a time fixed effect to absorb an aggregate trend, while not allowing for differential trends across the euro adopters and the synthetic group. Thus, the treatment effect that we estimate on risk sharing is a combination of changes in cyclical variations and changes in trends for the treated group relative to the counterfactual.
To check whether accounting for differential trends significantly alters our findings, we re-estimate equation \ref{eq:DiD} allowing for different time fixed effects for treated and control units. Note that this specification is equivalent to estimating our main regression separately for the treated and control group. In this sense we are now only using within group cross-sectional variation to compute our estimates of risk-sharing. Note however that this does not imply that we are only estimating within group risk-sharing.
 Table \ref{group_time_fe} reports the results. First, as in our main specification the pre-treatment effect is identical in the two groups, suggesting that our matching is successful and there are no differences due to different pre-treatment trends. In addition, the channels of risk sharing in the pre-treatment period are of comparable magnitude to our baseline specification. Second, when allowing for differential trends across the two groups, we find that the synthetic group faces a decrease in risk sharing due to lower international transfers and a decrease in the unsmoothed component, while at the same time facing an increase in risk sharing thorough private savings. Compared to the synthetic group, we find that the euro adopters saw an increase in smoothing thorough capital markets, a decrease through private savings, and an almost significant increase in the unsmoothed share of fluctuations.\footnote{Note that, when allowing for differential trends across the two groups, we have to estimate twice as many fixed effect coefficients, which significantly reduces our power to precisely estimate the effect on the smoothing coefficient.} 
%Two observations are worth noticing. First, estimating separate fixed effects for the two groups implies estimating risk sharing only within groups. To the contrary, estimating common fixed effects let us estimate risk sharing both within and across groups. So, while both interesting, the two specifications answer two different questions concerning how risk sharing has changed. 
%It is important to note that the baseline result includes fixed effects that are common to the treated and the synthetic group. This implies that the effects that we estimate on risk sharing are a combination of changes in cyclical variations and changes in trends for the treated group relative to the counterfactual. 
The finding that, when allowing for differential time fixed effects for the two groups, the euro effect on the unsmoothed component weakens, is thus in line with our interpretation of the results that we discussed in section \ref{mechanism}, that is that there is a component of the estimated change in risk sharing due to changes in trends. 
%The baseline effect that we estimate in the unsmoothed component is consistent with the interpretation that the euro has triggered a change in the trends in addition to a change in risk sharing. 
In summary, our results are consistent with the notion that the total effect of the introduction of the euro on risk sharing is a combination of changes in the cyclical variation and changes in the general trends, relative to the counterfactual. 

\subsection{Measurement Error}
\label{sec:me}
A legitimate concern regarding our methodology is that by using generated data through the SCM we may be including measurement error in our estimation. More precisely, by estimating our counterfactual, we may generate our data with some statistical error. This error is observationally equivalent to measurement error in our dependent and independent variable. The former is more troublesome due to the standard result that the presence of measurement error in the regressors implies a bias in the estimated coefficient.\\

We address this problem by following the solutions suggested by \cite{Ferrari_Garcia_2018}. They show that, under certain conditions, the bias can be either signed or corrected for. Section \ref{bias} formalises the bias problem and discusses potential solutions. In brief, there are two main ways to compute the bias. 
The first derives the bias from the aggregate macro variables of the countries that have never adopted the euro -- the ones used in our placebo test. For these countries, the only difference between the actual and the synthetic series both before and after the introduction of the euro is due to the statistical error. Hence, this difference can be computed and used to correct our baseline estimates. 
The second way to estimate the bias requires the assumption that the measurement error is time invariant and, more precisely, that it does not change after the adoption of the euro. If this is the case, the bias can be estimated from the difference between the actual and the synthetic macro variables before the adoption of the euro. 
Before the adoption of the euro, the only difference between the actual and the synthetic series is due to the minimisation procedure that the matching routine involves. This difference can be computed both for the placebo countries and the euro area countries and can be used to correct our baseline estimates for the period after the adoption of the euro. 
 \newline
\newline Table \ref{bias_correction} reports our bias-corrected estimates of the effect of the euro adoption on the risk sharing channels. The estimates are obtained by correcting for the bias the fourth row of coefficients (Post euro Actual) in Table \ref{ols}. Table \ref{bias_correction}  reports three different bias-corrected estimates. The row {\sl Placebo full sample} reports the bias-corrected estimates when the bias is computed using the difference between the actual and the synthetic series for placebo countries. The correction, while reducing the increase in the unsmoothed component of GDP variation from 22\% to 6\%, still confirms our baseline finding that the adoption of the euro has increased the unsmoothed component of the shock. The correction also implies a higher risk sharing through capital markets. The row {\sl Placebo pre euro} reports the bias-corrected estimates when the bias used for correction is computed using placebo countries only for the pre euro period. The corrected estimates are similar to the previous ones, except for the private saving channel, which remain substantial. The row {\sl Euro area pre euro} reports the bias-corrected estimates when the bias used for correction is computed only from the pre euro period using euro area countries. Again, the picture is similar and the unsmoothed component of the shock is now 8\%. Regardless of the way we correct for the bias, we confirm our baseline result that the adoption of the euro has decreased consumption smoothing for member states. The main difference between the baseline and the bias-corrected result is that  only the latter provides evidence of an increase in risk sharing through capital markets due to the adoption of the euro.

\section{Conclusion}
\label{sec: conclusion}
The adoption of the euro has increased trade and financial linkages among member states due to higher price stability and the absence of the exchange rate risk; it has, on the other hand, taken away from member states a policy tool to stabilise their economies against idiosyncratic shocks. Theory tells us that while deeper trade and financial integration increases risk sharing opportunities, the loss of monetary policy deprives countries of a policy tool to absorb shocks. Thus, it is not clear ex ante which of the two effects prevails with the adoption of the euro. 
\newline
\newline This paper has empirically evaluated whether the adoption of the common currency has changed risk sharing among euro area member states. We have tackled the major challenge of not having a natural counterfactual by building a synthetic one via the synthetic control method. We have then used this synthetic counterfactual to evaluate whether the adoption of the euro has changed member states' risk sharing. To isolate different risk sharing channels, we have carried out an output variance decomposition \`a la \cite{asdrubali_channels_1996}. With this decomposition, we have identified which portions of risk have been shared through capital markets, international transfers, public savings, and private savings.  
\newline
\newline Our main result is that the adoption of the euro has increased the unsmoothed component of the shock. At the same time, we do not find evidence of any change in international risk sharing through capital markets or international transfers. 
We show that there is heterogeneity across countries and that the aggregate result is mainly driven by periphery countries, which have witnessed a reduction in the portion of risk shared through private savings. This result is even stronger for the sample period excluding the Great Recession and the time thereafter.  We argue that the lower level of consumption smoothing is due to the higher GDP growth and volatility, which have followed the adoption of the euro. 
While we show that the euro adoption has increased the cyclical growth of output, our result is also compatible with the narrative that lower consumption smoothing might be due to a change in the trend growth of output, prompted by a policy regime change as significant as the euro adoption. In addition, we believe that our result that the periphery countries have seen a more robust contraction in consumption smoothing  may be the consequence of a stronger interest rate convergence in periphery than in core countries as argued theoretically for example in \cite{benigno_fornaro}.
\newline
\newline All in all, while the common currency has not changed shock absorption through international channels of risk sharing, it has severely affected consumption smoothing through private savings in the periphery countries. This result would call for a bigger effort on the policy side to make sure that private credit is provided uniformly throughout the area, especially in response to asymmetric shocks.

\bibliographystyle{apalike}
\bibliography{references.bib}

\begin{thebibliography}{}

\bibitem[Abadie, 2021]{abadie2021}
Abadie, A. (2021).
\newblock {Using Synthetic Controls: Feasibility, Data Requirements, and
  Methodological Aspects}.
\newblock {\em Journal of Economic Literature}, 59(2):391--425.

\bibitem[Abadie et~al., 2010]{abadie_synthetic_2010}
Abadie, A., Diamond, A., and Hainmueller, J. (2010).
\newblock Synthetic {Control} {Methods} for {Comparative} {Case} {Studies}:
  {Estimating} the {Effect} of {California}’s {Tobacco} {Control} {Program}.
\newblock {\em Journal of the American Statistical Association},
  105(490):493--505.

\bibitem[Abadie et~al., 2015]{abadie2015}
Abadie, A., Diamond, A., and Hainmueller, J. (2015).
\newblock Comparative politics and the synthetic control method.
\newblock {\em American Journal of Political Science}, 59(2):495--510.

\bibitem[Abadie and Gardeazabal, 2003]{abadie_economic_2003}
Abadie, A. and Gardeazabal, J. (2003).
\newblock The {Economic} {Costs} of {Conflict}: {A} {Case} {Study} of the
  {Basque} {Country}.
\newblock {\em The American Economic Review}, 93(1):113--132.

\bibitem[Aguiar and Gopinath, 2007]{aguiar_gopinath}
Aguiar, M. and Gopinath, G. (2007).
\newblock Emerging {Market} {Business} {Cycles}: The {Cycle} is the {Trend}.
\newblock {\em Journal of Political Economy}, 115(1):102.

\bibitem[Asdrubali et~al., 1996]{asdrubali_channels_1996}
Asdrubali, P., Sorensen, B., and Yosha, O. (1996).
\newblock Channels of {Interstate} {Risk} {Sharing}: {United} {States}
  1963-1990.
\newblock {\em The Quarterly Journal of Economics}, 111(4):1081--1110.

\bibitem[Baldwin and Giavazzi, 2015]{baldwin2015eurozone}
Baldwin, R.~E. and Giavazzi, F. (2015).
\newblock {The Eurozone Crisis: A Consensus View of the Causes and a Few
  Possible Remedies}.
\newblock {\em CEPR Press Londres}.

\bibitem[Balli et~al., 2012]{balli_etal_2012}
Balli, F., Kalemli-Ozcan, S., and Sørensen, B.~E. (2012).
\newblock {Risk Sharing through Capital Gains}.
\newblock {\em The Canadian Journal of Economics}, 45(2):472--492.

\bibitem[Baxter and Crucini, 1995]{baxter_business_1995}
Baxter, M. and Crucini, M. (1995).
\newblock Business {Cycles} and the {Asset} {Structure} of {Foreign} {Trade}.
\newblock {\em International Economic Review}, 36(4):821--854.

\bibitem[Benigno and Fornaro, 2014]{benigno_fornaro}
Benigno, G. and Fornaro, L. (2014).
\newblock The {Financial} {Resource} {Curse}.
\newblock {\em Scandinavial Journal of Economics}, (116):58--86.

\bibitem[Billmeier and Nannicini, 2012]{billmeier_assessing_2012}
Billmeier, A. and Nannicini, T. (2012).
\newblock Assessing {Economic} {Liberalization} {Episodes}: {A} {Synthetic}
  {Control} {Approach}.
\newblock {\em Review of Economics and Statistics}, 95(3):983--1001.

\bibitem[Born et~al., 2019a]{born_etal_2019a}
Born, B., M\"{u}ller, G., and Schularick, M. (2019a).
\newblock The {Costs} of {Economic} {Nationalism}: {Evidence} from the {Brexit}
  {Experiment}.
\newblock {\em Economic Journal}, 129(623):2722--2744.

\bibitem[Born et~al., 2019b]{born_etal_2019b}
Born, B., M\"{u}ller, G., Schularick, M., and Sedlacek, P. (2019b).
\newblock {Stable} {Genious}? {The} {Macroeconomic} {Impact} of {Trump}.
\newblock {\em Working Paper}.

\bibitem[Campos et~al., 2014]{campos_economic_2014}
Campos, N.~F., Coricelli, F., and Moretti, L. (2014).
\newblock Economic {Growth} and {Political} {Integration}: {Estimating} the
  {Benefits} from {Membership} in the {European} {Union} {Using} the
  {Synthetic} {Counterfactuals} {Method}.
\newblock {\em IZA Discussion Paper No. 8162}.

\bibitem[Cimadomo et~al., 2020]{cimadomo_etal_2020}
Cimadomo, J., Ciminelli, G., Furtuna, O., and Giuliodori, M. (2020).
\newblock {Private} and {Public} {Risk} {Sharing} in the {Euro} {Area}.
\newblock {\em European Economic Review}, 121:103347.

\bibitem[Duque~Gabriel and Pessoa, 2020]{duque_pessoa_2020}
Duque~Gabriel, R. and Pessoa, A.~S. (2020).
\newblock {Adopting} the {Euro}: a {Synthetic} {Control} {Approach}.
\newblock {\em Working Paper}.

\bibitem[Ferrari and Garcia~Galindo, 2018]{Ferrari_Garcia_2018}
Ferrari, A. and Garcia~Galindo, C. (2018).
\newblock {Note} on {Correction} for {Difference} in {Difference} {Estimation}
  with {Generated} {Data}.
\newblock {\em Working Paper}.

\bibitem[Frankel, 2010]{frankel}
Frankel, J. (2010).
\newblock {The Estimated Trade Effects of the Euro: Why Are They Below Those
  from Historical Monetary Unions among Smaller Countries?}
\newblock In {\em {Europe and the Euro}}, NBER Chapters, pages 169--212.
  National Bureau of Economic Research, Inc.

\bibitem[Furceri and Zdzienicka, 2015]{furceri_euro_2015}
Furceri, D. and Zdzienicka, A. (2015).
\newblock The {Euro} {Area} {Crisis}: {Need} for a {Supranational} {Fiscal}
  {Risk} {Sharing} {Mechanism}?
\newblock {\em Open Economies Review}, 26(4):683--710.

\bibitem[Hoffmann et~al., 2019]{Hoffmann_etal_2019}
Hoffmann, M., Maslov, E., S{\o}rensen, B., and Stewen, I. (2019).
\newblock Channels of {Risk} {Sharing} in the {Eurozone}: {What} {Can}
  {Banking} and {Capital} {Market} {Union} {Achieve}?
\newblock {\em IMF Economic Review}, 67:443--495.

\bibitem[Kalemli-Ozcan et~al., 2014]{kalemliozcan_debt_2014}
Kalemli-Ozcan, S., Luttini, E., and Sørensen, B. (2014).
\newblock Debt {Crises} and {Risk}-{Sharing}: {The} {Role} of {Markets} versus
  {Sovereigns}.
\newblock {\em The Scandinavian Journal of Economics}, 116(1):253--276.

\bibitem[Kalemli-Ozcan et~al., 2010]{Kalemli_etal_2010}
Kalemli-Ozcan, S., Papaioannou, E., and Peydro, J.-L. (2010).
\newblock {What} {Lies} {Beneath} the {Euro}'s {Effect} on {Financial}
  {Integration}? {Currency} {Risk}, {Legal} {Harmonization}, or {Trade}?
\newblock {\em Journal of International Economics}, 81(1):75 -- 88.

\bibitem[Poncela et~al., 2016]{Poncela_etal_2016}
Poncela, P., Pericoli, F., Manca, A.~R., and Nardo, M. (2016).
\newblock {Risk} {Sharing} in {Europe}.
\newblock {\em Working Paper}.

\bibitem[Puzzello and Gomis-Porqueras, 2018]{puzzello_gomis_2018}
Puzzello, L. and Gomis-Porqueras, P. (2018).
\newblock {Winners} and {Losers} from the {Euro}.
\newblock {\em European Economic Review}, 108:129--152.

\bibitem[Saia, 2017]{saia}
Saia, A. (2017).
\newblock Choosing the open sea: The cost to the uk of staying out of the euro.
\newblock {\em Journal of International Economics}, 108:82--98.

\bibitem[Terzi, 2020]{Terzi_2020}
Terzi, A. (2020).
\newblock Macroeconomic {Adjustment} in the {Euro} {Area}.
\newblock {\em European Economic Review}, 128:103516.

\bibitem[van Beers et~al., 2014]{van_beers_cross-country_2014}
van Beers, N., Bijlsma, M., and Zwart, G. (2014).
\newblock Cross-country {Insurance} {Mechanisms} in {Currency} {Unions}: an
  {Empirical} {Assessment}.
\newblock {\em Bruegel Working Paper}.

\end{thebibliography}
\clearpage

\newpage

\appendix

\renewcommand{\thesection}{A.\arabic{section}}
\renewcommand{\thefigure}{A.\arabic{figure}}
\renewcommand{\thetable}{A.\arabic{table}}

\setcounter{figure}{0}
\setcounter{table}{0}
{
\begin{center}
{\Huge {Appendix A} }
\end{center}}

\section{Before/After and Difference-in-Difference}

\subsection{Before/After}

\begin{table}[H]
\caption{}

\centering
\scriptsize
\begin{tabular}{lccccc}
\toprule
%          &&\multicolumn{1}{c}{(1)}&\multicolumn{1}{c}{(2)}&\multicolumn{1}{c}{(3)}&\multicolumn{1}{c}{(4)}&\multicolumn{1}{c}{(5)}\\
          &\multicolumn{1}{c}{Capital Markets}&\multicolumn{1}{c}{International Transfers}&\multicolumn{1}{c}{Public Savings}&\multicolumn{1}{c}{Private Savings}&\multicolumn{1}{c}{Unsmoothed}\\

\midrule
Pre euro &    -0.09         &     0.04         &     0.15\sym{*}  &     0.42\sym{***}&     0.48\sym{***}\\
          &  (-0.69)         &   (1.45)         &   (2.20)         &   (3.34)         &   (5.54)         \\
[1em]
Post euro &     0.47\sym{***}&    -0.04         &    -0.13\sym{*}  &    -0.29\sym{**} &    -0.01         \\
          &   (3.69)         &  (-1.31)         &  (-1.92)         &  (-2.52)         &  (-0.07)         \\
\midrule          
$N$    &      308         &      308         &      308         &      308         &      308         \\
$R^2$    &     0.35         &     0.15         &     0.56         &     0.45         &     0.71         \\
\bottomrule
\end{tabular}
\label{before_after_oecd}
\end{table}

\footnotesize
\justify \emph{Note:}  ***, **, and * denote significance at 1\%, 5\%,  10\% respectively. t-statistics are in parenthesis. The table displays OLS estimates with clustered standard errors over the period 1990-2018 using actual data before ({\sl Pre euro}) and after ({\sl Post euro}) the introduction of the euro. 
  \normalsize

\subsection{Difference-in-Differences}

\begin{table}[H]
\caption{}

\centering
\scriptsize
\begin{tabular}{llccccc}
\toprule
          &&\multicolumn{1}{c}{Capital Markets}&\multicolumn{1}{c}{International Transfers}&\multicolumn{1}{c}{Public Savings}&\multicolumn{1}{c}{Private Savings}&\multicolumn{1}{c}{Unsmoothed}\\

\midrule
Pre euro& Control &   -0.17\sym{**} &     0.03\sym{**} &     0.09\sym{***}&     0.14         &     0.91\sym{***}\\
       &   &  (-2.26)         &   (2.27)         &   (2.94)         &   (1.06)         &   (6.77)         \\
[1em]
&Treated&     0.12         &    -0.00         &     0.05         &     0.23         &    -0.39\sym{**} \\
       &   &   (0.76)         &  (-0.06)         &   (0.82)         &   (1.29)         &  (-2.57)         \\
[1em]
Post euro&Control&      0.24\sym{**} &    -0.04\sym{***}&    -0.01         &     0.18         &    -0.38\sym{**} \\
        &  &   (2.27)         &  (-3.45)         &  (-0.20)         &   (1.34)         &  (-2.65)         \\
[1em]
&Treated&      0.09         &     0.01         &    -0.09         &    -0.37\sym{**} &     0.36\sym{**} \\
         & &   (0.61)         &   (0.38)         &  (-1.58)         &  (-2.27)         &   (2.16)         \\
\midrule
$N$&     &      672         &      672         &      672         &      672         &      672         \\
$R^2$&        &     0.17         &     0.06         &     0.48         &     0.33         &     0.68         \\
\bottomrule
\end{tabular}
\label{dd}
\end{table}
\footnotesize
\raggedright \emph{Note:}  ***, **, and * denote significance at 1\%, 5\%,  10\% respectively. t-statistics are in parenthesis. The table displays OLS estimates with clustered standard errors over the period 1990-2018 for the treated (euro adopters) and the control (OECD non-euro adopters) group before ({\sl Pre euro}) and after ({\sl Post euro}) the introduction of the euro. 
  \normalsize

\normalsize

\begin{table}[H]
\caption{Parallel trend test}

\centering
\scriptsize
\begin{tabular}{llccccc}
\toprule
          &&\multicolumn{1}{c}{Capital Markets}&\multicolumn{1}{c}{International Transfers}&\multicolumn{1}{c}{Public Savings}&\multicolumn{1}{c}{Private Savings}&\multicolumn{1}{c}{Unsmoothed}\\
\midrule
          && \multicolumn{5}{c}{All countries}\\
    \midrule  
$\beta_2$& estimate &  .0002451  & -.0001272 &  .0001608  & .0004041 &  -.0006828  \\
[0.5em]
       &  p-value &  .665 &  .737  & .746 & .608   & .396        \\
       [1em]
$\beta_5$& estimate &  -8.57e-06 &  3.41e-07 &  -.0000112 &  1.79e-06 &  .0000176    \\
[0.5em]
       &  p-value &   .882   &   .983   &  .657   &   .974    &   .755     \\
\midrule
          && \multicolumn{5}{c}{Core countries}\\
    \midrule  
    $\beta_2$& estimate &  .0001936  & -.0005067 &  .0004753 &  .0006977 &  -.0008599  \\
[0.5em]
       &  p-value &  .690  & .298 &  .065  &  .438  & .066       \\
       [1em]
$\beta_5$& estimate &  -.0000256  & -4.00e-06 &  -8.21e-06  & .0000432 &  -5.45e-06     \\
[0.5em]
       &  p-value &    .697 &  .910 &  .788  & .686  & .928     \\
\midrule
          && \multicolumn{5}{c}{Periphery countries}\\
           \midrule  
    $\beta_2$& estimate &  .0004864 &   .0002968  &  -.0002719  &  -.0001311  &  -.0003803   \\
[0.5em]
       &  p-value &   .526&   .578 &   .699 &  .892 &   .782      \\
       [1em]
$\beta_5$& estimate &  .0000222 &   -5.91e-06 &   -.0000272 &   -.0000229 &   .0000338     \\
[0.5em]
       &  p-value &   .684 &  .679 &   .175 &   .559  &  .645    \\

\bottomrule
\end{tabular}
 \label{parallel_trend}
\end{table}
\footnotesize
\justify \emph{Note:} The table reports the estimated $\beta_2$ and $\beta_5$ of equation \ref{reg_parallel} and the respective p-values. The parallel trend assumption is always satisfied. 
  \normalsize

%  \begin{table}[H]
% \centering 
% \scriptsize
% \caption{Parallel trend test}
% \begin{tabular}{lccc}
% \toprule
% &{\bf All countries}  & {\bf  Core countries} &{\bf  Periphery countries}\\
% \midrule
% $\beta^m$ &  $0.22$ &   $0.51$   & $0.22$      \\
% [1em]
% $\beta^g$ &  $0.88$ &   $0.19$   & $0.49$      \\
% [1em]
% $\beta^p$ &  $0.61$ &   $0.63$   & $0.20$      \\
% [1em]
% $\beta^s$ &  $0.62$ &   $0.64$   & $0.62$      \\
% [1em]
% $\beta^u$ &  $0.95$ &   $0.57$   & $0.39$      \\
% \bottomrule
% \end{tabular}
% \label{parallel_trend}
% \end{table}
% \footnotesize
% \justify \emph{Note:} The table reports p-values for the parallel trend test. The test checks whether $H_0: \beta_1=\beta_2$ in Equation \ref{eq:DiD} is rejected.  The column {\sl All countries} runs the test for the full sample of countries, the column {\sl Core countries} for the subsample of core countries, and the column {\sl Periphery countries} for the subsample of periphery countries. The rows indicate the parallel trend test for the difference-in-differences regressions corresponding to Equations \ref{reg1} - \ref{reg5}. The parallel trend assumption is always satisfied. 
% \normalsize 

\bigskip

\section{Estimates of Risk Sharing Channels}

\subsection{Full sample period: 1990-2018}

\begin{table}[H]
\caption{All countries}

\centering
\scriptsize
\begin{tabular}{llccccc}
\toprule
          &&\multicolumn{1}{c}{Capital Markets}&\multicolumn{1}{c}{International Transfers}&\multicolumn{1}{c}{Public Savings}&\multicolumn{1}{c}{Private Savings}&\multicolumn{1}{c}{Unsmoothed}\\

\midrule
Pre euro&Synthetic&    -0.01         &     0.02         &     0.13\sym{***}&     0.40\sym{***}&     0.46\sym{***}\\
      &    &  (-0.12)         &   (1.21)         &   (6.02)         &  (19.45)         &   (7.03)         \\
[1em]
&Actual&    -0.05         &     0.01         &     0.01         &     0.00         &     0.03         \\
    &      &  (-0.37)         &   (0.25)         &   (0.22)         &   (0.00)         &   (0.38)         \\
[1em]
Post euro&Synthetic&     0.30         &    -0.05         &    -0.06         &     0.06         &    -0.24\sym{**} \\
     &     &   (1.40)         &  (-1.44)         &  (-1.18)         &   (0.36)         &  (-2.31)         \\
[1em]
&Actual&     0.05         &     0.02         &    -0.03         &    -0.26         &     0.22\sym{**} \\
      &    &   (0.27)         &   (0.58)         &  (-0.52)         &  (-1.56)         &   (2.65)         \\
\midrule
$N$&    &      616         &      616         &      616         &      616         &      616 \\
$R^2$& & 0.26         &     0.07         &     0.55         &     0.45         &     0.70  \\
\bottomrule
\end{tabular}
\label{ols}
\end{table}
\footnotesize
\justify \emph{Note:}  ***, **, and * denote significance at 1\%, 5\%,  10\% respectively. t-statistics are in parenthesis. The table displays OLS estimates with clustered standard errors over the period 1990-2018 for the actual and the synthetic series before ({\sl Pre euro}) and after ({\sl Post euro}) the introduction of the euro. The table displays three results: 1) As shown by the row {\sl Pre euro Synthetic}, before the adoption of the euro, shocks were absorbed mainly through the public and private savings channels (13\% and 40\% respectively) and a big part of shocks remained unsmoothed (46\%); 2) As displayed by the row {\sl Post euro Actual}, the adoption of the euro has increased the unsmoothed component of the shock by 22\%; 3) The quality of our matching is good. As a matter of fact, the row {\sl Pre euro Actual} shows that the channels of absorption in deviations from those computed using the synthetic data are never significantly different from zero in the pre euro period.
  \normalsize

\begin{table}[H]
\caption{Core countries}
\centering
\scriptsize
\begin{tabular}{llccccc}
\toprule
          &&\multicolumn{1}{c}{Capital Markets}&\multicolumn{1}{c}{International Transfers}&\multicolumn{1}{c}{Public Savings}&\multicolumn{1}{c}{Private Savings}&\multicolumn{1}{c}{Unsmoothed}\\  
    \midrule        
Pre euro&Synthetic&    -0.19\sym{**} &     0.08\sym{**} &     0.19\sym{***}&     0.46\sym{***}&     0.46\sym{***}\\
       &   &  (-2.33)         &   (2.97)         &   (4.83)         &   (3.75)         &   (3.93)         \\
[1em]
&Actual&     -0.05         &    -0.03         &     0.04         &     0.05         &    -0.01         \\
       &   &  (-0.49)         &  (-0.53)         &   (1.13)         &   (0.27)         &  (-0.07)         \\
[1em]
Post euro&Synthetic&    0.45\sym{***}&    -0.10\sym{**} &    -0.13\sym{**} &    -0.28         &     0.06         \\
         &   &   (3.43)         &  (-3.01)         &  (-2.87)         &  (-1.69)         &   (0.44)         \\
         [1em]
&Actual&  -0.31\sym{**} &     0.03         &     0.03         &     0.14         &     0.11         \\
      &    &  (-2.33)         &   (0.59)         &   (0.78)         &   (0.65)         &   (0.77)         \\
\midrule
N   &   &        336         &      336         &      336         &      336         &      336           \\
$R^2$  &    &      0.33         &     0.16         &     0.66         &     0.47         &     0.76          \\
\bottomrule
\end{tabular}
\label{core }
\end{table}
\footnotesize
\justify \emph{Note:} ***, **, and * denote significance at 1\%, 5\%, 10\% respectively. t-statistics are in parenthesis. The countries included as core  are: Austria, Belgium, Finland, France, Germany, Netherlands. 
The table displays OLS estimates with clustered standard errors over the period 1990-2018 for the actual and the synthetic series before ({\sl Pre euro}) and after ({\sl Post euro}) the introduction of the euro. 
The table shows the following two results. First, as displayed by the row {\sl Pre euro Synthetic}, before the adoption of the euro core countries were able to smooth output variations mainly through public and private savings (19\% and 46\% respectively). Only  a minor share (8\%) was smoothed through international transfers. Capital markets were actually amplifying the output variation. A share of 46\% of the output variation remained unsmoothed. 
Second, as displayed by the row {\sl Post euro Actual}, the adoption of the euro has not affected consumption smoothing in the core countries in a significant way, while it has increased the amplification of output variation through capital markets. \normalsize

\begin{table}[H]
\caption{Periphery  countries}
\centering
\scriptsize
\begin{tabular}{llccccc}
\toprule
          &&\multicolumn{1}{c}{ Capital Markets}&\multicolumn{1}{c}{ International Transfers}&\multicolumn{1}{c}{ Public Savings}&\multicolumn{1}{c}{ Private Savings}&\multicolumn{1}{c}{ Unsmoothed}\\
 \midrule
Pre euro&Synthetic&    0.03         &     0.01         &     0.11\sym{***}&     0.40\sym{***}&     0.45\sym{***}\\
      &    &   (0.71)         &   (0.37)         &   (5.03)         &   (8.77)         &   (5.66)         \\
[1em]
&Actual&     0.11\sym{*}  &     0.01         &    -0.05         &    -0.10         &     0.03         \\
    &      &  (1.89)         &   (0.24)         &  (-1.28)         &  (-1.61)         &   (0.30)         \\
[1em]
Post euro&Synthetic&     0.15         &    -0.06         &     0.00         &     0.30\sym{**} &    -0.40\sym{***}\\
       &   &  0.80)         &  (-0.98)         &   (0.11)         &   (2.72)         &  (-3.56)         \\
 [1em]
&Actual&         0.09         &     0.04         &    -0.03         &    -0.44\sym{***}&     0.34\sym{***}\\
      &    &   (0.54)         &   (0.77)         &  (-0.63)         &  (-3.97)         &   (3.58)         \\
 \midrule
N   &   &     280         &      280         &      280         &      280         &      280         \\
$R^2$  &   & 0.41         &     0.11         &     0.58         &     0.61         &     0.74         \\
\bottomrule
\end{tabular}
\label{periphery }
\end{table}

\footnotesize
\justify\emph{Note:} 
***, **, and * denote significance at 1\%, 5\%, 10\% respectively. t-statistics are in parenthesis. The countries included as periphery  are: Greece, Ireland, Italy, Portugal, Spain.  
The tables display OLS estimates with clustered standard errors over the period 1990-2018 for the actual and the synthetic series before ({\sl Pre euro}) and after ({\sl Post euro}) the introduction of the euro. 
The tables show the following results. First, as displayed by the row {\sl Pre euro Synthetic}, before the adoption of the euro periphery countries were able to smooth
output variations mainly through public and private savings (11\% and 40\% respectively).  A share of 45\% of the output variation remained unsmoothed. 
Second, as displayed by the row {\sl Post euro Actual}, the adoption of the euro has decreased consumption smoothing in the periphery countries by 34\%, compared to a counterfactual scenario without the adoption of the euro. Moreover, it has decreased risk sharing through private saving by 44\%. 
\normalsize

\subsection{Sample period up to the Great Recession: 1990-2007}

\begin{table}[H]
\caption{All countries}
\centering
\scriptsize
\begin{tabular}{llccccc}
\toprule
          &&\multicolumn{1}{c}{Capital Markets}&\multicolumn{1}{c}{International Transfers}&\multicolumn{1}{c}{Public Savings}&\multicolumn{1}{c}{Private Savings}&\multicolumn{1}{c}{Unsmoothed}\\
\midrule
Pre euro&Synthetic&     -0.01         &     0.02         &     0.13\sym{***}&     0.40\sym{***}&     0.46\sym{***}\\
       &   &  (-0.15)         &   (1.21)         &   (5.97)         &  (19.13)         &   (7.01)         \\
       [1em]
&Actual&     -0.05         &     0.01         &     0.01         &     0.00         &     0.03         \\
        &  &  (-0.35)         &   (0.25)         &   (0.22)         &   (0.00)         &   (0.36)         \\
[1em]
Post euro&Synthetic&     0.18         &    -0.05         &    -0.04         &     0.04         &    -0.13\sym{**} \\
                     &        &   (1.12)         &  (-1.42)         &  (-0.77)         &   (0.25)         &  (-2.40)         \\
[1em]
&Actual&     -0.02         &     0.02         &    -0.07         &    -0.24\sym{*}  &     0.30\sym{***}\\
          &&  (-0.11)         &   (0.59)         &  (-1.33)         &  (-1.78)         &   (4.78)         \\
\midrule
N & &$      374         $&$      374         $&$      374         $&$      374         $&$      374   $      \\
$R^2$&        & 0.12         &     0.06         &     0.43         &     0.43         &     0.74         \\
\bottomrule
\end{tabular}
\label{nocrisis}
\end{table}
\footnotesize
\justify\emph{Note:} ***, **, and * denote significance at 1\%, 5\%, 10\% respectively. t-statistics are in parenthesis. The table reports OLS estimates with clustered standard errors over the period 1990-2007 for the actual and the synthetic series before ({\sl Pre euro}) and after ({\sl Post euro}) the introduction of the euro. As displayed by the row {\sl Post euro Actual}, when excluding the period after 2007, the unsmoothed component of the shock increases even further (30\%) as compared to the baseline result (22\%). This shows that the decrease in risk sharing is not due to the turbulences of the financial crisis or the period thereafter.
\normalsize

\begin{table}[H]
\caption{Core  countries}
\centering
\scriptsize
\begin{tabular}{llccccc}
\toprule
          &&\multicolumn{1}{c}{Capital Markets}&\multicolumn{1}{c}{International Transfers}&\multicolumn{1}{c}{Public Savings}&\multicolumn{1}{c}{Private Savings}&\multicolumn{1}{c}{Unsmoothed}\\
\midrule
Pre euro&Synthetic& -0.19\sym{**} &     0.08\sym{**} &     0.18\sym{***}&     0.46\sym{***}&     0.47\sym{***}\\
      &    &  (-2.28)         &   (2.96)         &   (4.72)         &   (3.81)         &   (4.17)         \\
[1em]
&Actual&  -0.06         &    -0.03         &     0.05         &     0.06         &    -0.02         \\
      &    &  (-0.51)         &  (-0.50)         &   (1.18)         &   (0.31)         &  (-0.16)         \\
[1em]
Post euro&Synthetic&  0.48\sym{***}&    -0.13\sym{**} &    -0.13\sym{**} &    -0.34\sym{*}  &     0.13         \\
     &     &   (3.45)         &  (-2.78)         &  (-2.76)         &  (-1.91)         &   (0.92)         \\
[1em]
&Actual&  -0.32\sym{*}  &     0.07         &    -0.03         &     0.12         &     0.16         \\
       &   &  (-1.88)         &   (1.00)         &  (-0.51)         &   (0.56)         &   (1.20)         \\
\midrule
N & & 204         &      204         &      204         &      204         &      204         \\
$R^2$&    & 0.32         &     0.15         &     0.57         &     0.46         &     0.75         \\
\bottomrule
\end{tabular}
\label{nocrisis_core}
\end{table}
\smallskip
\footnotesize
\justify\emph{Note:} ***, **, and * denote significance at 1\%, 5\%, 10\% respectively. t-statistics are in parenthesis. The countries included as core  are: Austria, Belgium, Finland, France, Germany, Netherlands. The table reports OLS estimates with clustered standard errors over the period 1990-2007 for the actual and the synthetic series before ({\sl Pre euro}) and after ({\sl Post euro}) the introduction of the euro. As displayed by the row {\sl Post euro Actual}, also when excluding the period from 2008 onward, the adoption of the euro has not affected consumption smoothing in the core countries in a significant way, while it has increased the amplification of output variation through capital markets. \normalsize

\normalsize

\begin{table}[H]
\caption{Periphery countries}
\centering
\scriptsize
\begin{tabular}{llccccc}
\toprule
          &&\multicolumn{1}{c}{Capital Markets}&\multicolumn{1}{c}{International Transfers}&\multicolumn{1}{c}{Public Savings}&\multicolumn{1}{c}{Private Savings}&\multicolumn{1}{c}{Unsmoothed}\\
\midrule
Pre euro&Synthetic& 0.03         &     0.01         &     0.11\sym{***}&     0.40\sym{***}&     0.45\sym{***}\\
     &     &  (0.71)         &   (0.37)         &   (4.99)         &   (8.71)         &   (5.63)         \\
[1em]
&Actual&    0.11\sym{*}  &     0.01         &    -0.05         &    -0.10         &     0.03         \\
      &    &  (1.87)         &   (0.24)         &  (-1.27)         &  (-1.60)         &   (0.30)         \\
[1em]
Post euro&Synthetic&  0.01         &    -0.04         &     0.04         &     0.24\sym{**} &    -0.26\sym{***}\\
       &   &   (0.13)         &  (-0.59)         &   (1.10)         &   (2.53)         &  (-4.06)         \\
       [1em]
&Actual&     0.05         &     0.01         &    -0.07\sym{*}  &    -0.42\sym{***}&     0.42\sym{***}\\
     &     & (0.45)         &   (0.23)         &  (-1.96)         &  (-4.35)         &   (4.77)         \\
\midrule
N & &  170         &      170         &      170         &      170         &      170         \\
$R^2$&    &0.23         &     0.11         &     0.44         &     0.62         &     0.78         \\
\bottomrule
\end{tabular}
\label{nocrisis_periphery}
\end{table}
\smallskip
\footnotesize
\justify\emph{Note:} ***, **, and * denote significance at 1\%, 5\%, 10\% respectively. t-statistics are in parenthesis. The countries included as periphery are: Greece, Ireland, Italy, Portugal, Spain. The table displays OLS estimates with clustered standard errors over the period 1990-2007 for the actual and the synthetic series before ({\sl Pre euro}) and after ({\sl Post euro}) the introduction of the euro. As displayed by the row {\sl Post euro Actual}, when excluding the period from 2008 onward, the adoption of the euro has decreased consumption smoothing in the periphery countries by 42\% (which is even more than the baseline result of 34\% for periphery countries). Moreover, it has decreased risk sharing through private saving by 42\%. \normalsize

\subsection{ World Bank data}

\begin{table}[H]
\caption{World Bank Data Analysis}

\centering
\scriptsize
\begin{tabular}{llccc}
\toprule
          &&\multicolumn{1}{c}{Before/After}&\multicolumn{1}{c}{difference-in-differences}& SCM \\
          [1em]
          &&\multicolumn{1}{c}{Unsmoothed}&\multicolumn{1}{c}{Unsmoothed}&{Unsmoothed}\\
\midrule
Pre euro& Control &    &     0.66\sym{***} & 0.47\sym{***} \\
       &   &          &   (8.42)        &(6.67)   \\
[1em]
&Treated&     0.46\sym{***}        &      -0.13     & -0.04      \\
       &   &    (6.12)            &  (-1.29)     &(-0.57)      \\
[1em]
Post euro&Control&      &      0.03 &-0.47\sym{***} \\
        &  &           &    (0.36)            &(-3.70)    \\
[1em]
&Treated&      0.01             &        0.06    &0.56\sym{***}     \\
         & &   (0.04)         &    (0.34)        &(3.86)      \\
\midrule
$N$&     &      308         &      3220          & 616      \\
$R^2$&        &     0.71         &     0.32           &0.56     \\
\bottomrule
\end{tabular}
\label{worldbank}
\end{table}
\footnotesize
\justify \emph{Note:}  ***, **, and * denote significance at 1\%, 5\%,  10\% respectively. t-statistics are in parenthesis. The table displays OLS estimates with clustered standard errors over the period 1990-2018 for the treated and the control series before ({\sl Pre euro}) and after ({\sl Post euro}) the introduction of the euro. The data used is from the World Bank. This allows us to have a much bigger group of control countries, but at the cost of estimating only the unsmoothed component and not the other coefficients. 
  \normalsize

\begin{figure}[H]
\centering\footnotesize
\caption{Permutation Test for $\beta^u$}
\subfigure{\includegraphics[width=0.7\textwidth]{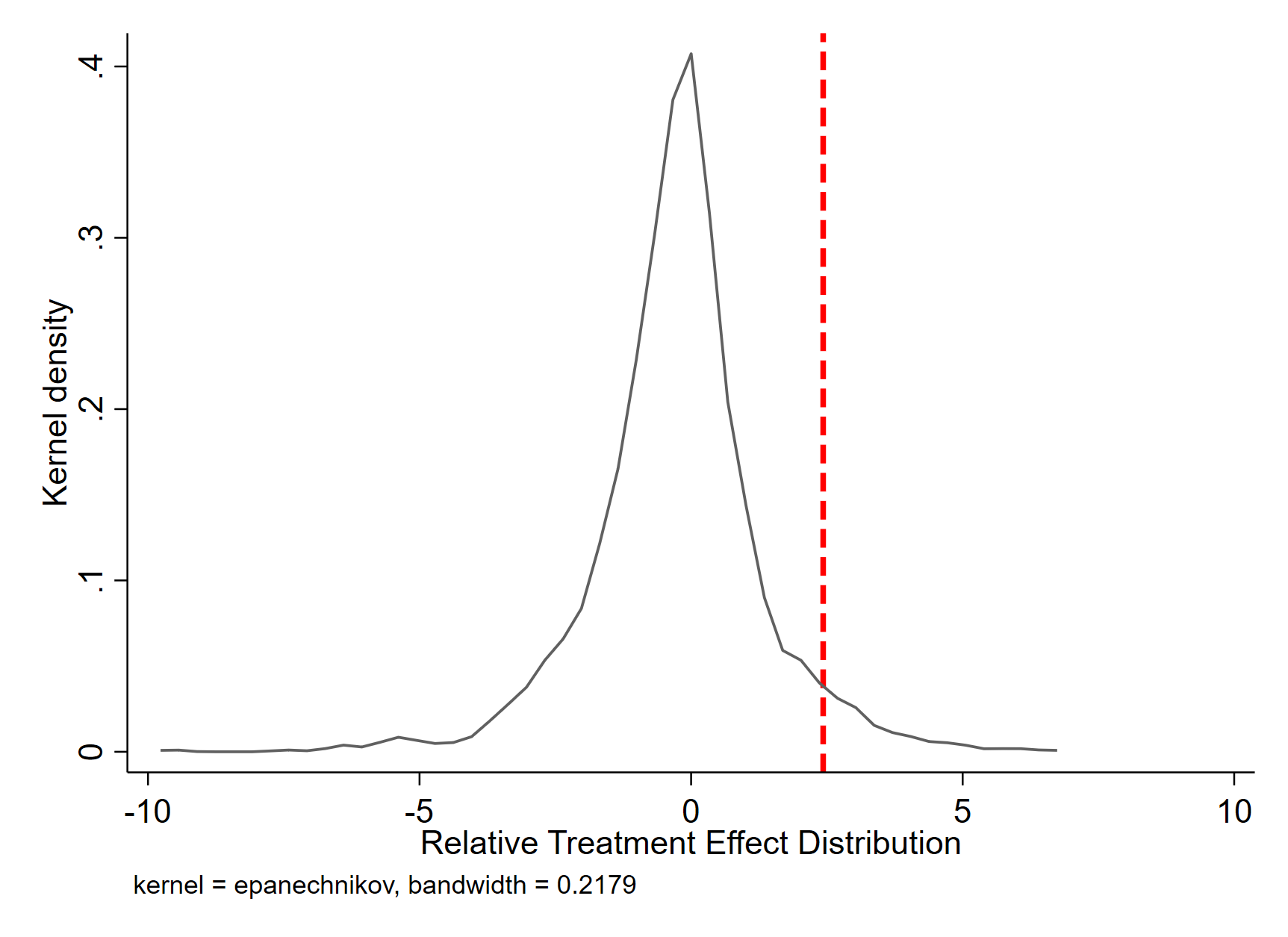}}
 \label{permutation}
\end{figure} 
\footnotesize
\justify \emph{Note:} The figure plots the treatment effect distribution from 1500 permutations for the unsmoothed component. We re-assign treatment to 11 countries, generate synthetic series for GDP, government consumption and consumption and re-estimate equation \ref{reg5}. The graph plots the distribution of standardized treatment effects ($\beta_4$ in equation \ref{eq:DiD}) as described in the main body. The dashed line represents the estimated effect when the assignment is the one in the data.
\normalsize

\subsection{GDP Growth and Volatility}

 \begin{figure}[H]
\centering\footnotesize
\caption{GDP in euro area countries}
\subfigure[Cross-country average]{\includegraphics[width=0.49\textwidth]{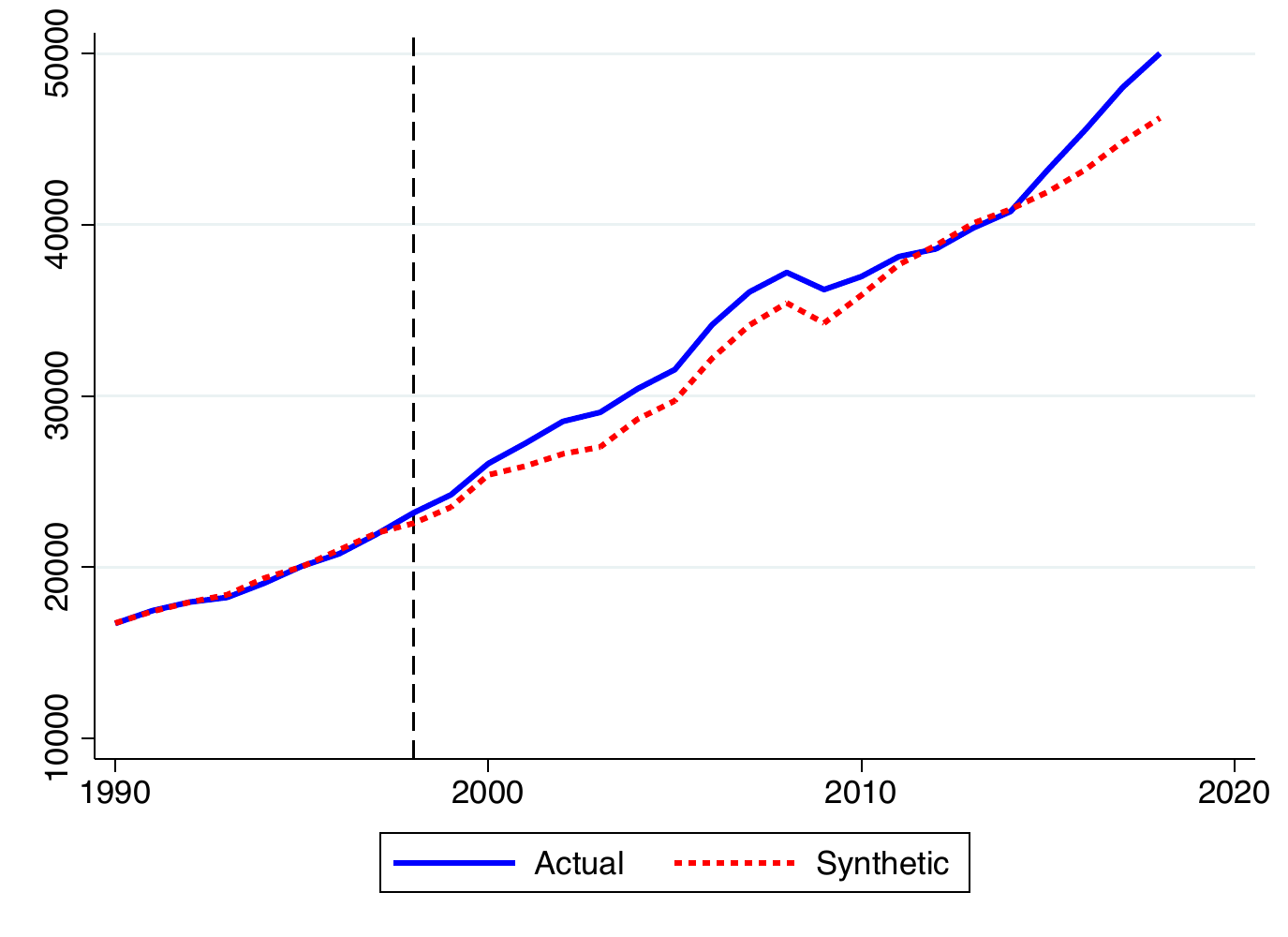}}
\subfigure[Percentage difference]{\includegraphics[width=0.49\textwidth]{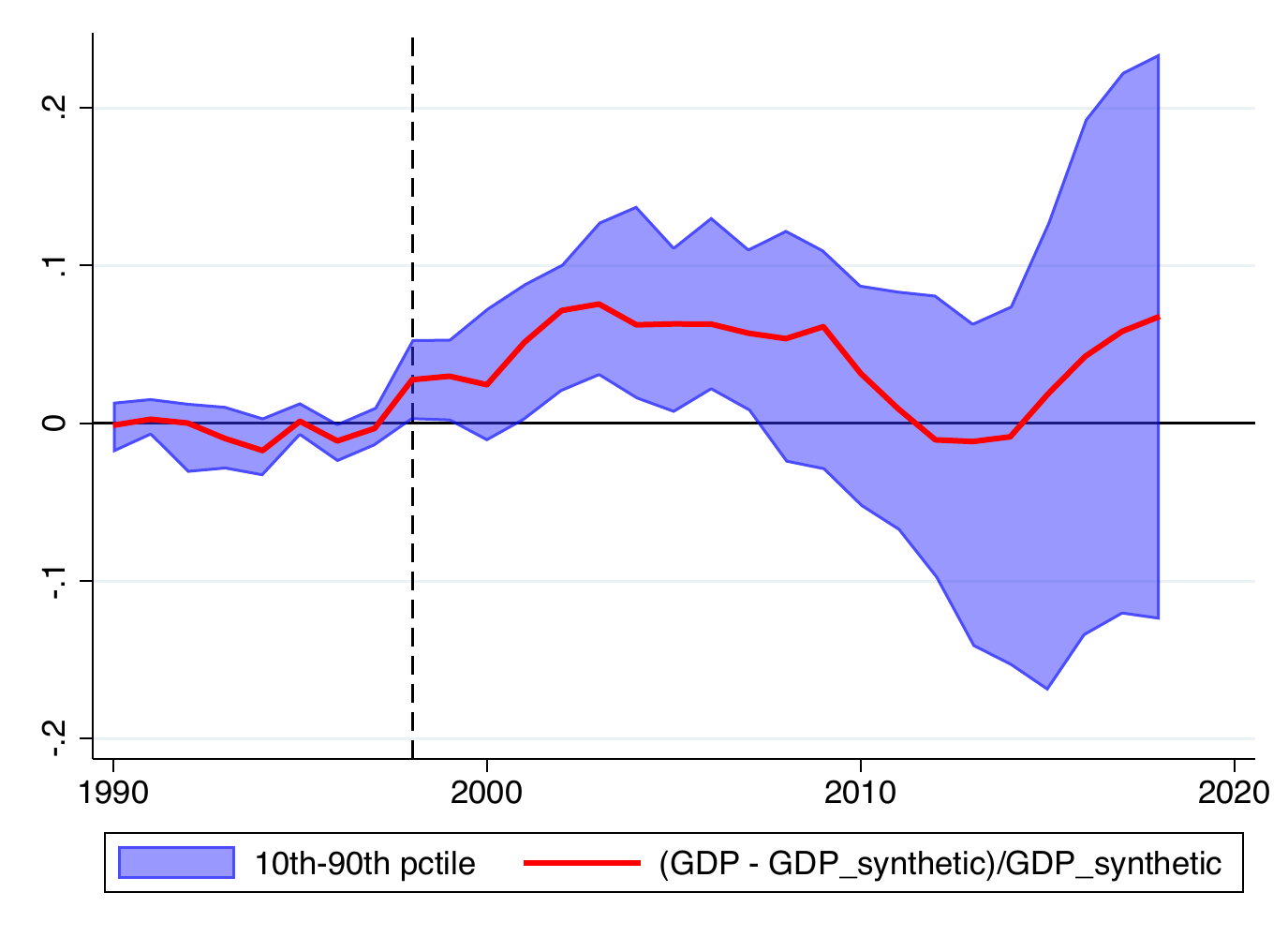}}
 \label{fig:growth}
\end{figure} 
\footnotesize
\justify \emph{Note:} The figure shows that GDP has increased more with the adoption of the euro than it would have without it.
As illustrated by panel (a), countries display on average higher GDP per capita  with the euro (blue line) than without the euro (red line).
Panel (b) plots the cross-country average of the actual GDP change from its synthetic counterpart (red line), along with the 80 central percentiles (blue area). This change is significant for the period until the Great Recession.\normalsize

\bigskip
\begin{table}[H]
\captionsetup{width=0.7\textwidth}
\caption{GDP growth and volatility}
\centering
\scriptsize
\begin{tabular}{llcccc}
\toprule
         & &\multicolumn{2}{c}{1990-2018}&\multicolumn{2}{c}{1990-2007}\\%
          &&\multicolumn{1}{c}{GDP Growth}&\multicolumn{1}{c}{GDP Variance}&\multicolumn{1}{c}{GDP Growth}&\multicolumn{1}{c}{GDP Variance}\\
\midrule
 Pre euro  & Synthetic  &  -0.02\sym{***}&     12.57\sym{***}& -0.00 &12.57\sym{***}\\
          &     &    (-3.87)       &  (37.28) & (-0.46)       &  (39.97)    \\
 [1em]
   &Actual&   0.03\sym{***}& 1.56\sym{***}      & 0.02\sym{***} & 1.07\sym{***}  \\
   &       & (3.29)    &   (3.28)      &(3.18)   &   (2.41)            \\
 [1em]
 Post euro &Synthetic&   -0.00    &    2.89\sym{***} &    -0.00      &     0.86\sym{*} \\
  &         &   (-0.49)         &   (6.07)    &(-0.49)     &   (1.92)        \\
 [1em]

 &Actual&0.03\sym{***} &    0.02\sym{**} & 0.03\sym{***} &    1.21\sym{***}  \\
 &          &(3.28)      &  (2.45)             &    (3.30)         & (2.72)      \\
 \midrule
N&     &      616         &       44     & 374 & 44                  \\
$R^2$&       & 0.40  &         0.50   & 0.50 & 0.18                   \\

\bottomrule
\end{tabular}
\label{growth_did}
\end{table}
\footnotesize
\justify \emph{Note:} GDP is detrended using a linear quadratic trend. 
 ***, **, and * denote significance at 1\%, 5\%,  10\% respectively. t-statistics are in parenthesis. Sample period is 1990-2018.
The table displays regressions of GDP growth and logged variance on a set of dummies spanning the possible combinations of pre euro/post euro and actual/synthetic data. The row {\sl Post euro Actual} shows that the adoption of the euro has increased both GDP growth and volatility.
\normalsize

%  \begin{figure}[H]
% \centering\footnotesize
% \caption{Euro area convergence and consumption smoothing}
% \subfigure[Spread convergence]{\includegraphics[width=0.49\textwidth]{input/convergence_nocrisis_ci.png}}
% \subfigure[GDP growth]{\includegraphics[width=0.49\textwidth]{input/convergence_gdp2_ci.png}}
%  \label{fig:convergence}
% \end{figure} 
% \footnotesize
% \justify \emph{Note:} The figure shows that countries that face a lower consumption smoothing before the Great Recession are also those with a stronger interest rate convergence and a greater GDP growth. In both panels, the x-axis plots the  post-euro effect on $\beta^u$  estimated at country-level. In panel (a) the y-axis plots the change in interest rate spreads against Germany between 1995 and 2007. The spreads are computed by using the EMU convergence criterion bond yields from Eurostat. In panel (b) the y-axis plots the GDP growth over its synthetic counterpart.
% \normalsize

\pagebreak

\normalsize
\justify

\setcounter{figure}{0}
\setcounter{table}{0}
\setcounter{section}{0}
\renewcommand{\thesection}{B.\arabic{section}}
\renewcommand{\thefigure}{B.\arabic{figure}}
\renewcommand{\thetable}{B.\arabic{table}}
\begin{center}
    
{\Huge {Appendix B} \\\normalsize Not for Publication}
\end{center}

\section{Matrices of weights}
\begin{table}[H]
\centering
\caption{Matrix of weights: GDP}
\tiny
\begin{tabular}{lccccccccccc} \toprule
Non euro area & Austria & Belgium & Finland & France & Germany & Greece & Ireland & Italy & Netherlands & Portugal & Spain \\ \midrule
Australia & . & . & . & . & . & . & . & . & . & . & . \\
Canada & . & . & . & . & . & . & . & . & . & . & . \\
Denmark & 42.10 & . & . & . & . & . & . & 5.800 & 50.20 & . & 19.20 \\
Israel & 1 & 31.50 & . & 18 & 3.900 & . & . & 28.50 & . & . & . \\
Japan & 24.50 & 15.60 & . & . & 73.60 & 3.200 & . & 12.60 & . & . & . \\
Korea & . & . & . & . & . & . & 40.40 & . & . & 13.70 & 6 \\
Mexico & 4 & 10.70 & 14.30 & 14.80 & 4.200 & 39.80 & . & 7.100 & 4 & 41.10 & 31.60 \\
New Zealand & . & . & . & . & . & . & . & . & . & . & . \\
Norway & . & . & . & . & . & . & 59 & . & 1.900 & . & . \\
Sweden & . & . & 72.70 & 40.80 & . & 29.30 & . & 16.90 & . & 45.20 & 43.30 \\
Switzerland & 12.50 & 42.20 & . & 26.30 & 18.30 & 6.600 & . & 29.20 & . & . & . \\
United Kingdom & . & . & 13 & . & . & 21.10 & 0.700 & . & 7.800 & . & . \\
 United States & 15.90 & . & . & . & . & . & . & . & 36 & . & . \\ \bottomrule
\end{tabular}
\label{oecd}
\end{table}

\begin{table}[H]
\centering
\caption{Matrix of weights: consumption}
\tiny
\begin{tabular}{lccccccccccc} \toprule
Non euro area & Austria & Belgium & Finland & France & Germany & Greece & Ireland & Italy & Netherlands & Portugal & Spain \\ \midrule
Australia & . & . & . & . & . & 29.20 & 33 & 33 & 69 & 12.80 & 34.10 \\
Canada & . & . & . & . & . & . & . & . & 3.200 & . & . \\
Denmark & . & 0.700 & . & . & . & . & . & 28.50 & . & . & 0.100 \\
Israel & 12.80 & . & . & 1.400 & . & . & . & . & . & . & . \\
Japan & 7.600 & 13.60 & . & . & 33 & 2.600 & . & . & . & 40.40 & . \\
Korea & . & . & . & . & . & 1.200 & 17.70 & . & . & 3.800 & . \\
Mexico & 3.200 & . & 22.20 & 1.400 & 2.700 & 25.90 & 5.800 & 4.900 & 7.200 & 36 & 23.70 \\
New Zealand & . & . & 31.80 & 22.70 & . & 22 & 29.10 & . & 13.50 & . & 14.40 \\
Norway & 17.10 & . & . & . & . & . & 14.40 & . & 7.200 & 7 & . \\
Sweden & 23.40 & 76.80 & 30.20 & 63.70 & 33 & . & . & 2.600 & . & . & 27.60 \\
 Switzerland & 35.90 & 9 & 15.80 & 10.80 & 31.30 & 19.10 & . & 30.90 & . & . & . \\ 
 United Kingdom & . & . & . & . & . & . & . & . & . & . & . \\
 United States & . & . & . & . & . & . & . & . & . & . & . \\ \bottomrule
\end{tabular}
\end{table}

\begin{table}[H]
\centering
\caption{Matrix of weights: net disposable national income}
\tiny
\begin{tabular}{lccccccccccc} \toprule
Non euro area & Austria & Belgium & Finland & France & Germany & Greece & Ireland & Italy & Netherlands & Portugal & Spain \\ \midrule
Australia & . & . & . & 26.50 & . & . & . & . & 1.400 & . & 24.80 \\
Canada & . & 6.500 & . & . & . & . & . & . & . & . & . \\
Denmark & . & . & . & . & . & . & . & 15.20 & 51.40 & . & 29 \\
Israel & 50.10 & 38.40 & . & . & . & 42.40 & . & 33.20 & . & 49.60 & 0.100 \\
Japan & 5.800 & 26.10 & . & . & 70.90 & 3.100 & . & . & . & 1.400 & . \\
Korea & . & . & . & . & . & . & 21.90 & . & . & . & . \\
Mexico & 3.400 & 3.400 & 3.600 & 12.40 & 3.400 & 28.50 & 11.70 & 5.200 & 2.500 & 38.20 & 25.60 \\
New Zealand & . & . & 56.70 & 6.200 & . & . & 1 & . & . & . & . \\
Norway & . & . & . & . & . & . & 65.40 & . & 10.40 & . & . \\
Sweden & 0.600 & . & 39.70 & 23.90 & . & 12.90 & . & 21.30 & . & 0.700 & 20.50 \\
Switzerland & 40.10 & 25.50 & . & 31.10 & 25.60 & 13.10 & . & 25.10 & . & 10 & . \\
United Kingdom & . & . & . & . & . & . & . & . & . & . & . \\
 United States & . & . & . & . & . & . & . & . & 34.30 & . & . \\  \bottomrule
\end{tabular}
\end{table}

\begin{table}[H]
\centering
\caption{Matrix of weights: net national income}
\tiny
\begin{tabular}{lccccccccccc} \toprule
Non euro area & Austria & Belgium & Finland & France & Germany & Greece & Ireland & Italy & Netherlands & Portugal & Spain \\ \midrule
Australia & . & . & . & 24.50 & . & . & . & . & . & . & 27.30 \\
Canada & . & 16.30 & . & . & . & . & . & . & . & . & . \\
Denmark & . & . & . & . & . & . & . & . & 54.60 & . & 16 \\
Israel & 36 & 23.90 & . & 0.800 & . & 28.30 & . & 37.40 & . & 57.30 & 6.600 \\
Japan & 15.80 & 33.40 & . & . & 77.60 & 6.500 & . & 1.100 & . & . & . \\
Korea & . & . & . & . & . & . & 27.10 & . & . & . & . \\
Mexico & 2.500 & 1.700 & 3.400 & 11.90 & 1.200 & 30.40 & 9 & 5.700 & 1.500 & 34.50 & 25.10 \\
New Zealand & . & . & 53 & 6.600 & . & . & 4.800 & . & . & . & . \\
Norway & . & . & . & . & . & . & 59.10 & . & 10.30 & . & . \\
Sweden & . & . & 43.70 & 22.50 & . & 21.40 & . & 19.20 & 1.300 & 4.500 & 25.10 \\
Switzerland & 45.70 & 24.70 & . & 33.70 & 21.20 & 13.50 & . & 36.60 & 0.600 & 3.600 & . \\
United Kingdom & . & . & . & . & . & . & . & . & . & . & . \\
 United States & . & . & . & . & . & . & . & . & 31.70 & . & . \\ \bottomrule
\end{tabular}
\end{table}

\begin{table}[H]
\centering
\caption{Matrix of weights: government consumption}
\tiny
\begin{tabular}{lccccccccccc} \toprule
Non euro area & Austria & Belgium & Finland & France & Germany & Greece & Ireland & Italy & Netherlands & Portugal & Spain \\ \midrule
Australia & . & . & . & . & . & . & 7.100 & . & 9.700 & 28.60 & . \\
Canada & . & 32.30 & . & 17.20 & 31.30 & . & . & 39.40 & 23.10 & 9.300 & 48.80 \\
Denmark & 45.80 & . & . & 32.20 & . & . & . & . & 45.20 & . & . \\
Israel & . & 26.90 & . & 12.80 & 20.40 & 21.30 & 19 & . & 3.400 & . & . \\
Japan & 39.40 & 20.20 & . & 23.30 & 17 & . & . & . & . & . & . \\
Korea & . & . & . & . & . & 30 & 24.60 & . & . & 37.30 & 36.10 \\
Mexico & . & . & . & . & . & . & 6.200 & . & . & 7.900 & 5.400 \\
New Zealand & . & . & 30.20 & . & . & 44.70 & . & 34.80 & . & . & . \\
Norway & 14.80 & 18.20 & . & 14.40 & 13.10 & . & 43.10 & . & 18.60 & 17 & 9.800 \\
Sweden & . & . & 38.20 & . & . & . & . & 9.200 & . & . & . \\
Switzerland & . & 2.400 & . & . & 18.20 & . & . & . & . & . & . \\
United Kingdom & . & . & . & . & . & . & . & 16.50 & . & . & . \\
 United States & . & . & 31.70 & . & . & 4.100 & . & . & . & . & . \\ \bottomrule
\end{tabular}
\label{oecd_g}
\end{table}

\footnotesize
\justify \emph{Note:} Table \ref{oecd} to \ref{oecd_g} show the matrix of weights that generate the convex combination of non euro area countries' macro variables that best reproduce those of euro area countries over the matching window 1990-1998. For~example, Finnish GDP is best reproduced by a vector of Mexican, Swedish, and British macro variables in the percentages of 14.3, 72.7, and 13.0. The weights are computed using the synthetic control method of  \cite{abadie_economic_2003}.
\normalsize

\section{Actual and synthetic series}
\graphicspath{{Figures/}} 
\begin{figure}[H]
  \caption{Actual and synthetic series for GDP}
\centering
  \begin{tabular}{@{}ccc@{}}
    \includegraphics[width=0.33\textwidth]{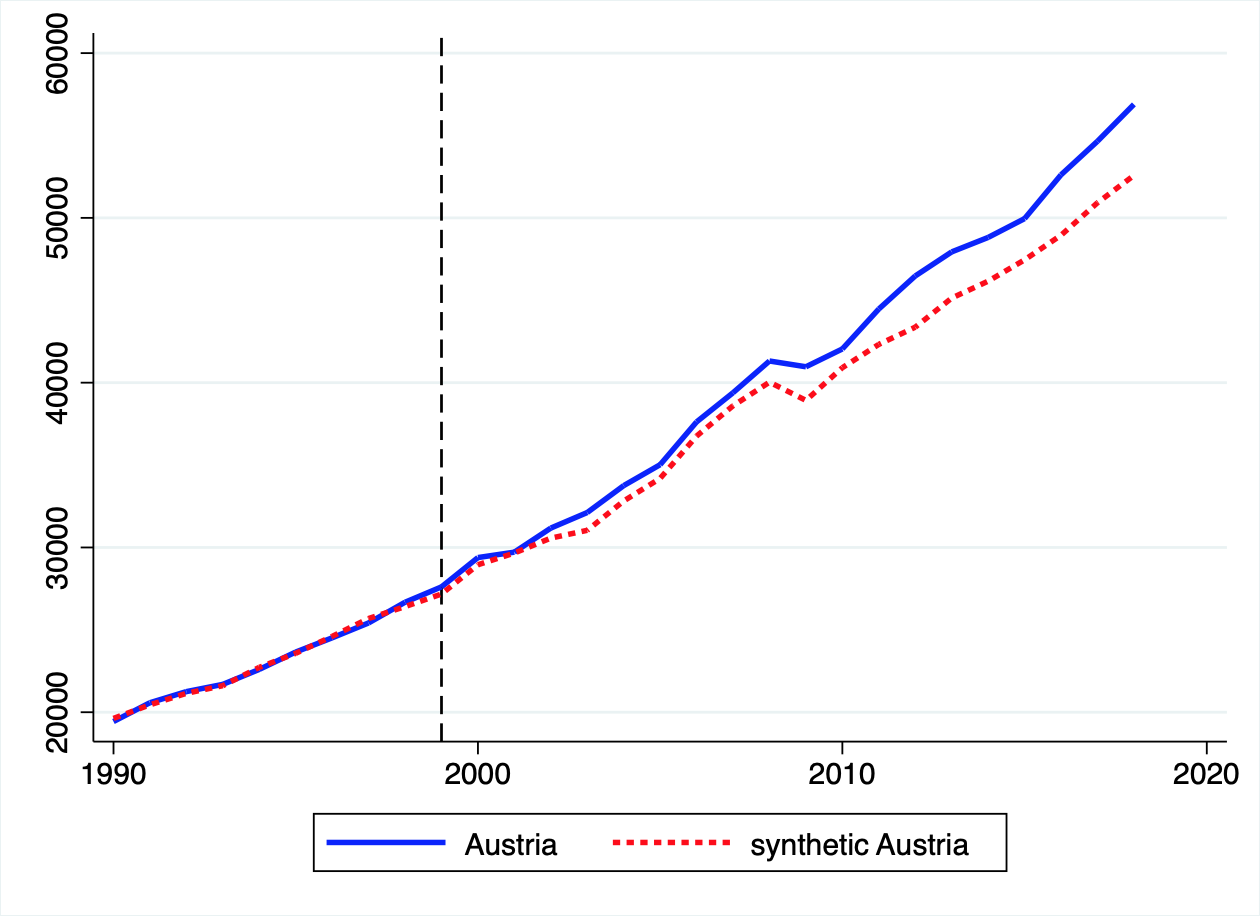} &
    \includegraphics[width=0.33\textwidth]{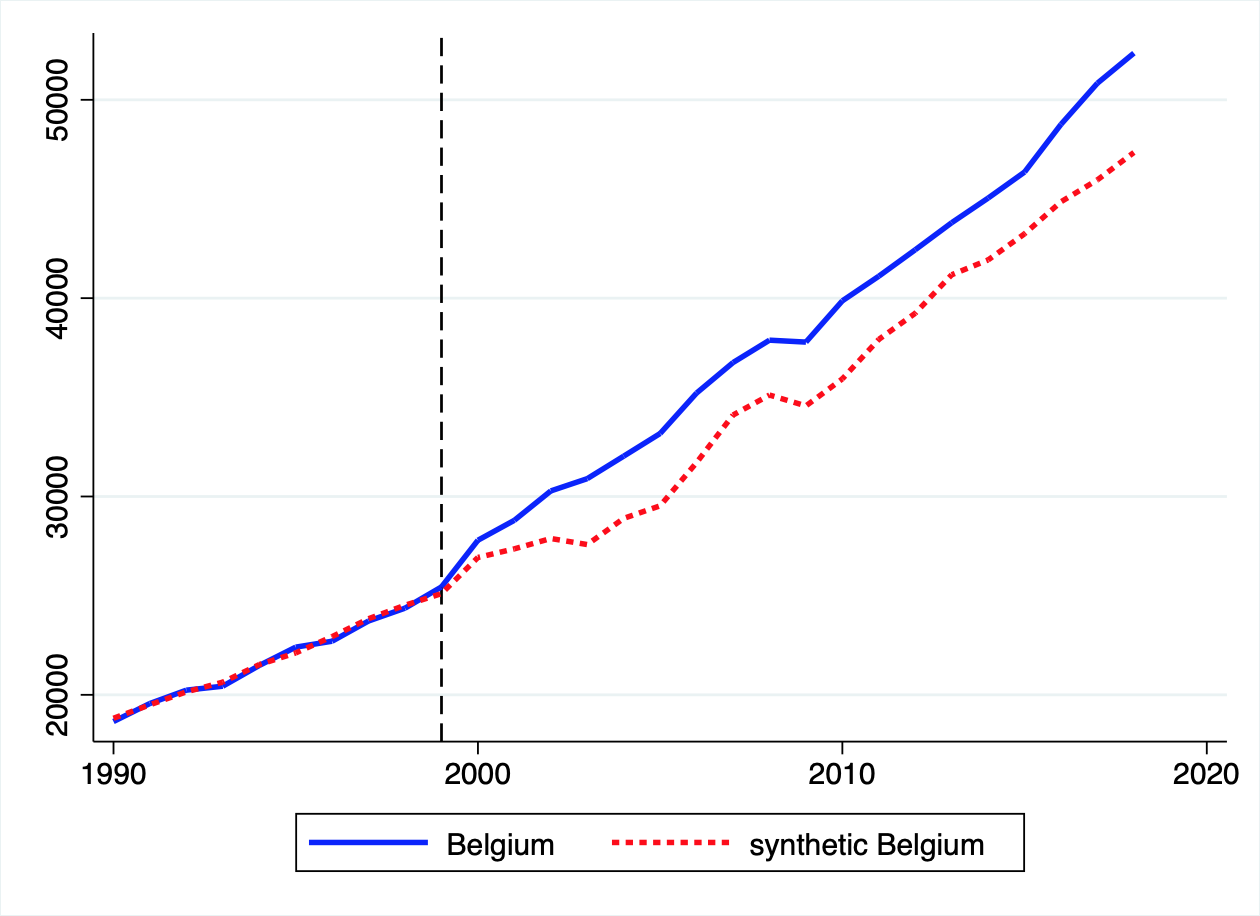} &
    \includegraphics[width=0.33\textwidth]{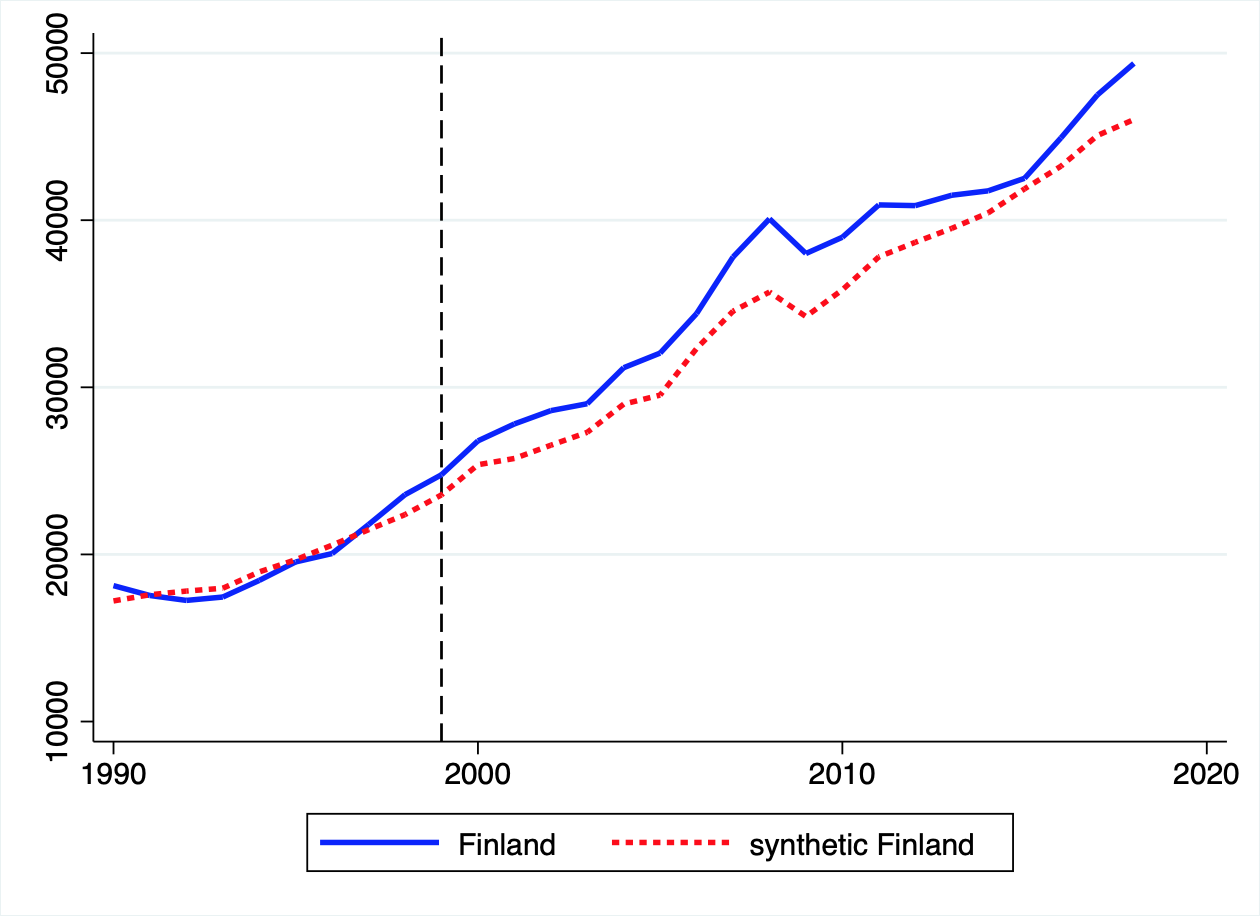} \\
   \includegraphics[width=.33\textwidth]{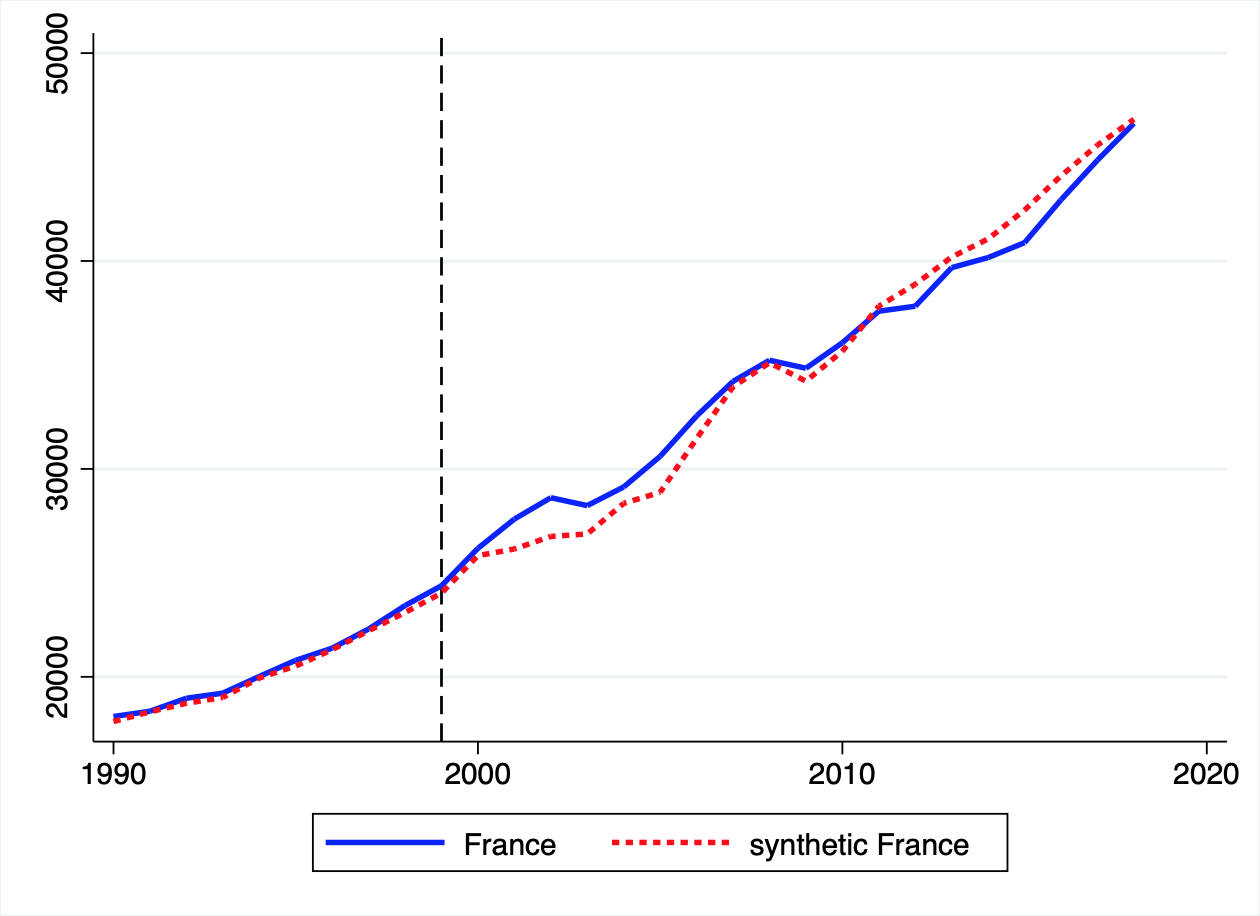}  & 
    \includegraphics[width=.33\textwidth]{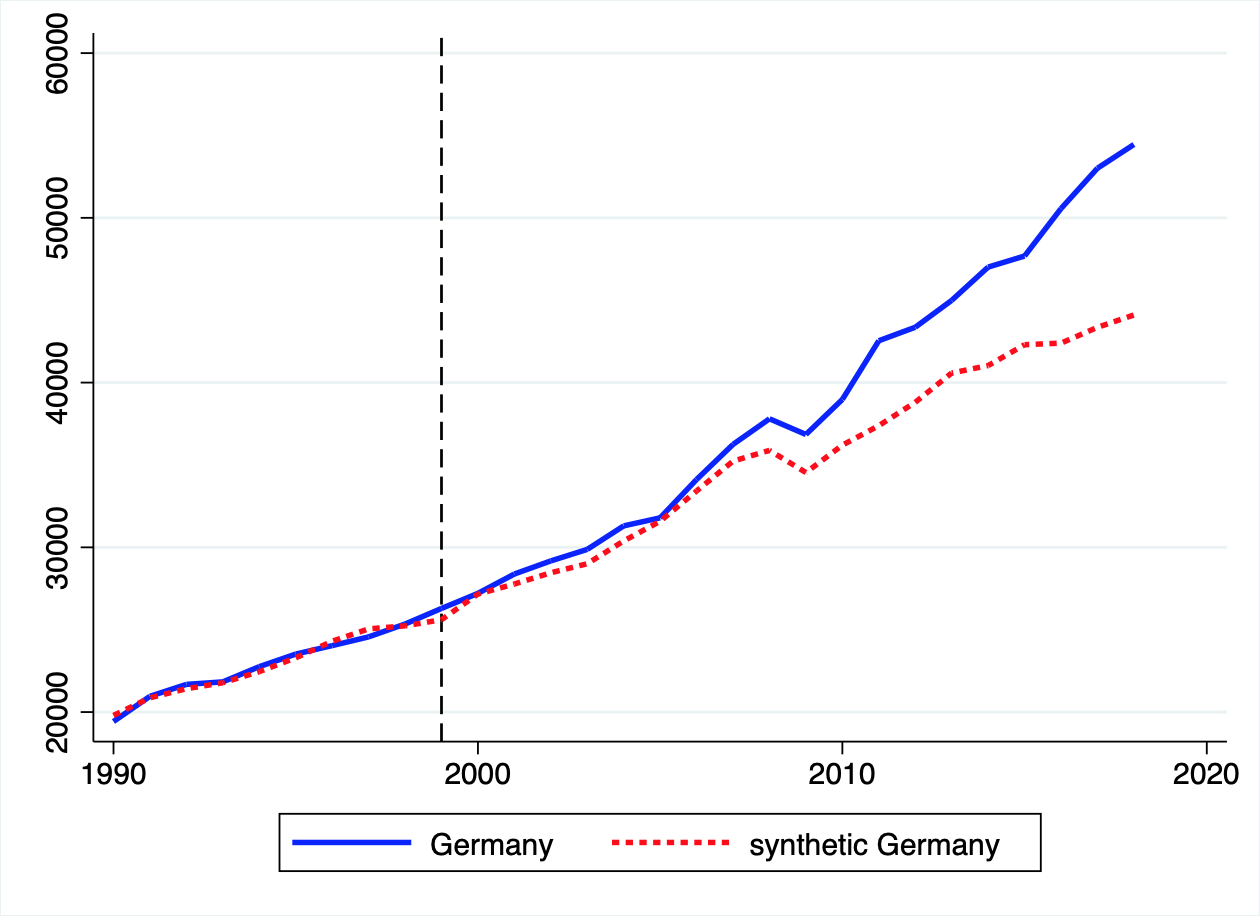} &
    \includegraphics[width=.33\textwidth]{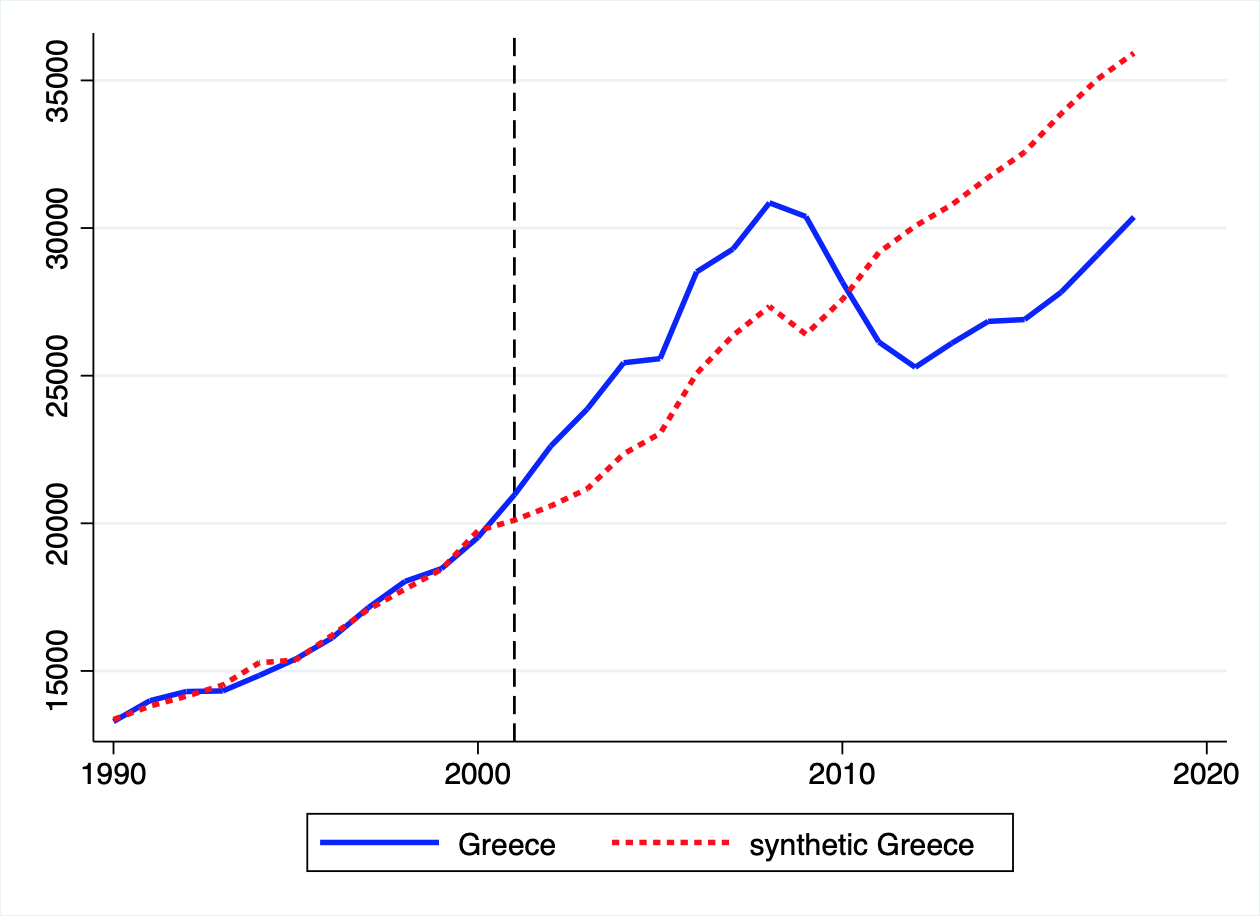} \\
    \includegraphics[width=.33\textwidth]{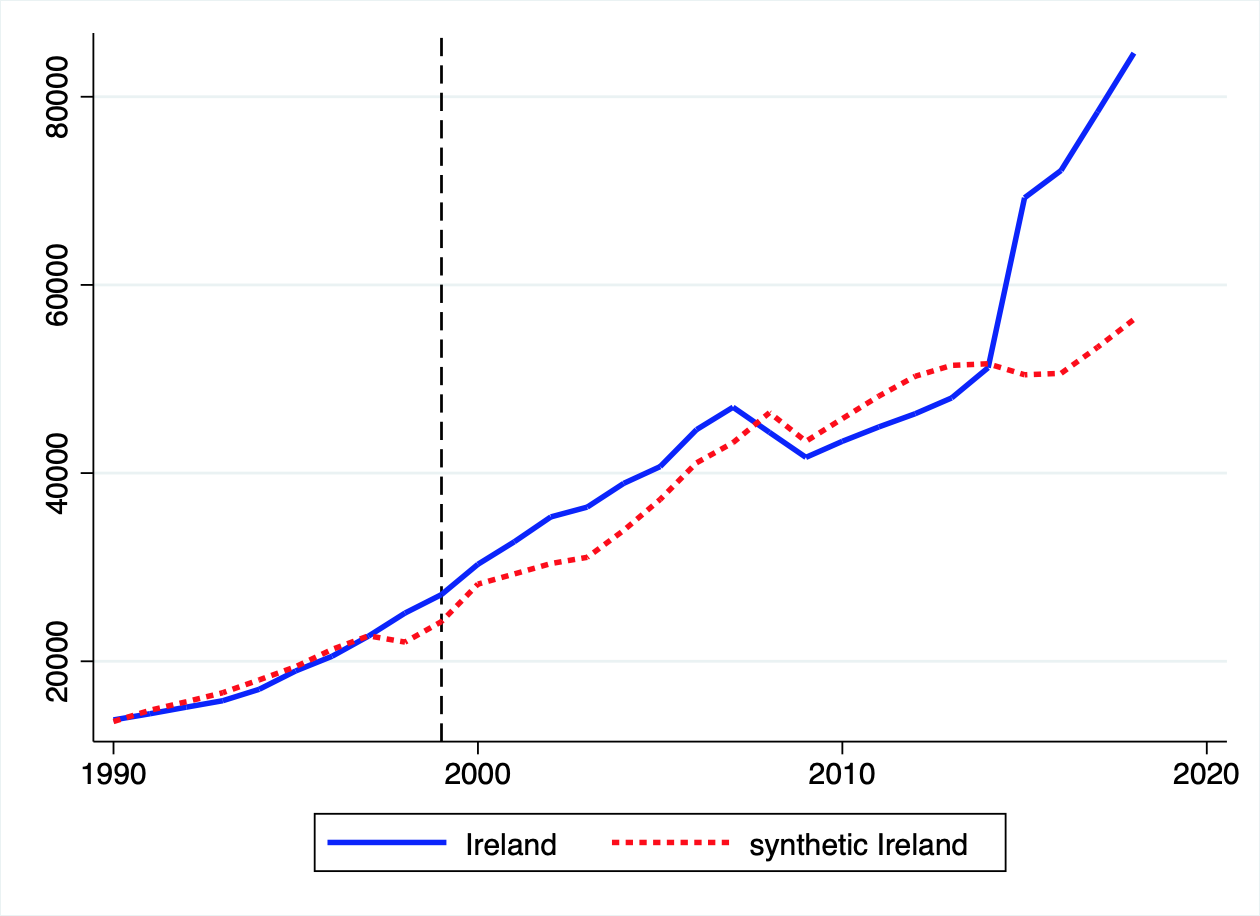} &
    \includegraphics[width=.33\textwidth]{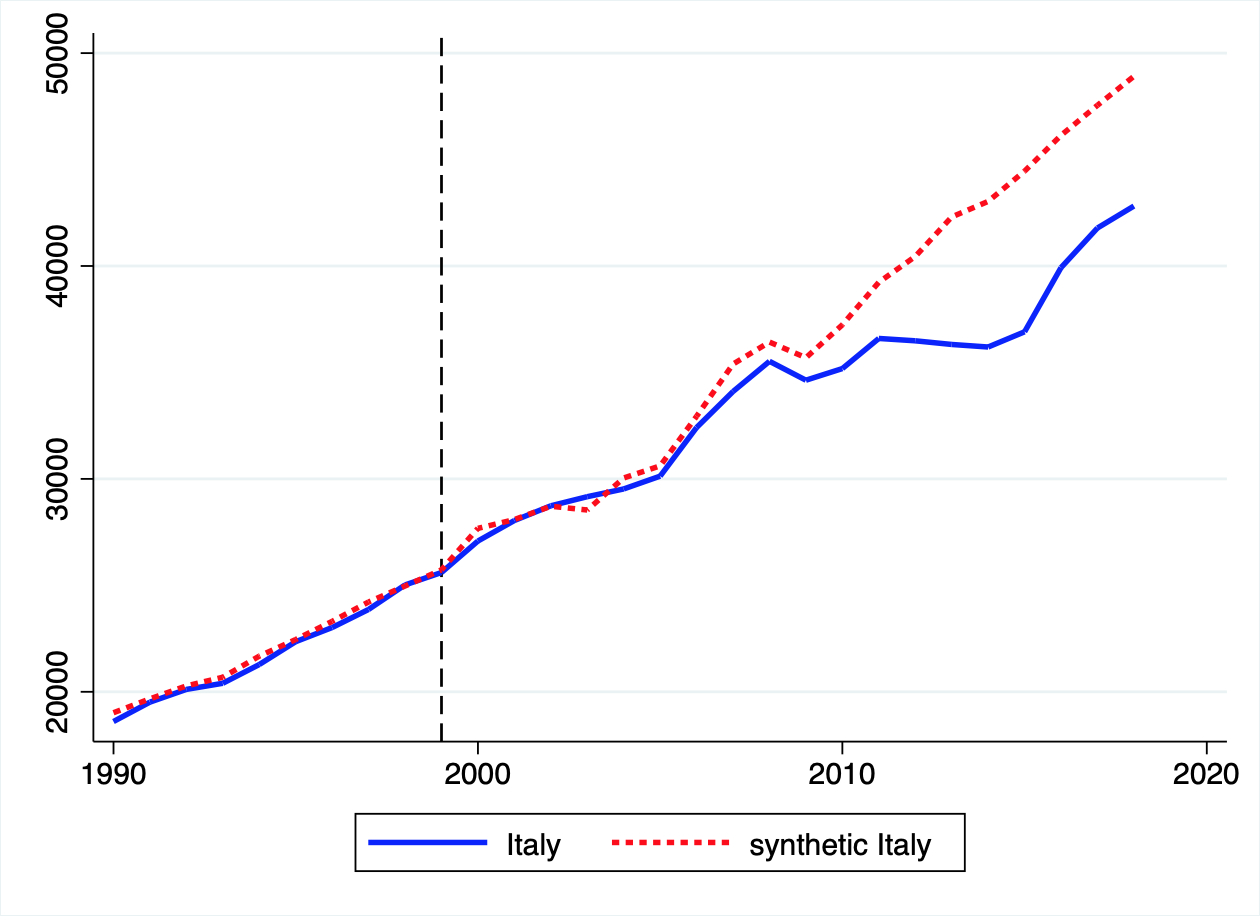}   &
    \includegraphics[width=.33\textwidth]{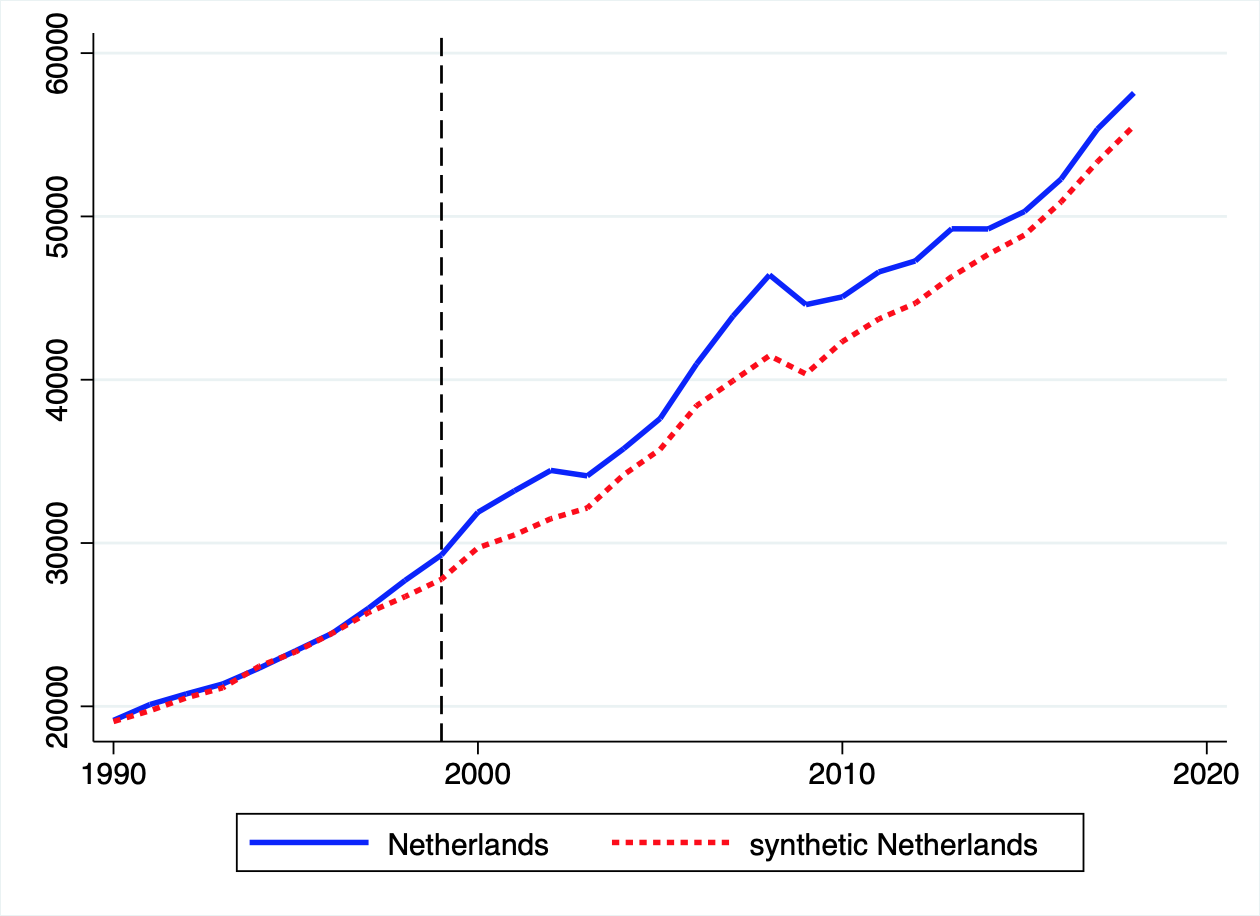} \\
    \includegraphics[width=.33\textwidth]{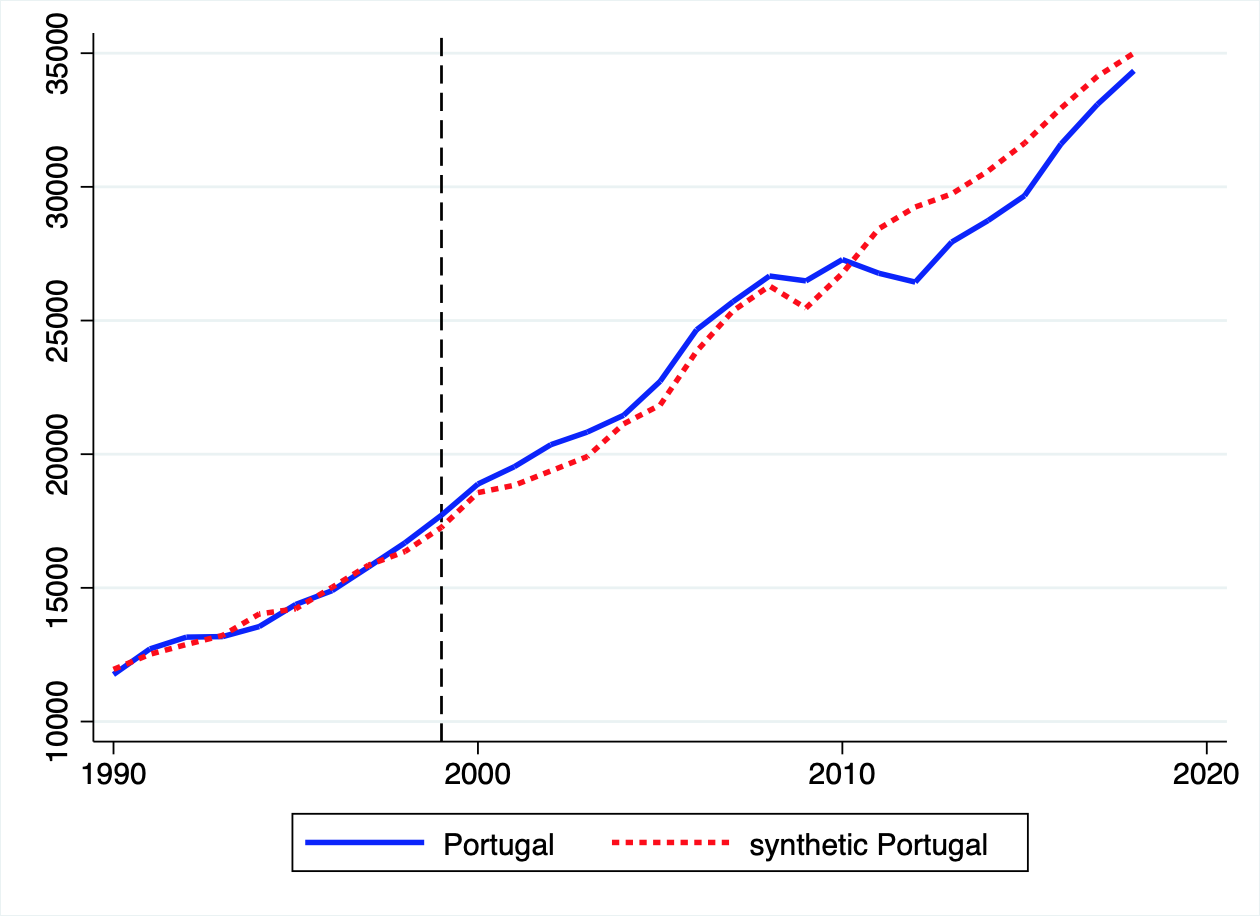} &
   \includegraphics[width=.33\textwidth]{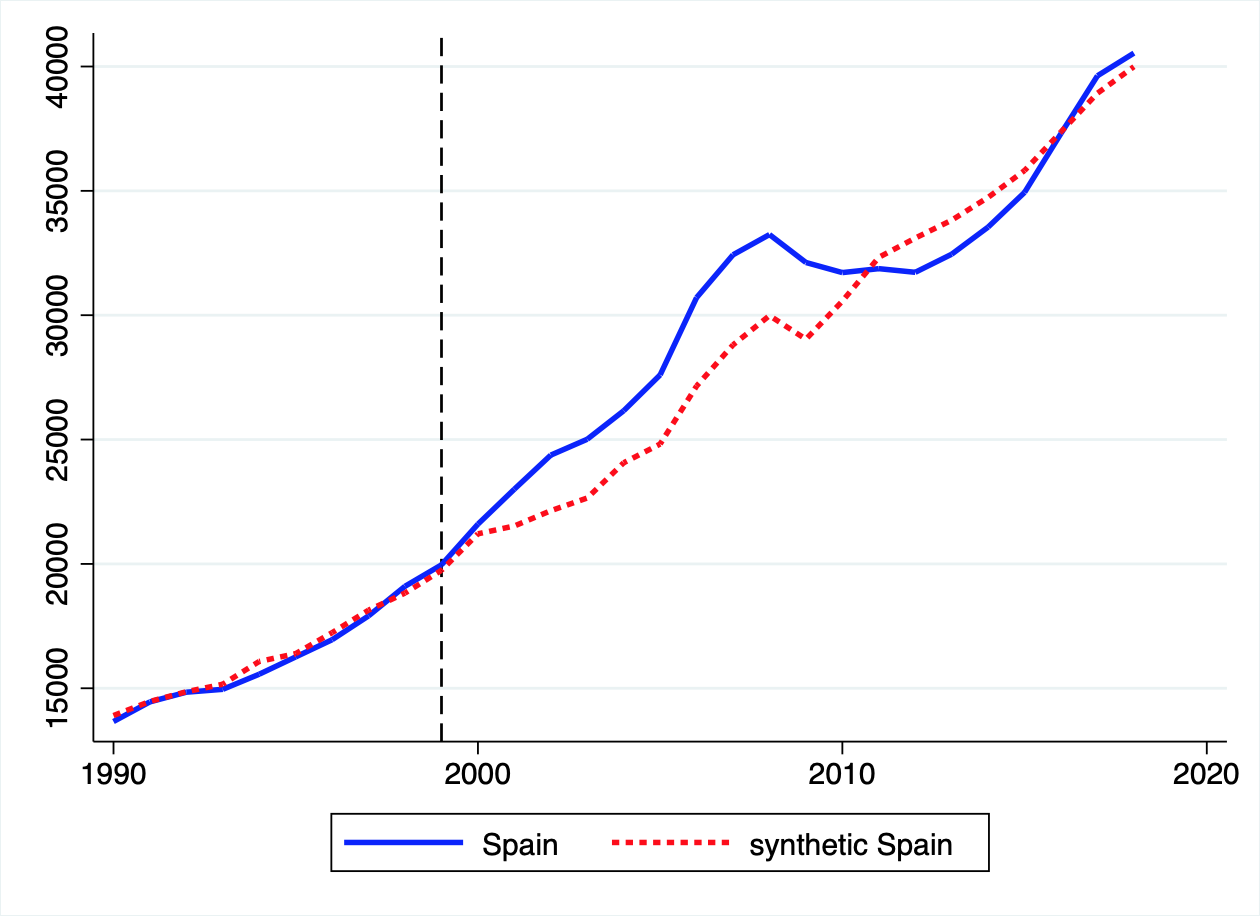} &
  \end{tabular}
   \label{fig:matching_GDP}
\end{figure}
\footnotesize
\justify\emph{Note:} The figure shows the actual  (blue solid) and the synthetic (red dashed) GDP series computed using the synthetic control method of \cite{abadie_economic_2003}. The matching window is 1990-1998. The series are in euros per capita. \normalsize

\begin{figure}[H]
  \caption{Actual and synthetic series for consumption}
\centering
  \begin{tabular}{@{}ccc@{}}
    \includegraphics[width=0.33\textwidth]{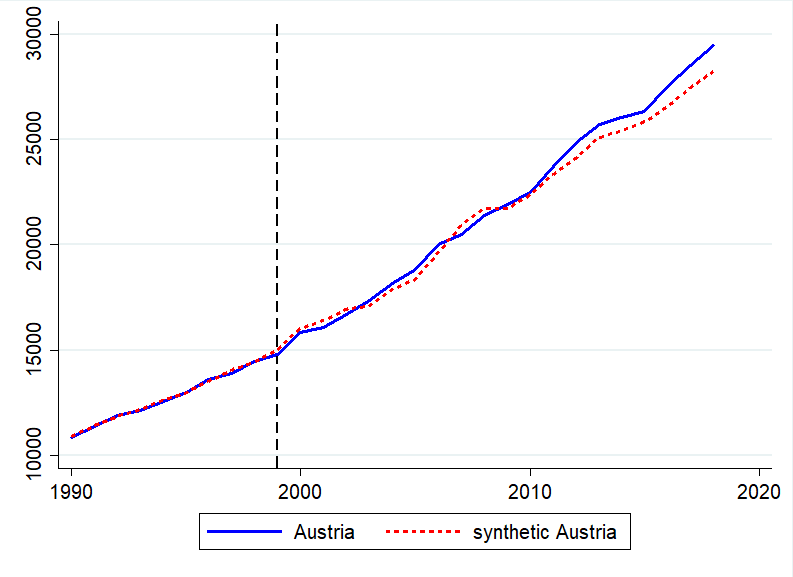} &
    \includegraphics[width=0.33\textwidth]{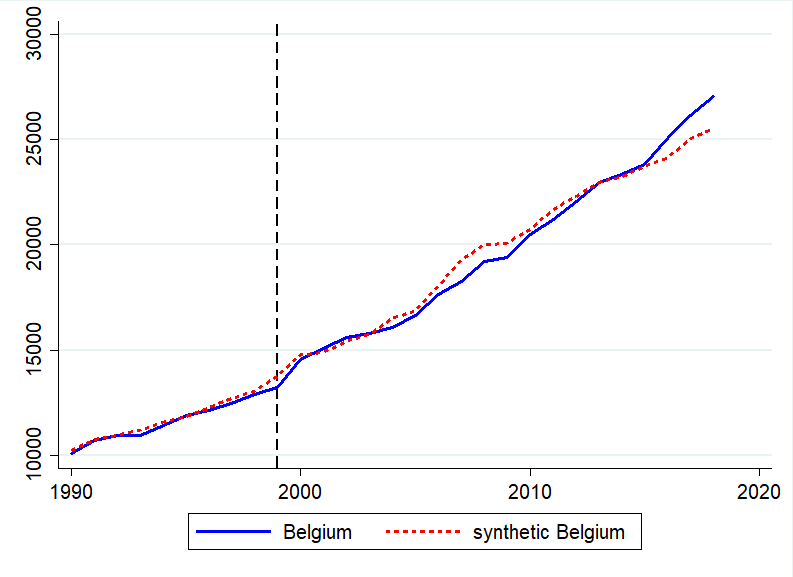} &
    \includegraphics[width=0.33\textwidth]{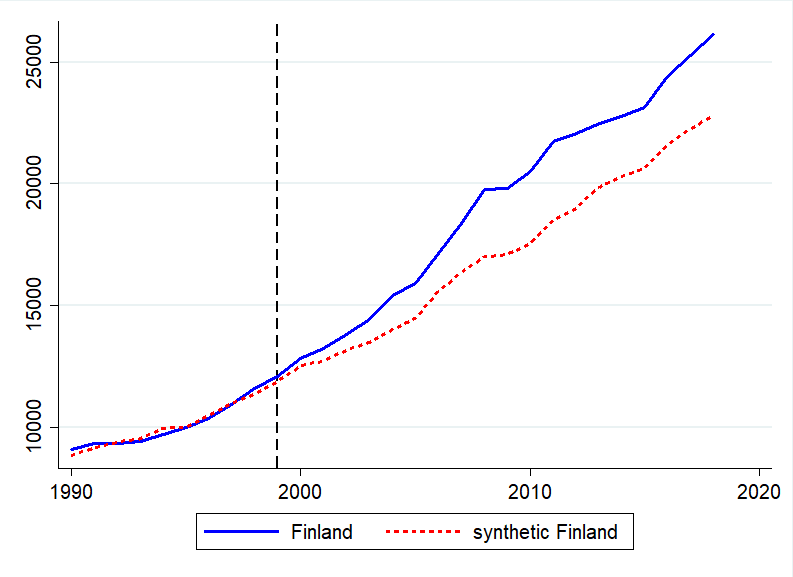} \\
   \includegraphics[width=.33\textwidth]{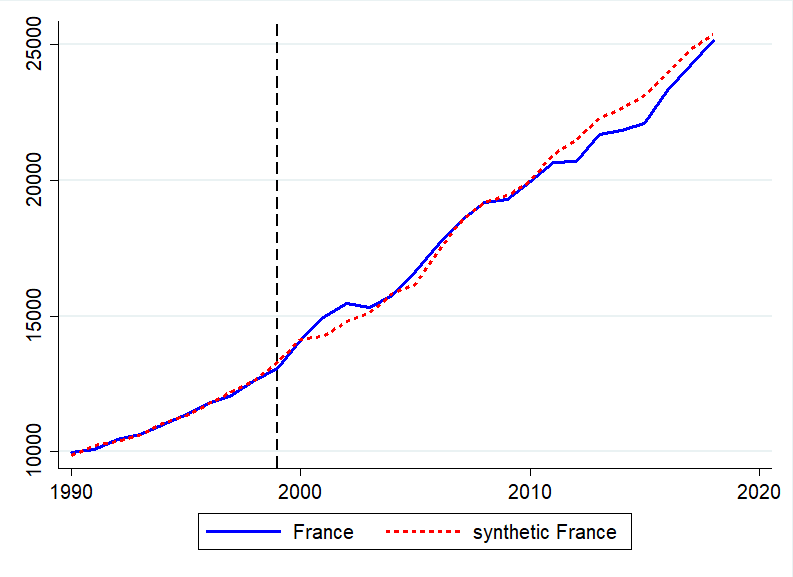}  & 
    \includegraphics[width=.33\textwidth]{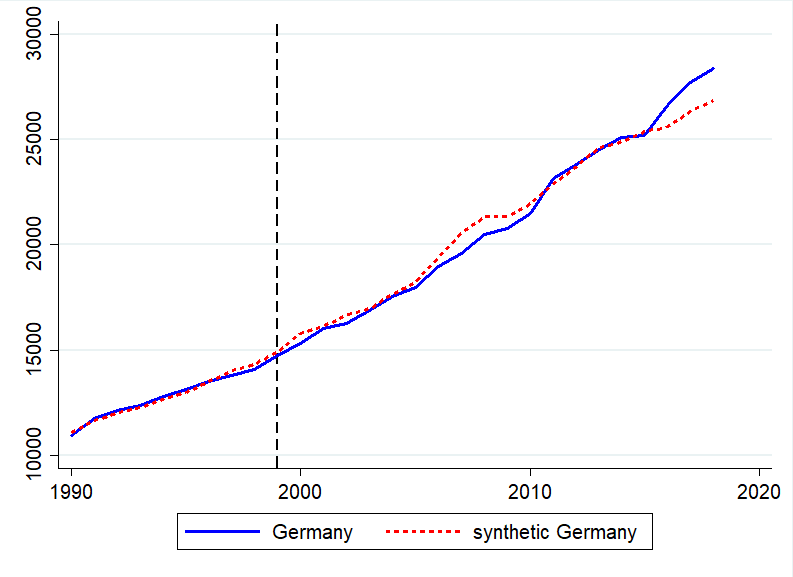} &
    \includegraphics[width=.33\textwidth]{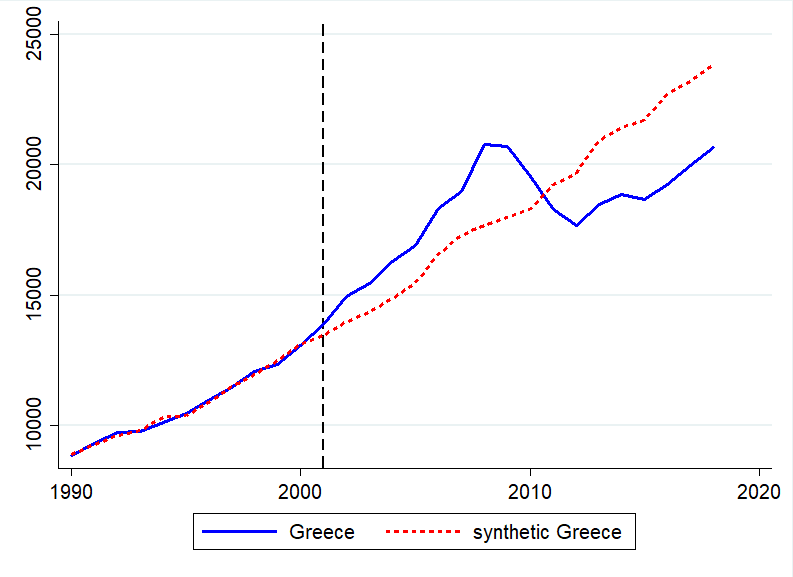} \\
    \includegraphics[width=.33\textwidth]{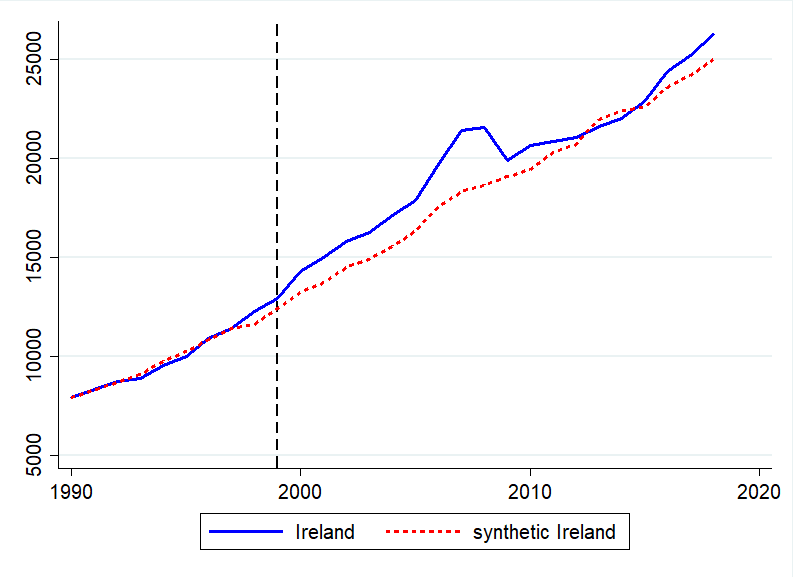} &
    \includegraphics[width=.33\textwidth]{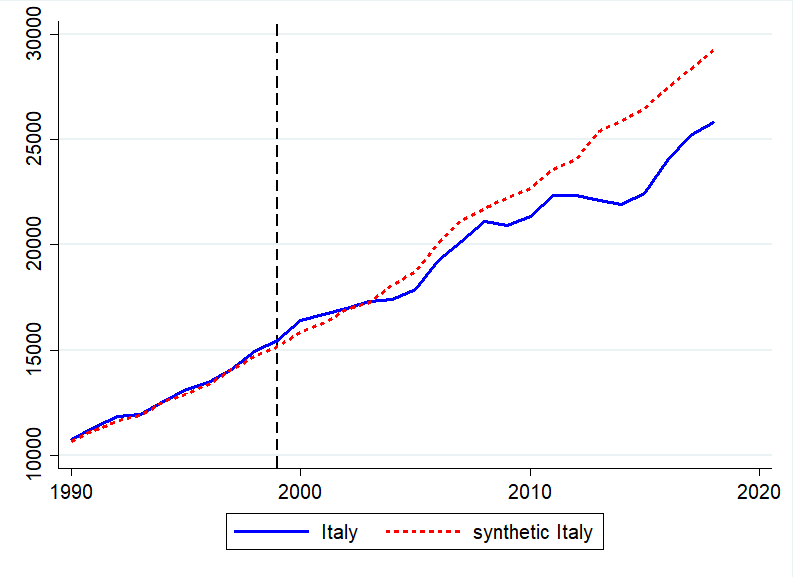}   &
    \includegraphics[width=.33\textwidth]{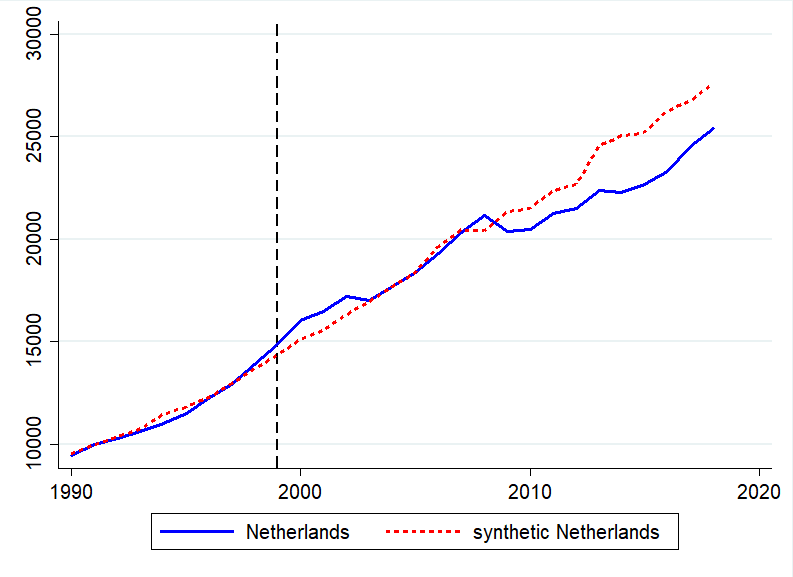} \\
    \includegraphics[width=.33\textwidth]{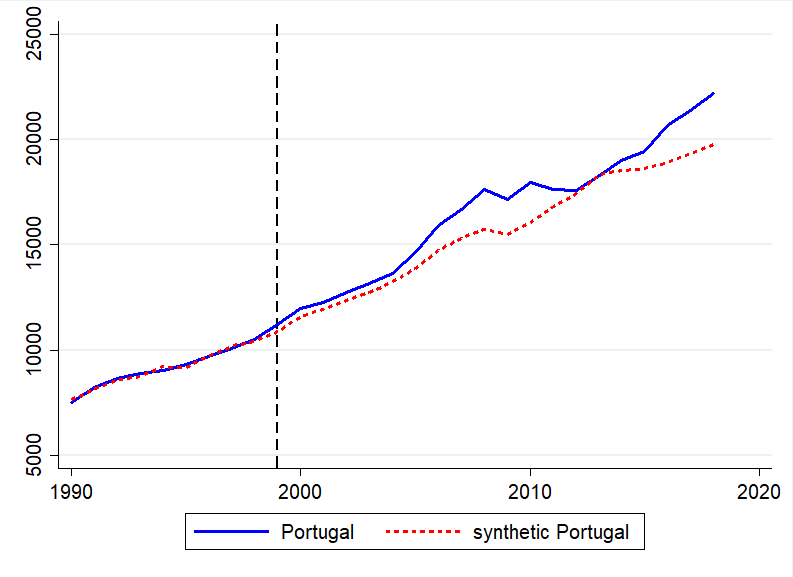} &
   \includegraphics[width=.33\textwidth]{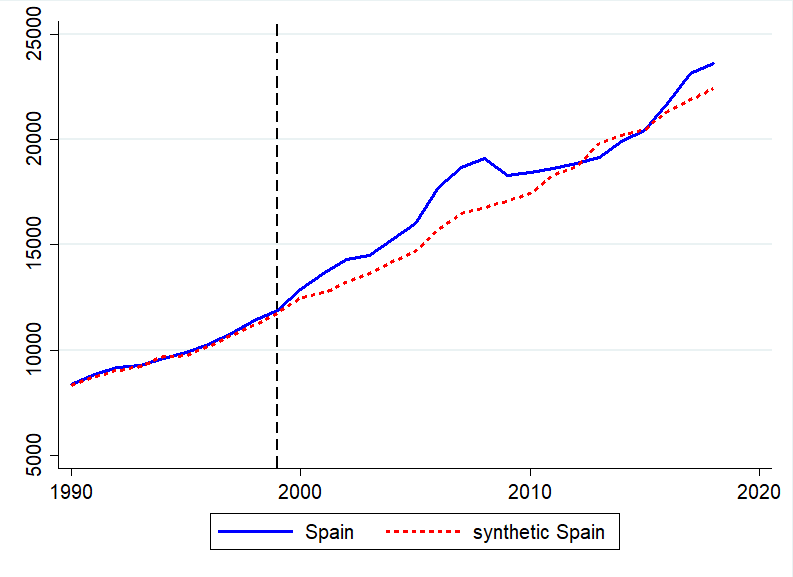} &
  \end{tabular}
   \label{fig:matching_C}
\end{figure}
\footnotesize
\justify\emph{Note:} The figure shows the actual  (blue solid) and the synthetic (red dashed) consumption series computed using the synthetic control method of \cite{abadie_economic_2003}. The matching window is 1990-1998. The series are in euros per capita. \normalsize

\begin{figure}[H]
  \caption{Actual and synthetic series for net national income}
\centering
  \begin{tabular}{@{}ccc@{}}
    \includegraphics[width=0.33\textwidth]{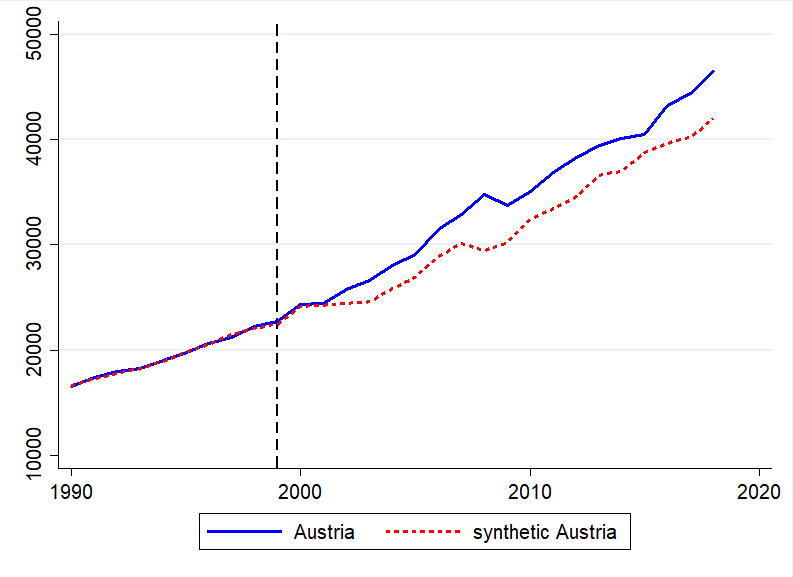} &
    \includegraphics[width=0.33\textwidth]{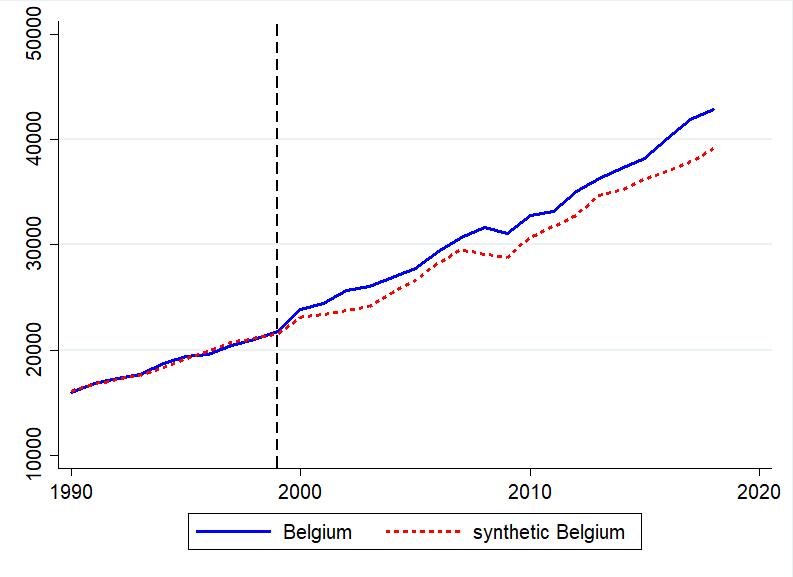} &
    \includegraphics[width=0.33\textwidth]{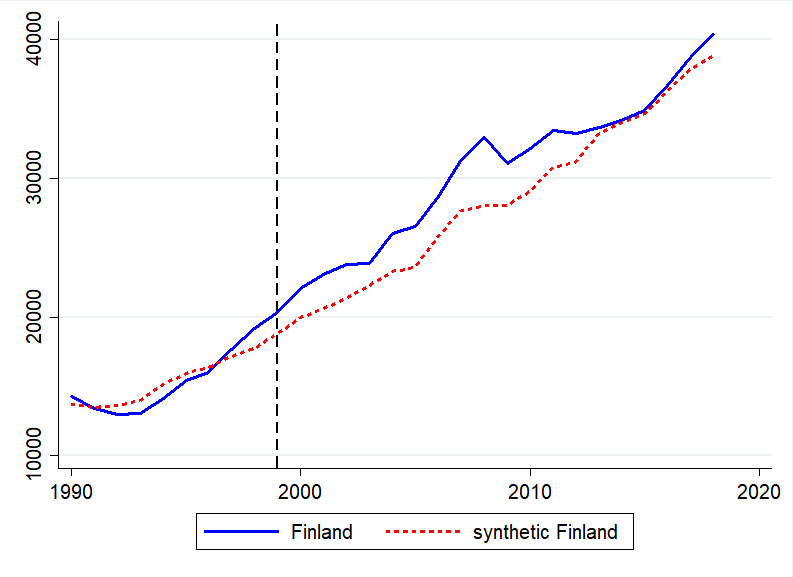} \\
   \includegraphics[width=.33\textwidth]{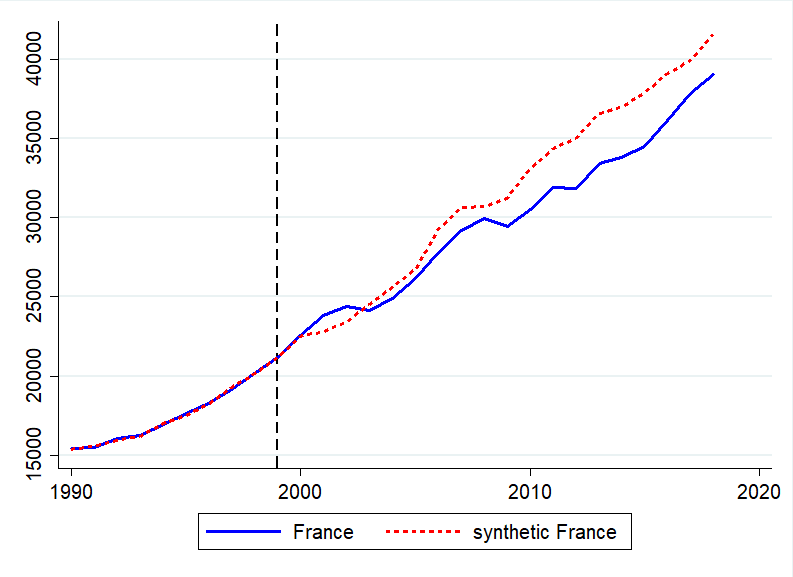}  & 
    \includegraphics[width=.33\textwidth]{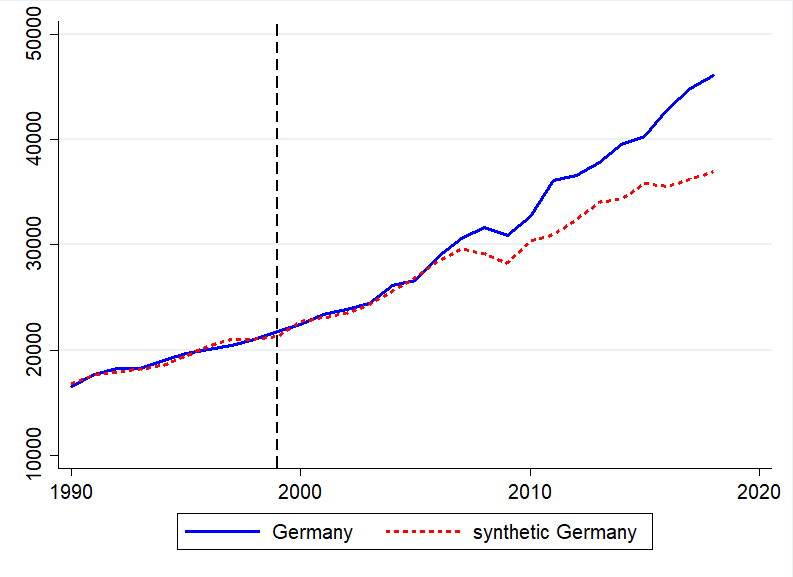} &
    \includegraphics[width=.33\textwidth]{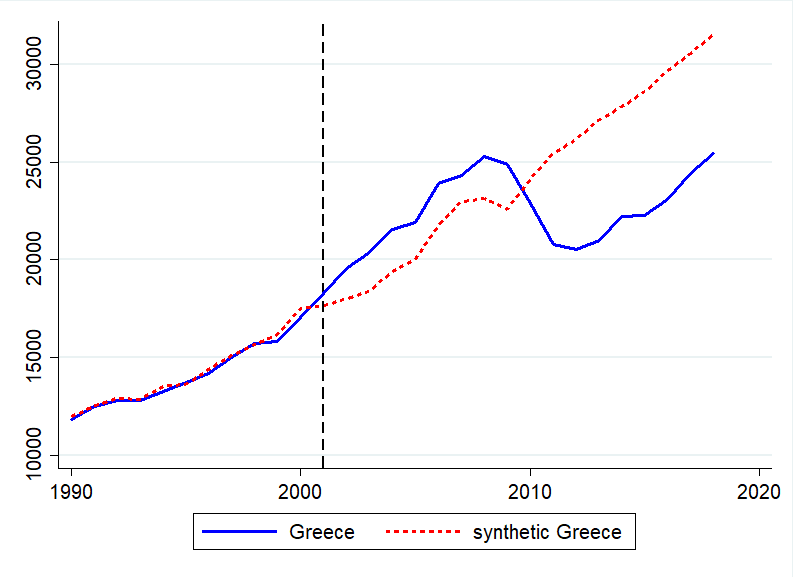} \\
    \includegraphics[width=.33\textwidth]{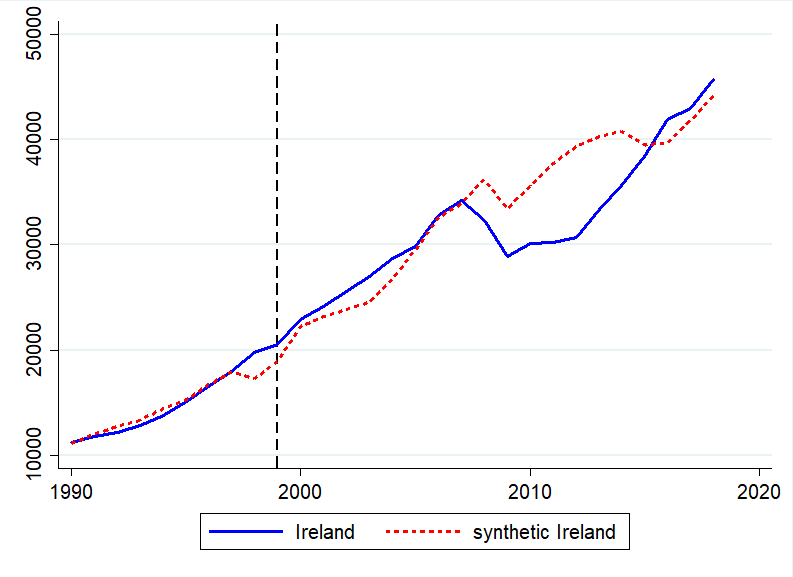} &
    \includegraphics[width=.33\textwidth]{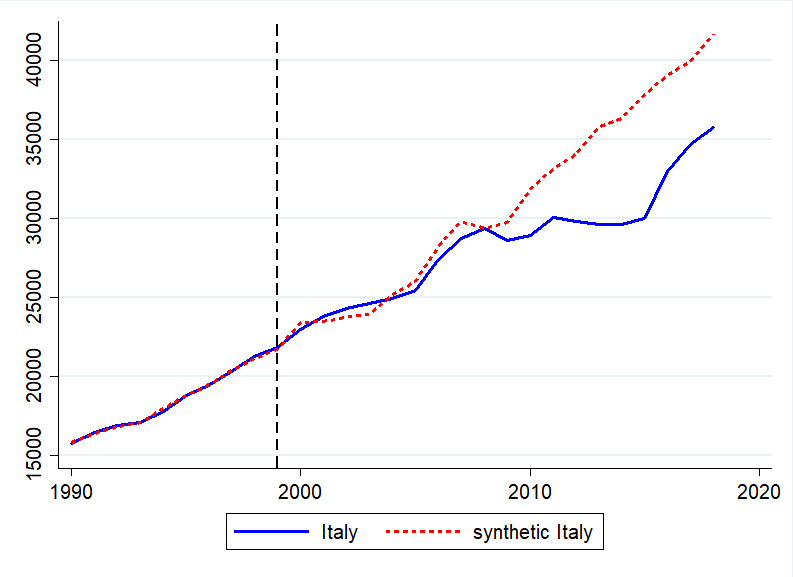}   &
    \includegraphics[width=.33\textwidth]{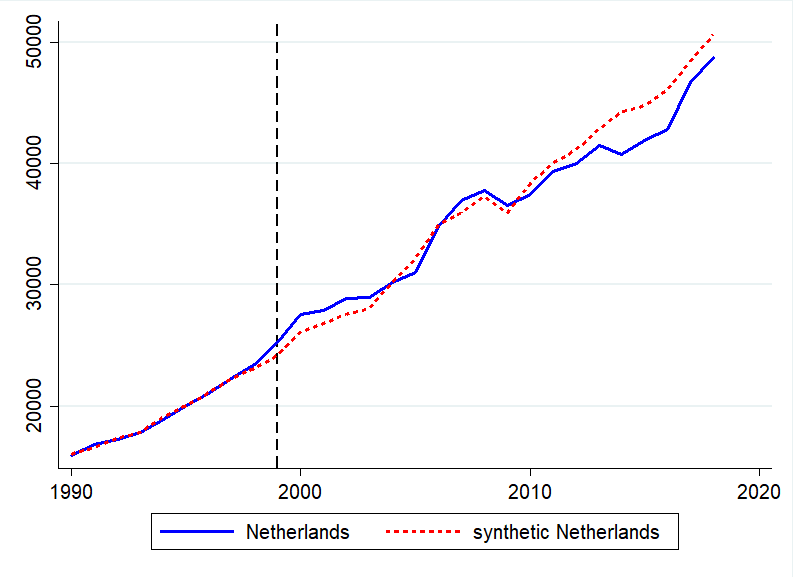} \\
    \includegraphics[width=.33\textwidth]{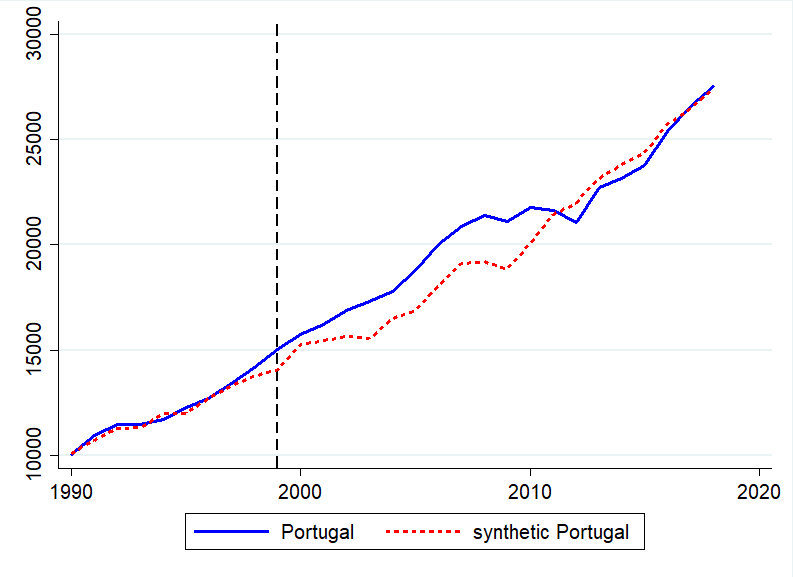} &
   \includegraphics[width=.33\textwidth]{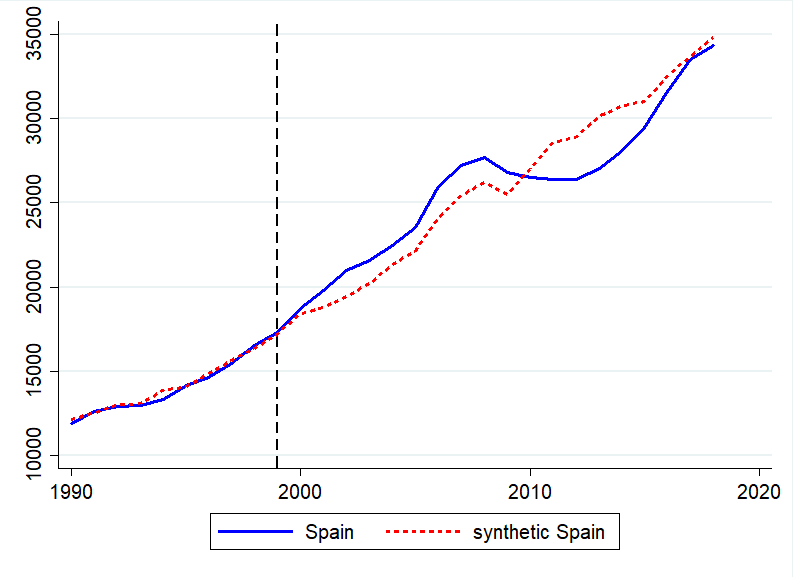} &

  \end{tabular}
   \label{fig:matching_NNI}
\end{figure}
\footnotesize
\justify\emph{Note:} The figure shows the actual  (blue solid) and the synthetic (red dashed) net national income series computed using the synthetic control method of \cite{abadie_economic_2003}. The matching window is 1990-1998. The series are in euros per capita. \normalsize

\begin{figure}[H]
  \caption{Actual and synthetic series for disposable national income}
\centering
  \begin{tabular}{@{}ccc@{}}
    \includegraphics[width=0.33\textwidth]{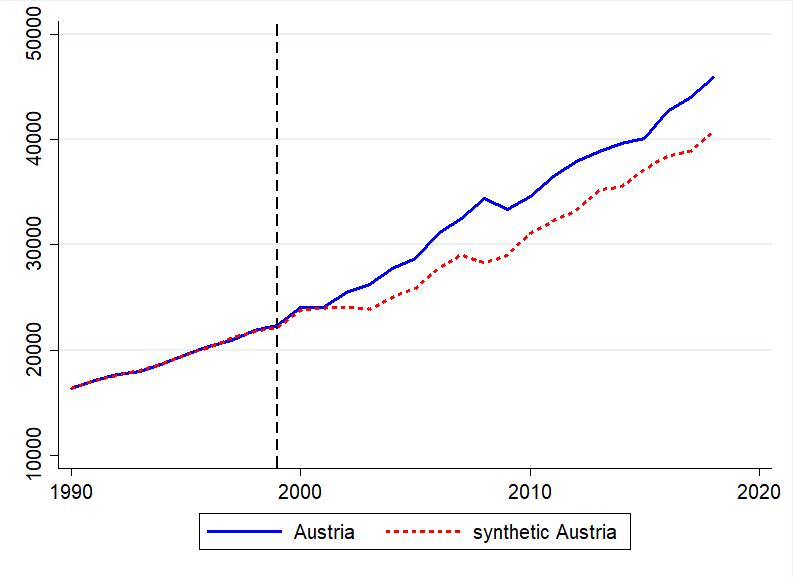} &
    \includegraphics[width=0.33\textwidth]{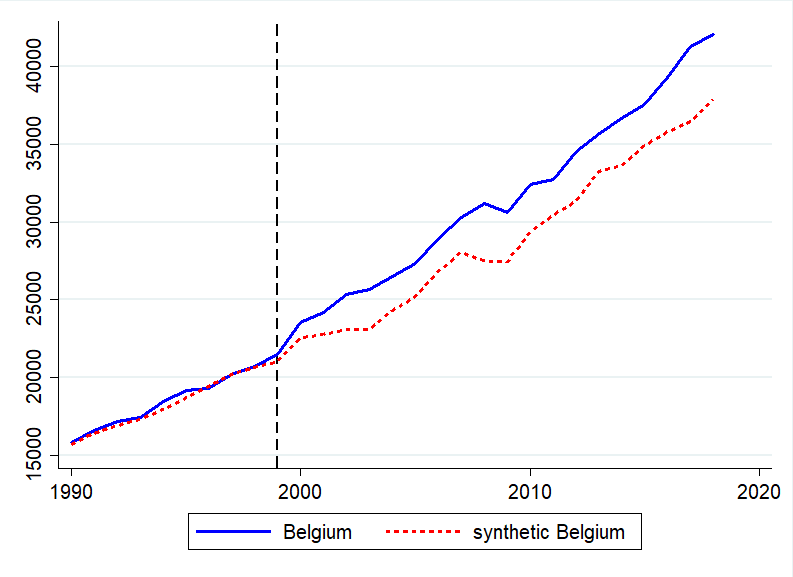} &
    \includegraphics[width=0.33\textwidth]{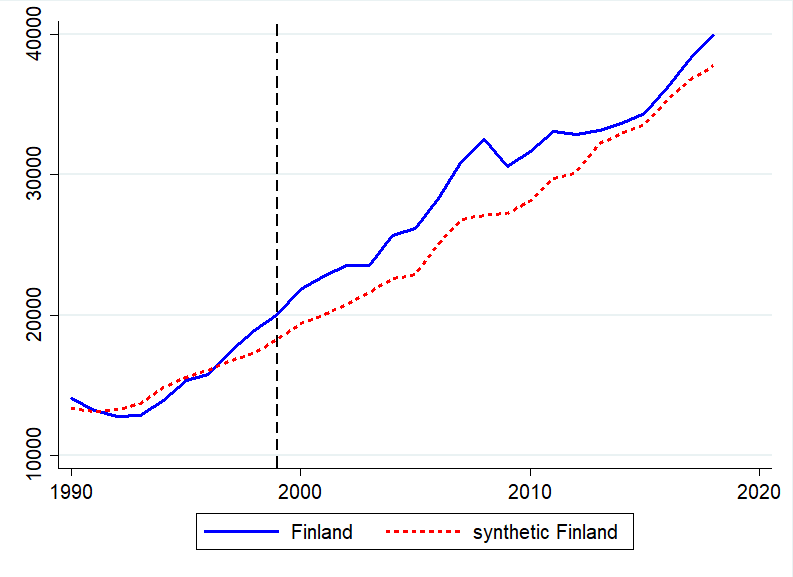} \\
   \includegraphics[width=.33\textwidth]{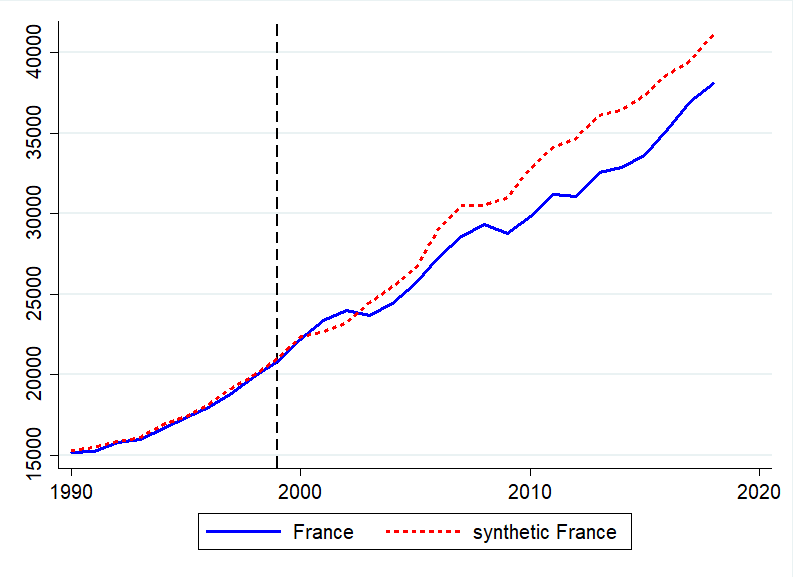}  & 
    \includegraphics[width=.33\textwidth]{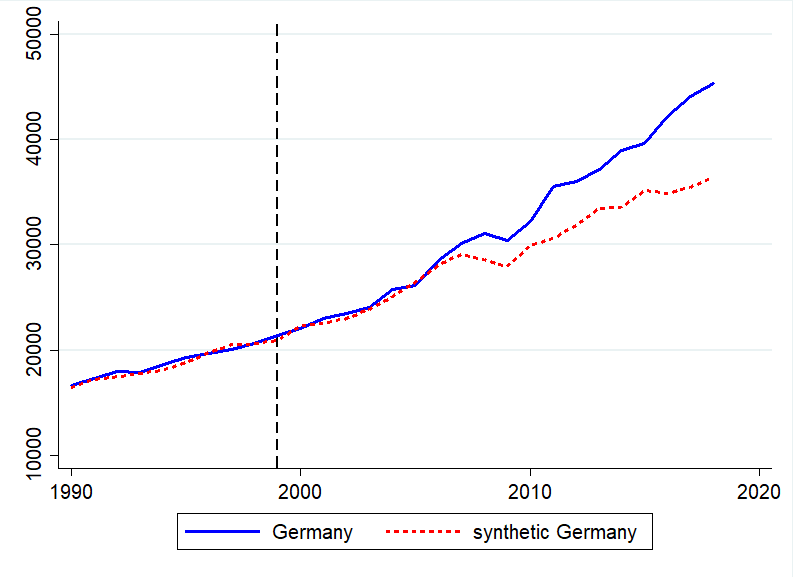} &
    \includegraphics[width=.33\textwidth]{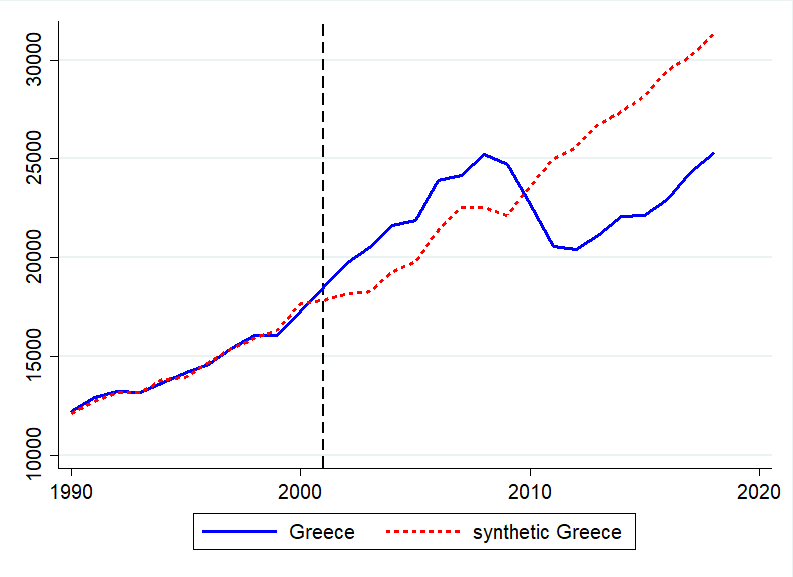} \\
    \includegraphics[width=.33\textwidth]{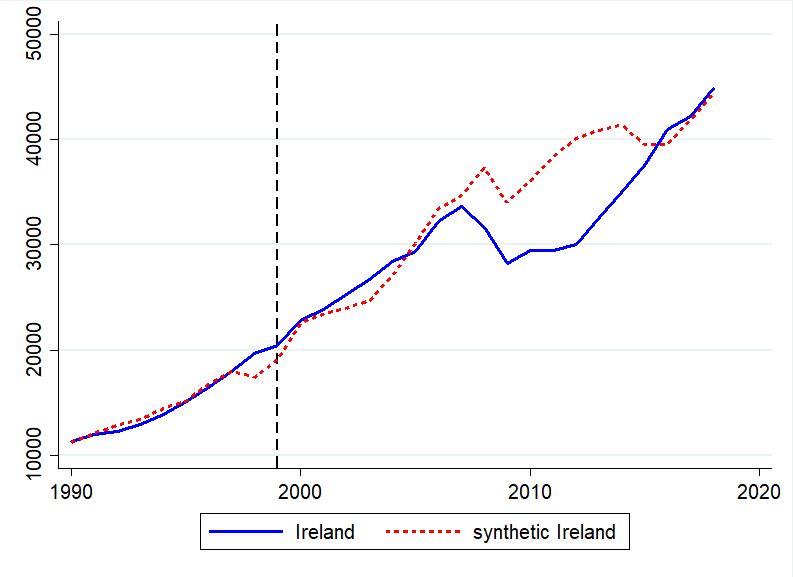} &
    \includegraphics[width=.33\textwidth]{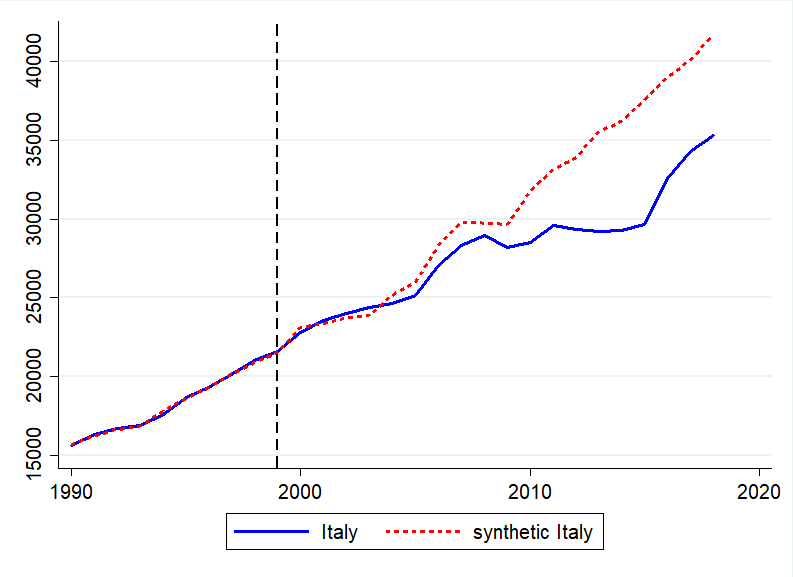}   &
    \includegraphics[width=.33\textwidth]{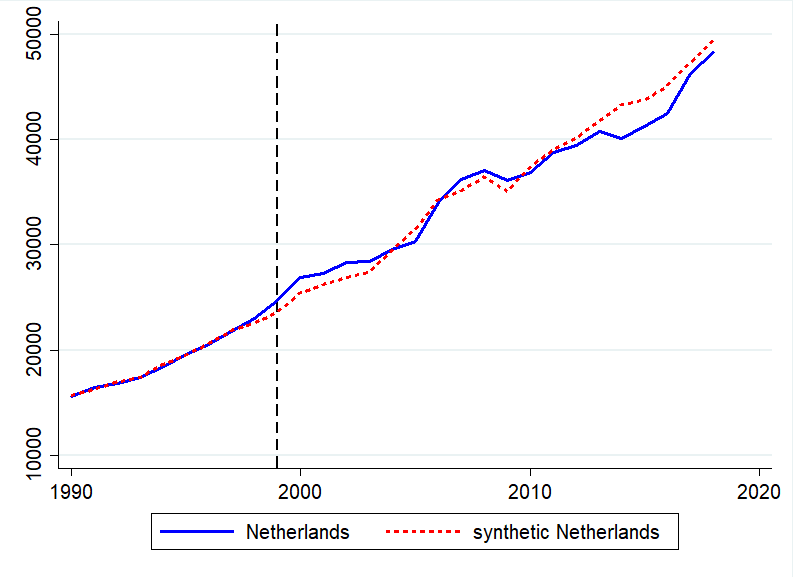} \\
    \includegraphics[width=.33\textwidth]{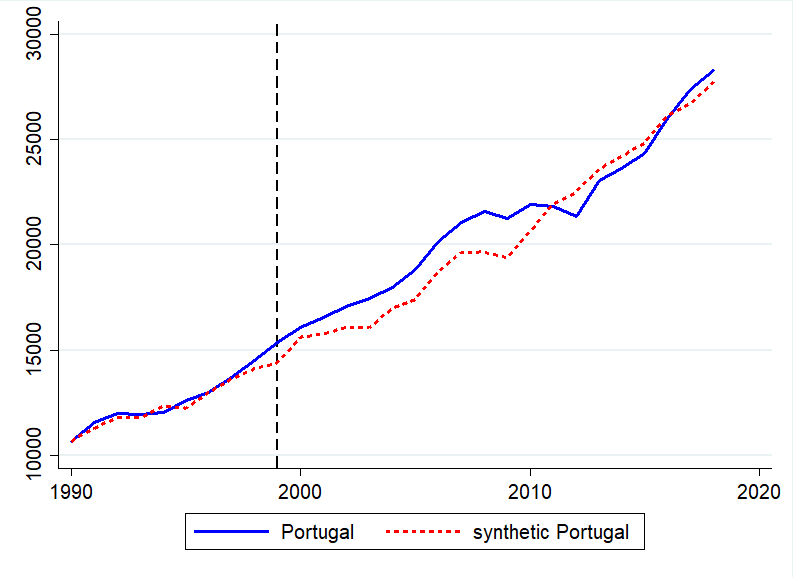} &
   \includegraphics[width=.33\textwidth]{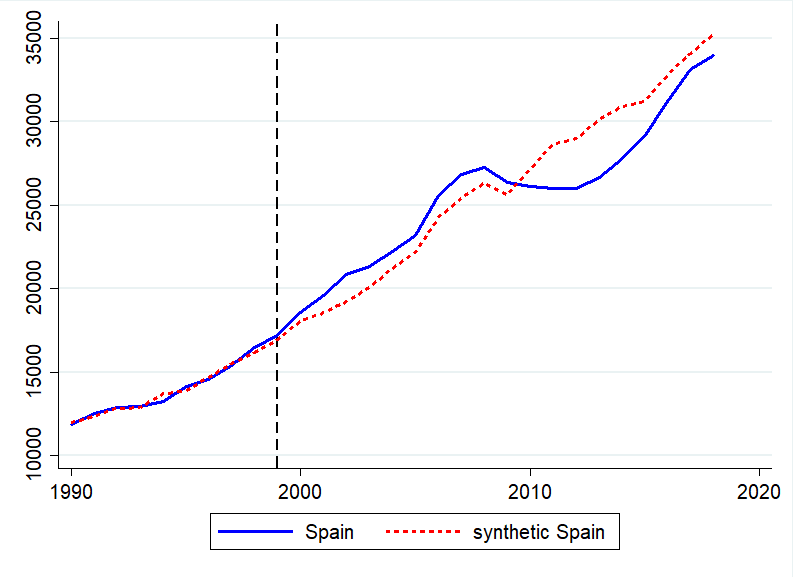} &

  \end{tabular}
   \label{fig:matching_DNI}
\end{figure}
\footnotesize
\justify\emph{Note:} The figure shows the actual  (blue solid) and the synthetic (red dashed) disposable national income series computed using the synthetic control method of \cite{abadie_economic_2003}. The matching window is 1990-1998. The series are in euros per capita. \normalsize

\begin{figure}[H]
  \caption{Actual and synthetic series for government consumption}
\centering
  \begin{tabular}{@{}ccc@{}}
    \includegraphics[width=0.33\textwidth]{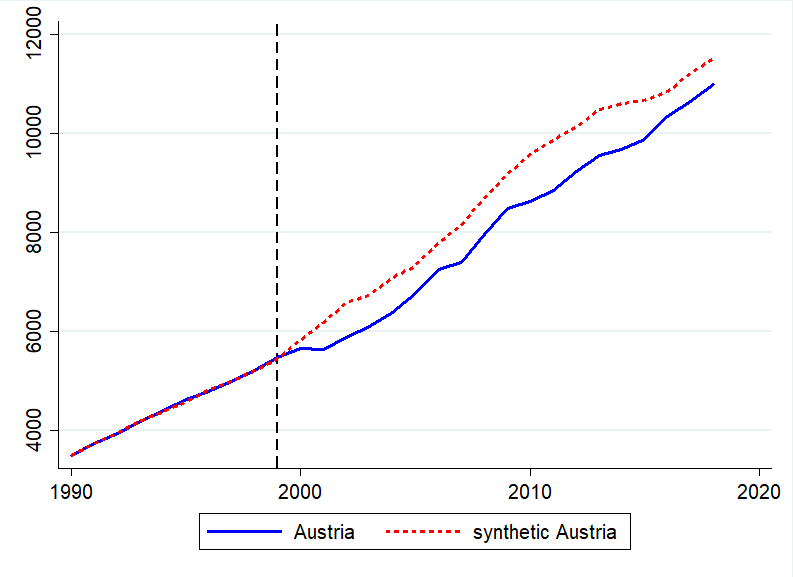} &
    \includegraphics[width=0.33\textwidth]{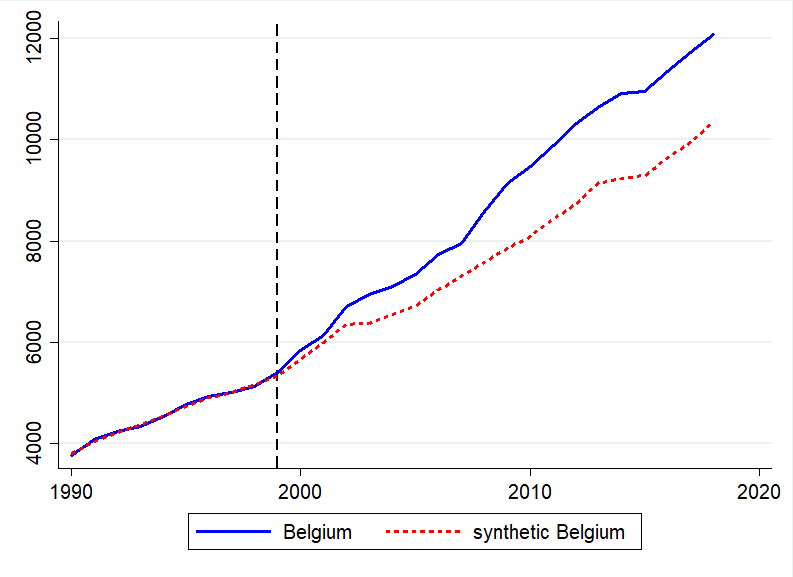} &
    \includegraphics[width=0.33\textwidth]{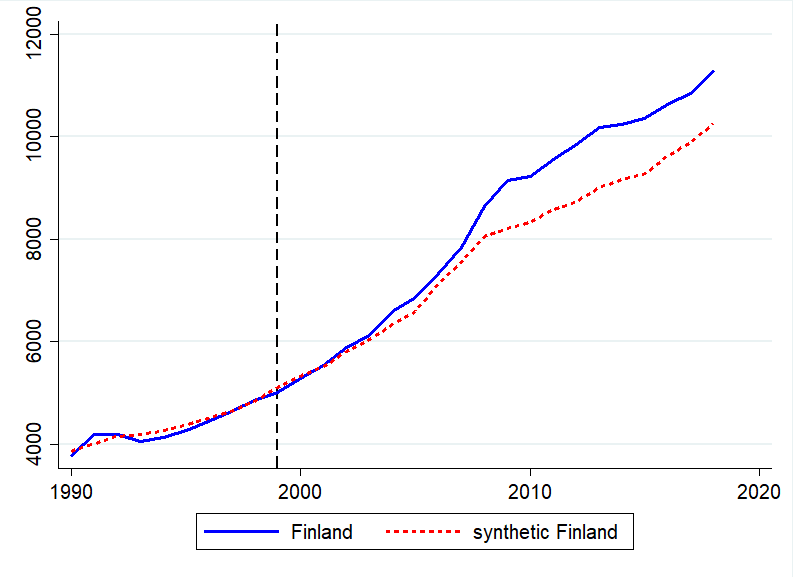} \\
   \includegraphics[width=.33\textwidth]{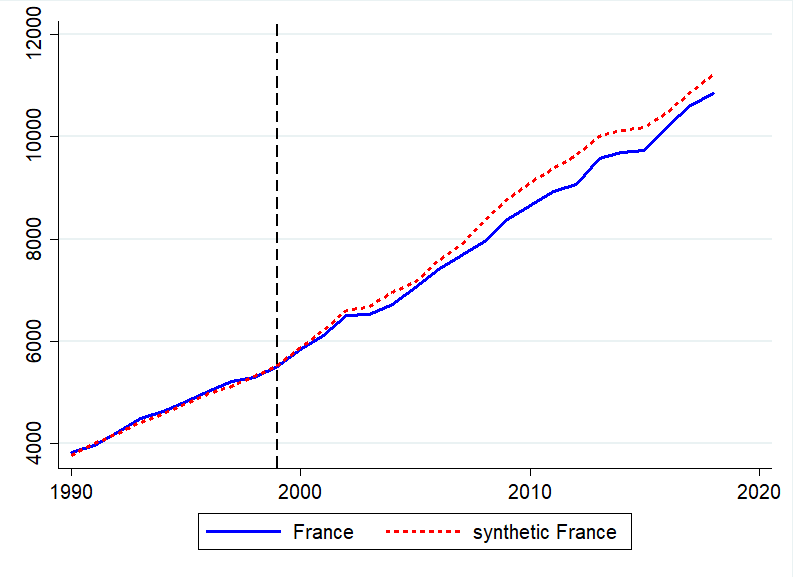}  & 
    \includegraphics[width=.33\textwidth]{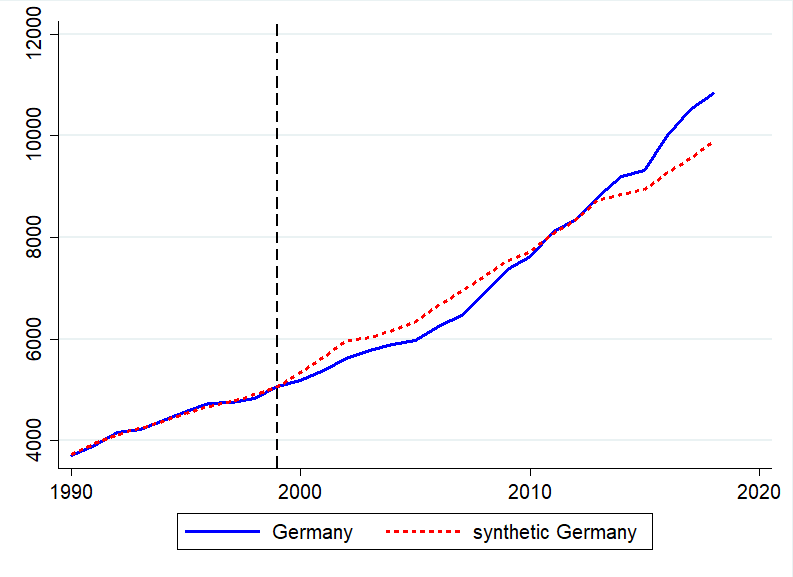} &
    \includegraphics[width=.33\textwidth]{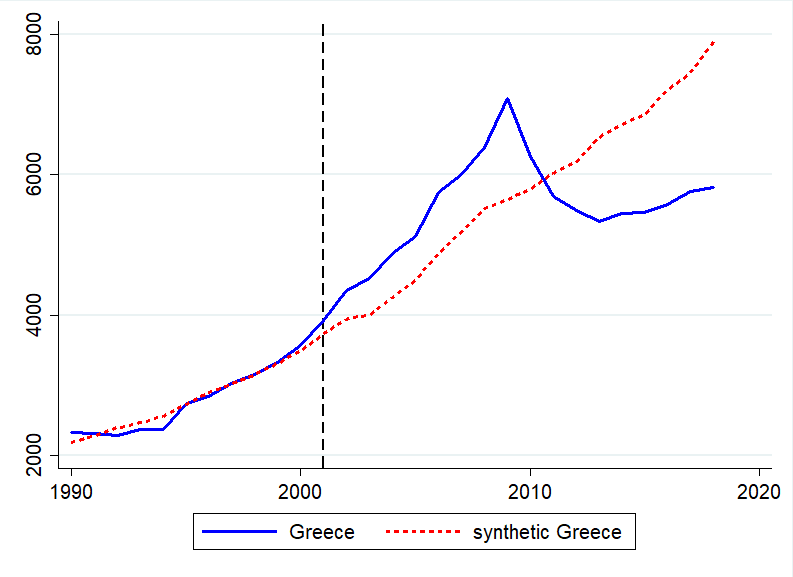} \\
    \includegraphics[width=.33\textwidth]{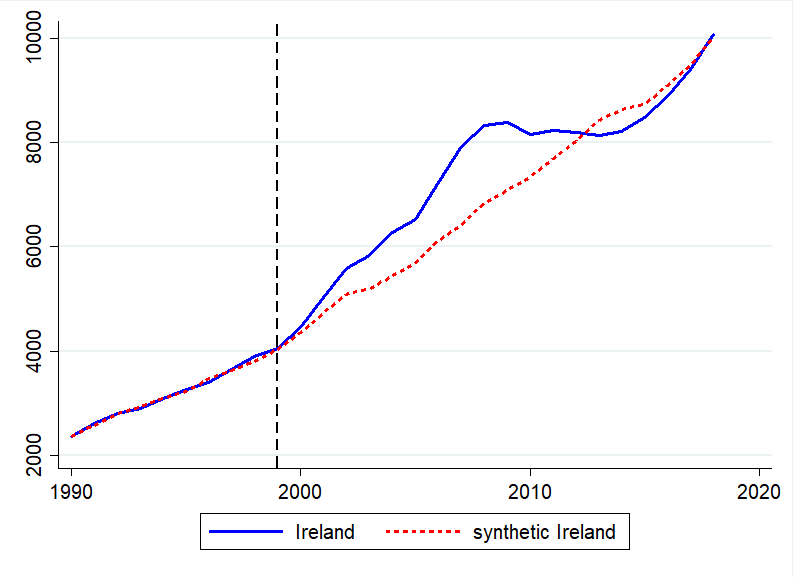} &
    \includegraphics[width=.33\textwidth]{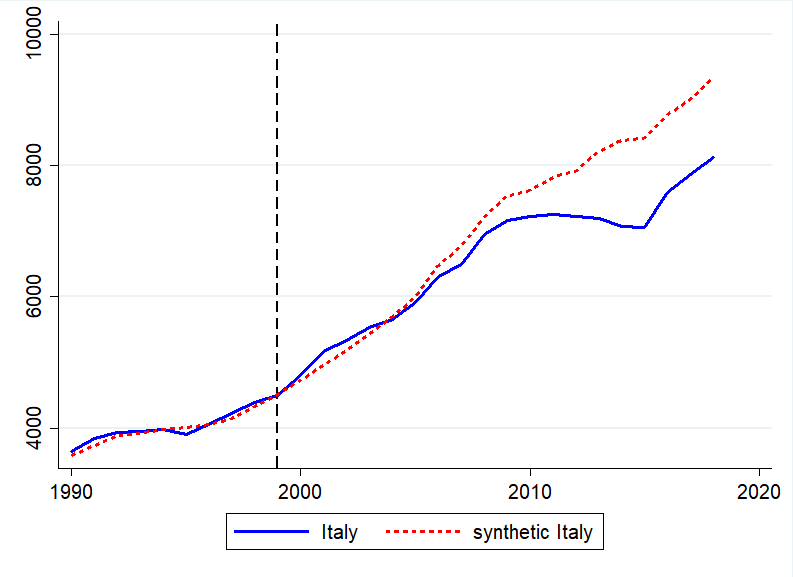}   &
    \includegraphics[width=.33\textwidth]{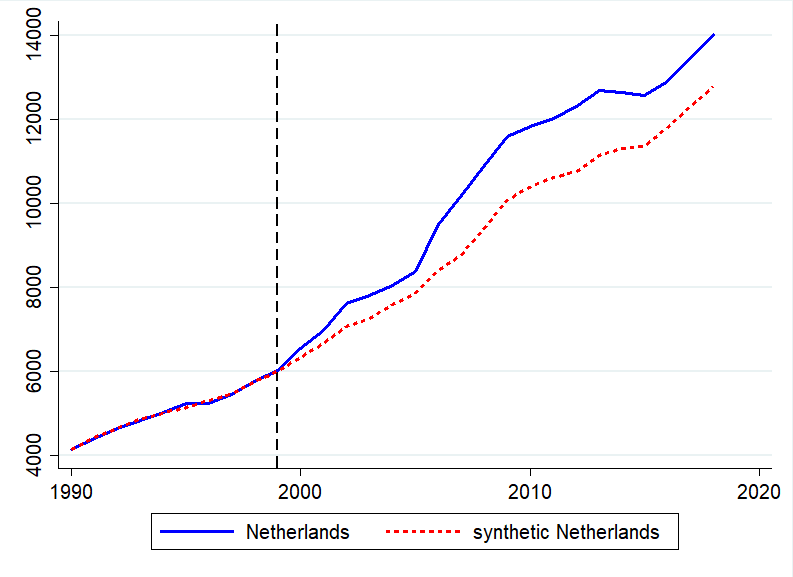} \\
    \includegraphics[width=.33\textwidth]{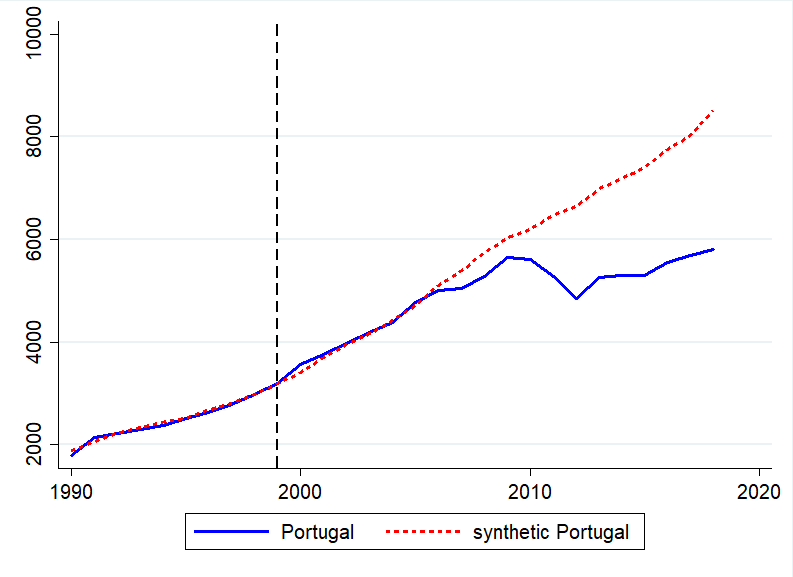} &
   \includegraphics[width=.33\textwidth]{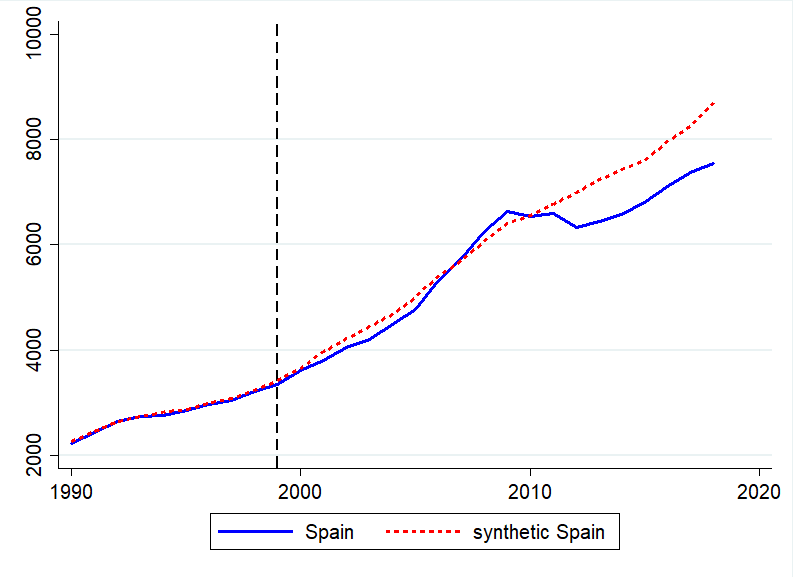} &

  \end{tabular}
   \label{fig:matching_G}
\end{figure}
\footnotesize
\justify\emph{Note:} The figure shows the actual  (blue solid) and the synthetic (red dashed) government consumption series computed using the synthetic control method of \cite{abadie_economic_2003}. The matching window is 1990-1998. The series are in euros per capita. \normalsize

 \begin{figure}[H]
\centering\footnotesize
\caption{Distribution of country-level $\hat\beta_2$}
\subfigure[$\hat\beta_2$]{\includegraphics[width=0.4\textwidth]{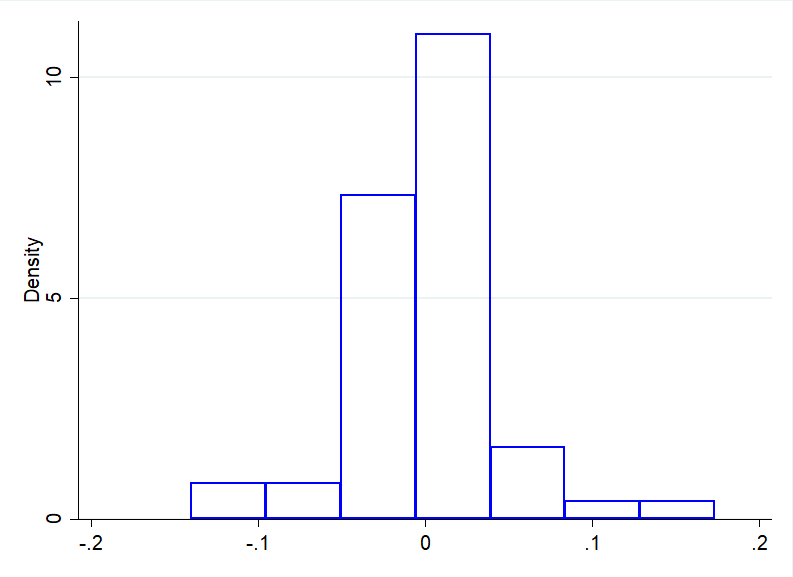}}
\subfigure[p-value]{\includegraphics[width=0.4\textwidth]{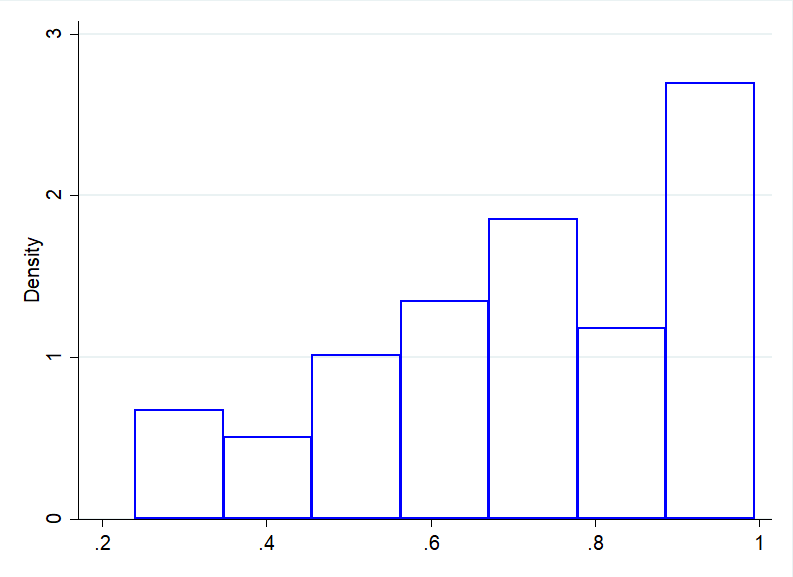}}
 \label{fig:beta_distrib}
\end{figure} 
\footnotesize
\justify \emph{Note:} The figure shows the distribution of the $\hat\beta_2$ computed at country-level and the corresponding p-values. The distribution of the $\hat\beta_2$ is concentrated around zero and the p-values are all above 0.2, showing that the country-level $\hat\beta_2$ are not significantly different from zero in the pre-treatment period. \normalsize

\section{Proof for Variance Decomposition}\label{proofdecomp}

We want to show that equation \ref{eq:DiD} still constitutes a variance decomposition. We do so in the context of a simplified decomposition and sample splitting but the logic readily extends to our full problem. 

Consider the identity $Y=\frac{Y}{C}C$. Applying the same steps as in Section \ref{sec: method}, we obtain the identity
\begin{align}
    Var(\Delta \ln Y)=Cov(\Delta \ln Y,\Delta \ln Y-\Delta \ln C)+Cov(\Delta \ln Y,\Delta \ln C),
\end{align}
which implies 
\begin{align}
    1=\frac{Cov(\Delta \ln Y,\Delta \ln Y-\Delta \ln C)}{Var(\Delta \ln Y)}+\frac{Cov(\Delta \ln Y,\Delta \ln C)}{Var(\Delta \ln Y)}\equiv \beta^R+\beta^U.
\end{align}
Now consider the following sample split, according to a dummy variable denoted $D$. Denote the $\beta^R_0,\beta^U_0$ and $\beta^R_1,\beta^U_1$ the coefficients obtained by estimating 
\begin{align}
    &\Delta \ln Y_D-\Delta \ln C_D=\beta^R_D\Delta \ln Y_D+u_D,\\
    &\Delta \ln C_D=\beta^U_D\Delta \ln Y_D+\epsilon_D,\quad \text{for } D=\{0,1\}.
\end{align}
Note first that the variance decomposition holds in each subsample, so that $1=\beta^R_D+\beta^U_D,\,\forall D$. Next note that we can estimate this alternative equations
\begin{align}
    &\Delta \ln Y-\Delta \ln C=\gamma_1\Delta \ln Y+\gamma_2 D\times \Delta \ln Y+ u,  \label{decomp1}\\
    &\Delta \ln C=\delta_1\Delta \ln Y+\delta_2 D\times \Delta \ln Y +\epsilon.
    \label{decomp2}
\end{align}
By the Frisch-Waugh-Lovell theorem we obtain $\hat\gamma_1=\beta^R_0$, $\hat\gamma_2=\beta^R_1-\beta^R_0$, $\hat\delta_1=\beta^U_0$ and $\hat\delta_2=\beta_1^U-\beta_0^U$. Since the decomposition holds in each subsample we know that $1=\beta_0^R+\beta_0^U$ and $1=\beta_1^R+\beta^U_1$. We can then sum up the coefficients obtained by estimating \ref{decomp1} and \ref{decomp2} as 
\begin{align}
    \hat\gamma_1+\hat\gamma_2+\hat\delta_1+\hat\delta_2=\beta^R_0+\beta^R_1-\beta^R_0+\beta^U_0+\beta^U_1-\beta^U_0=\beta^R_1+\beta^U_1=1,
\end{align}
which proves that a variance decomposition obtains. Our main estimation is an extension of this where $D=\{\text{Treated}, \text{Euro}, \text{Treated Euro}\}$ and our decomposition is given by equations \ref{reg1}-\ref{reg5}.  In summary, estimating equation \ref{eq:DiD} is equivalent to estimating the problem in each subsample and then taking the appropriate differences.\hfill$\blacksquare$
\section{Robustness Checks}

\subsection{Panel Correlated Standard Errors}

\begin{table}[H]
\caption{OLS estimates with panel correlated standard errors: 1990-2018}
\centering
\scriptsize
\begin{tabular}{llccccc}
\toprule
          &&\multicolumn{1}{c}{Capital Markets}&\multicolumn{1}{c}{International Transfers}&\multicolumn{1}{c}{Public Savings}&\multicolumn{1}{c}{Private Savings}&\multicolumn{1}{c}{Unsmoothed}\\
          \midrule
          && \multicolumn{5}{c}{All countries}\\
\midrule
Pre euro&Synthetic &    -0.01         &     0.02         &     0.13\sym{***}&     0.40\sym{***}&     0.46\sym{***}\\
       &   &  (-0.10)         &   (0.66)         &   (4.64)         &   (3.75)         &   (7.34)         \\
[1em]
Post euro&Actual&  0.05         &     0.02         &    -0.03         &    -0.26\sym{*}  &     0.22\sym{**} \\
    &      &   (0.36)         &   (0.64)         &  (-0.77)         &  (-1.95)         &   (2.01)         \\

\midrule
$R^2$&        &   0.26         &     0.07         &     0.55         &     0.45         &     0.70         \\
\midrule
          && \multicolumn{5}{c}{Core countries}\\
    \midrule      
   Pre euro&Synthetic &         -0.19         &     0.08\sym{**} &     0.19\sym{***}&     0.46\sym{***}&     0.46\sym{***}\\
     &     &  (-1.33)         &   (2.02)         &   (4.04)         &   (3.04)         &   (4.87)         \\
[1em]
Post euro&Actual&  -0.31\sym{*}  &     0.03         &     0.03         &     0.14         &     0.11         \\
     &     &  (-1.91)         &   (0.71)         &   (0.52)         &   (0.80)         &   (0.89)         \\
\midrule
$R^2$&      &      0.33         &     0.16         &     0.66         &     0.47         &     0.76         \\
\midrule
          && \multicolumn{5}{c}{Periphery countries}\\
    \midrule      
   Pre euro&Synthetic & 0.03         &     0.01         &     0.11\sym{***}&     0.40\sym{***}&     0.45\sym{***}\\
   &       &   (0.36)         &   (0.28)         &   (3.66)         &   (3.94)         &   (5.92)         \\
[1em]
Post euro&Actual& 0.09         &     0.04         &    -0.03         &    -0.44\sym{***}&     0.34\sym{**} \\
     &     &   (0.46)         &   (0.84)         &  (-0.57)         &  (-2.95)         &   (2.19)         \\

\midrule
$R^2$&  &       0.41         &     0.11         &     0.58         &     0.61         &     0.74         \\\bottomrule
\end{tabular}
\label{pcse1}
\end{table}
\footnotesize
\justify \emph{Note:} ***, **, and * denote significance at 1\%, 5\%,  10\% respectively. t-statistics are in parenthesis. The table displays OLS estimates with panel correlated standard errors over the period 1990-2011 for the actual and the synthetic series before ({\sl Pre euro}) and after ({\sl Post euro}) the introduction of the euro. The row {\sl Post euro Actual} displays the effect of the introduction of the euro. The adoption of the euro increases the unsmoothed component of the shock by 22\%, while it decreases private consumption smoothing by 26\%. 
  \normalsize

  \begin{table}[H]
\caption{OLS estimates with panel correlated standard errors: 1990-2007}
\centering
\scriptsize
\begin{tabular}{llccccc}
\toprule
          &&\multicolumn{1}{c}{Capital Markets}&\multicolumn{1}{c}{International Transfers}&\multicolumn{1}{c}{Public Savings}&\multicolumn{1}{c}{Private Savings}&\multicolumn{1}{c}{Unsmoothed}\\
          \midrule
          && \multicolumn{5}{c}{All countries}\\
\midrule
Pre euro&Synthetic &     -0.01         &     0.02         &     0.13\sym{***}&     0.40\sym{***}&     0.46\sym{***}\\
     &     &  (-0.17)         &   (0.69)         &   (5.31)         &   (4.44)         &   (8.00)         \\
[1em]
Post euro&Actual&   -0.02         &     0.02         &    -0.07         &    -0.24\sym{*}  &     0.30\sym{***}\\
   &       &  (-0.17)         &   (0.65)         &  (-1.63)         &  (-1.85)         &   (3.13)         \\

\midrule
$R^2$&        &   0.12         &     0.06         &     0.43         &     0.43         &     0.74         \\

\midrule
          && \multicolumn{5}{c}{Core countries}\\
    \midrule      
   Pre euro&Synthetic &          -0.19         &     0.08\sym{*}  &     0.18\sym{***}&     0.46\sym{***}&     0.47\sym{***}\\
     &     &  (-1.55)         &   (1.88)         &   (4.08)         &   (3.30)         &   (5.15)         \\
[1em]
Post euro&Actual&   -0.32\sym{**} &     0.07         &    -0.03         &     0.12         &     0.16         \\
     &     &  (-2.03)         &   (1.30)         &  (-0.40)         &   (0.66)         &   (1.18)         \\
\midrule
$R^2$&      &       0.32         &     0.15         &     0.57         &     0.46         &     0.75         \\
\midrule
          && \multicolumn{5}{c}{Periphery countries}\\
    \midrule      
   Pre euro&Synthetic &  0.03         &     0.01         &     0.11\sym{***}&     0.40\sym{***}&     0.45\sym{***}\\
      &    &   (0.53)         &   (0.28)         &   (4.45)         &   (4.87)         &   (6.84)         \\
[1em]
Post euro&Actual&   0.05         &     0.01         &    -0.07         &    -0.42\sym{***}&     0.42\sym{***}\\
   &       &   (0.45)         &   (0.26)         &  (-1.36)         &  (-3.12)         &   (3.36)         \\

\midrule
$R^2$&  &    0.23         &     0.11         &     0.44         &     0.62         &     0.78         \\\bottomrule
\end{tabular}
\label{pcse2}
\end{table}
\footnotesize
\justify \emph{Note:} ***, **, and * denote significance at 1\%, 5\%,  10\% respectively. t-statistics are in parenthesis. The table displays OLS estimates with panel correlated standard errors over the period 1990-2011 for the actual and the synthetic series before ({\sl Pre euro}) and after ({\sl Post euro}) the introduction of the euro. The row {\sl Post euro Actual} displays the effect of the introduction of the euro. The adoption of the euro increases the unsmoothed component of the shock by 30\%, while it decreases private consumption smoothing by 24\%. 
  \normalsize
  
\subsection{Matching using natural resources}
\begin{table}[H]
\caption{Matching using also natural resources}
\centering
\scriptsize
\begin{tabular}{llccccc}
\toprule
          &&\multicolumn{1}{c}{Capital Markets}&\multicolumn{1}{c}{International Transfers}&\multicolumn{1}{c}{Public Savings}&\multicolumn{1}{c}{Private Savings}&\multicolumn{1}{c}{Unsmoothed}\\
          \midrule
          && \multicolumn{5}{c}{All countries}\\
\midrule
Pre euro&Synthetic &    -0.03         &     0.05         &     0.12\sym{***}&     0.38\sym{***}&     0.48\sym{***}\\
       &     &  (-0.36)         &   (1.45)         &   (6.58)         &  (12.56)         &   (6.16)         \\

[1em]
Post euro&Actual&  0.02         &     0.03         &    -0.03         &    -0.25         &     0.23\sym{**} \\
    &       &   (0.12)         &   (0.98)         &  (-0.54)         &  (-1.51)         &   (2.70)         \\

\midrule
$R^2$&        &   0.26         &     0.07         &     0.55         &     0.45         &     0.70         \\
\midrule
          && \multicolumn{5}{c}{Core countries}\\
    \midrule      
   Pre euro&Synthetic &         -0.19\sym{**} &     0.07\sym{**} &     0.19\sym{***}&     0.42\sym{***}&     0.50\sym{***}\\
     &     &  (-2.23)         &   (2.45)         &   (4.00)         &   (3.67)         &   (4.09)         \\
[1em]
Post euro&Actual&     -0.30\sym{*}  &     0.02         &     0.04         &     0.10         &     0.14         \\
     &   &  (-2.16)         &   (0.48)         &   (0.98)         &   (0.47)         &   (1.00)         \\
\midrule
$R^2$&      &       0.33         &     0.15         &     0.64         &     0.47         &     0.76         \\
\midrule
          && \multicolumn{5}{c}{Periphery countries}\\
    \midrule      
   Pre euro&Synthetic & 0.01         &     0.04         &     0.11\sym{***}&     0.40\sym{***}&     0.45\sym{***}\\
   &       &   (0.09)         &   (1.03)         &   (4.99)         &   (8.62)         &   (5.40)              \\
[1em]
Post euro&Actual&   0.10         &     0.04         &    -0.03         &    -0.42\sym{**} &     0.31\sym{***}\\
     &     &     (0.52)         &   (0.90)         &  (-0.50)         &  (-2.96)         &   (3.47)         \\       \\

\midrule
$R^2$&  &         0.42         &     0.07         &     0.56         &     0.55         &     0.74        \\\bottomrule
\end{tabular}
\label{nat_resources}
\end{table}
\footnotesize
\justify \emph{Note:} ***, **, and * denote significance at 1\%, 5\%,  10\% respectively. t-statistics are in parenthesis. The table displays OLS estimates with panel correlated standard errors over the period 1990-2011 for the actual and the synthetic series before ({\sl Pre euro}) and after ({\sl Post euro}) the introduction of the euro. The row {\sl Post euro Actual} displays the effect of the introduction of the euro. The matching is done using total natural resources rents (\% of GDP) taken from the Word Development Indicators, in addition to the other variables used in the baseline specification (log per capita consumption, government consumption, national income, and disposable national income).
  \normalsize

\subsection{Match on First Differences}

\begin{table}[H]
\caption{Match on first differences: 1990-2018}
\centering
\scriptsize
\begin{tabular}{llccccc}
\toprule
          &&\multicolumn{1}{c}{Capital Markets}&\multicolumn{1}{c}{International Transfers}&\multicolumn{1}{c}{Public Savings}&\multicolumn{1}{c}{Private Savings}&\multicolumn{1}{c}{Unsmoothed}\\
          \midrule
          && \multicolumn{5}{c}{All countries}\\
\midrule
Pre euro&Synthetic &    -0.04         &     0.07         &     $0.15^{***}$&     $0.30^{***}$&     $0.52^{***}$\\
 &         &  (-0.39)         &   (1.53)         &   (4.94)         &   (4.33)         &   (7.79)         \\
[1em]
Post euro&Actual&  -0.13         &     0.06         &     0.03         &    -0.07         &     0.11         \\
 &         &  (-0.99)         &   (1.26)         &   (0.60)         &  (-0.59)         &   (1.41)         \\

\midrule
$R^2$&        &   $  0.20   $      &  $   0.06      $   &  $   0.47     $    &  $   0.53     $    &    $ 0.95     $    \\
\midrule
          && \multicolumn{5}{c}{Core countries}\\
    \midrule      
   Pre euro&Synthetic &         -0.12         &     0.09\sym{**} &     0.19\sym{***}&     0.40\sym{***}&     0.44\sym{***}\\
& &  (-0.74)         &   (2.27)         &   (4.24)         &   (3.32)         &   (5.61)         \\
[1em]
Post euro&Actual&   -0.24         &     0.07         &     0.02         &     0.16         &    -0.01         \\
&&  (-1.31)         &   (1.69)         &   (0.52)         &   (0.89)         &  (-0.10)         \\
\midrule
$R^2$&      &     0.39         &     0.19         &     0.69         &     0.55         &     0.82         \\
\midrule
          && \multicolumn{5}{c}{Periphery countries}\\
    \midrule      
   Pre euro&Synthetic &  -0.05         &     0.10         &     0.12\sym{***}&     0.28\sym{***}&     0.55\sym{***}\\
& &  (-0.43)         &   (0.92)         &   (6.13)         &   (4.19)         &   (5.87)         \\
[1em]
Post euro&Actual&-0.17         &     0.10         &     0.02         &    -0.24\sym{***}&     0.29\sym{***}\\
& &  (-1.69)         &   (0.90)         &   (0.59)         &  (-3.27)         &   (3.96)         \\
\midrule
$R^2$&  &     0.44         &     0.09         &     0.58         &     0.50         &     0.76         \\
\bottomrule
\end{tabular}
\label{firstdiff1}
\end{table}

\footnotesize
\justify\emph{Note:} ***, **, and * denote significance at 1\%, 5\%,  10\% respectively. t-statistics are in parenthesis.
The table displays OLS estimates with clustered standard errors over the period 1990-2018 for the actual and the synthetic series before ({\sl Pre euro}) and after ({\sl Post euro}) the introduction of the euro. Differently from our baseline results, synthetic data are generated by matching over first differenced data instead of data in levels. We confirm our baseline result that periphery countries have decreased their consumption smoothing, and this is due to a reduction in the private saving absorption channel. \normalsize

\begin{table}[H]
\caption{Match on first differences: 1990-2007}
\centering
\scriptsize
\begin{tabular}{llccccc}
\toprule
          &&\multicolumn{1}{c}{Capital Markets}&\multicolumn{1}{c}{International Transfers}&\multicolumn{1}{c}{Public Savings}&\multicolumn{1}{c}{Private Savings}&\multicolumn{1}{c}{Unsmoothed}\\
          \midrule
          && \multicolumn{5}{c}{All countries}\\
\midrule
Pre euro&Synthetic &    -0.05         &     0.07         &     0.15\sym{***}&     0.30\sym{***}&     0.52\sym{***}\\
 &         &  (-0.42)         &   (1.52)         &   (4.92)         &   (4.32)         &   (8.32)         \\
[1em]
Post euro&Actual&   -0.11         &     0.05         &    -0.02         &    -0.11         &     0.19\sym{**} \\

 &         &  (-0.76)         &   (0.80)         &  (-0.46)         &  (-0.93)         &   (2.71)         \\
\midrule
$R^2$&        &    0.15         &     0.09         &     0.40         &     0.35         &     0.77         \\
\midrule
          && \multicolumn{5}{c}{Core countries}\\
    \midrule      
   Pre euro&Synthetic &         -0.11         &     0.08\sym{*}  &     0.18\sym{***}&     0.39\sym{***}&     0.45\sym{***}\\
& &  (-0.68)         &   (2.07)         &   (3.93)         &   (3.17)         &   (6.17)         \\
[1em]
Post euro&Actual&  -0.20         &     0.06         &    -0.03         &     0.11         &     0.07         \\
&&  (-0.97)         &   (0.94)         &  (-0.68)         &   (0.63)         &   (0.66)         \\
\midrule
$R^2$&      0.34         &     0.19         &     0.60         &     0.50         &     0.79         \\
\midrule
          && \multicolumn{5}{c}{Periphery countries}\\
    \midrule      
   Pre euro&Synthetic &   -0.05         &     0.10         &     0.12\sym{***}&     0.28\sym{***}&     0.55\sym{***}\\

& &  (-0.43)         &   (0.91)         &   (6.09)         &   (4.16)         &   (5.83)         \\
[1em]
Post euro&Actual& -0.09         &     0.06         &    -0.04         &    -0.28\sym{**} &     0.36\sym{***}\\
& &  (-0.70)         &   (0.41)         &  (-0.91)         &  (-3.20)         &   (4.67)         \\
\midrule
$R^2$&  &     0.22         &     0.09         &     0.35         &     0.35         &     0.79         \\
\bottomrule
\end{tabular}
\label{firstdiff2}
\end{table}

\footnotesize
\justify\emph{Note:} ***, **, and * denote significance at 1\%, 5\%,  10\% respectively. t-statistics are in parenthesis.
The table displays OLS estimates with clustered standard errors over the period 1990-2007 for the actual and the synthetic series before ({\sl Pre euro}) and after ({\sl Post euro}) the introduction of the euro. Differently from our baseline results, synthetic data are generated by matching over first differenced data instead of data in levels. We confirm our baseline result  that countries have decreased their consumption smoothing. This result is driven by periphery countries, which decrease risk sharing through private savings. \normalsize

\subsection{Matching Window: 1990-1995}

\begin{table}[H]
\caption{Matching window: 1990-1995}
\centering
\scriptsize
\begin{tabular}{llccccc}
\toprule
          &&\multicolumn{1}{c}{Capital Markets}&\multicolumn{1}{c}{International Transfers}&\multicolumn{1}{c}{Public Savings}&\multicolumn{1}{c}{Private Savings}&\multicolumn{1}{c}{Unsmoothed}\\
          \midrule
          && \multicolumn{5}{c}{All countries}\\
\midrule
Pre euro&Synthetic &  0.05         &     0.06         &     0.11\sym{***}&     0.31\sym{***}&     0.47\sym{***}\\
       &   &   (0.47)         &   (1.37)         &   (4.40)         &   (6.12)         &   (7.95)         \\
[1em]
Post euro&Actual&  0.19         &     0.06         &    -0.06         &    -0.31\sym{*}  &     0.12         \\
         & &   (1.04)         &   (1.37)         &  (-0.87)         &  (-1.76)         &   (1.59)         \\

\midrule
$R^2$&        &  0.26         &     0.09         &     0.57         &     0.47         &     0.73         \\
\midrule
          && \multicolumn{5}{c}{Core countries}\\
    \midrule      
   Pre euro&Synthetic &         -0.25\sym{*}  &     0.14\sym{*}  &     0.19\sym{***}&     0.55\sym{***}&     0.37\sym{***}\\
       &   &  (-2.13)         &   (1.87)         &   (4.07)         &   (6.45)         &   (4.45)         \\[1em]
Post euro&Actual&  -0.22         &     0.05         &     0.01         &     0.22         &    -0.07         \\
   &       &  (-1.26)         &   (0.88)         &   (0.29)         &   (1.30)         &  (-0.76)         \\
\midrule
$R^2$&    &   0.32         &     0.15         &     0.70         &     0.57         &     0.81         \\
\midrule
          && \multicolumn{5}{c}{Periphery countries}\\
    \midrule      
   Pre euro&Synthetic &  0.13         &     0.02         &     0.08\sym{***}&     0.28\sym{***}&     0.50\sym{***}\\
       &   &   (1.54)         &   (0.94)         &   (4.30)         &   (4.74)         &   (5.46)         \\
[1em]
Post euro&Actual&0.23         &     0.09         &    -0.07         &    -0.56\sym{***}&     0.31\sym{***}\\
      &    &   (1.36)         &   (1.32)         &  (-1.25)         &  (-4.31)         &   (3.25)         \\

\midrule
$R^2$&  &     0.41         &     0.20         &     0.58         &     0.58         &     0.74         \\
\bottomrule
\end{tabular}
\label{ols1995}
\end{table}

\footnotesize
\justify\emph{Note:} ***, **, and * denote significance at 1\%, 5\%,  10\% respectively. t-statistics are in parenthesis.
The table displays OLS estimates with clustered standard errors over the period 1990-2018 for the actual and the synthetic series before ({\sl Pre euro}) and after ({\sl Post euro}) the introduction of the euro. Differently from our baseline results, synthetic data are generated by matching over the period 1990-1995. We no longer find any change in consumption smoothing of euro area members for the full sample of countries. However, we confirm our result that periphery countries have decreased their consumption smoothing, and this is due to a reduction in the private saving absorption channel. \normalsize

\begin{table}[H]
\caption{Matching window: 1990-1995. Exclude period from the Great Recession}
\centering
\scriptsize
\begin{tabular}{llccccc}
\toprule
          &&\multicolumn{1}{c}{Capital Markets}&\multicolumn{1}{c}{International Transfers}&\multicolumn{1}{c}{Public Savings}&\multicolumn{1}{c}{Private Savings}&\multicolumn{1}{c}{Unsmoothed}\\
          \midrule
          && \multicolumn{5}{c}{All countries}\\
\midrule
Pre euro&Synthetic &    0.05         &     0.06         &     0.11\sym{***}&     0.31\sym{***}&     0.47\sym{***}\\
     &     &   (0.45)         &   (1.37)         &   (4.40)         &   (6.09)         &   (7.97)         \\

[1em]
Post euro&Actual&   0.03         &     0.06         &    -0.07         &    -0.24         &     0.22\sym{***}\\
     &     &   (0.16)         &   (1.18)         &  (-1.12)         &  (-1.34)         &   (3.38)         \\
\midrule
$R^2$&        &    0.15         &     0.09         &     0.40         &     0.37         &     0.75         \\
\midrule
          && \multicolumn{5}{c}{Core countries}\\
    \midrule      
   Pre euro&Synthetic &       -0.25\sym{*}  &     0.13\sym{*}  &     0.18\sym{***}&     0.54\sym{***}&     0.38\sym{***}\\
     &     &  (-2.02)         &   (1.89)         &   (3.68)         &   (5.84)         &   (4.99)         \\
          [1em]
Post euro&Actual&  -0.31         &     0.04         &    -0.01         &     0.28         &    -0.00         \\
    &      &  (-1.46)         &   (0.37)         &  (-0.12)         &   (1.62)         &  (-0.03)         \\
\midrule
$R^2$&   &    0.31         &     0.21         &     0.57         &     0.49         &     0.79         \\
\midrule
          && \multicolumn{5}{c}{Periphery countries}\\
    \midrule      
   Pre euro&Synthetic &    0.13         &     0.02         &     0.08\sym{***}&     0.28\sym{***}&     0.50\sym{***}\\
       &   &   (1.53)         &   (0.94)         &   (4.27)         &   (4.70)         &   (5.43)         \\
          [1em]
Post euro&Actual& 0.10         &     0.10\sym{*}  &    -0.10\sym{*}  &    -0.50\sym{***}&     0.39\sym{***}\\
       &   &   (0.56)         &   (1.91)         &  (-2.18)         &  (-3.35)         &   (4.26)         \\
\midrule
$R^2$&  &     0.29         &     0.18         &     0.38         &     0.53         &     0.78         \\\bottomrule
\end{tabular}
\label{ols1995_nocrisis}
\end{table}

\footnotesize
\justify\emph{Note:} ***, **, and * denote significance at 1\%, 5\%,  10\% respectively. t-statistics are in parenthesis.
The table displays OLS estimates with clustered standard errors over the period 1990-2007 for the actual and the synthetic series before ({\sl Pre euro}) and after ({\sl Post euro}) the introduction of the euro. Differently from our baseline results, synthetic data are generated by matching over the period 1990-1995. We confirm our baseline result  that countries have decreased their consumption smoothing. This result is driven by periphery countries, which reduce shock absorption through private savings. \normalsize

\subsection{Accounting Identity and fixed weights}
\begin{figure}[H]
        \caption{Test of GDP Accounting Identity}\centering
    \includegraphics[width=1\textwidth]{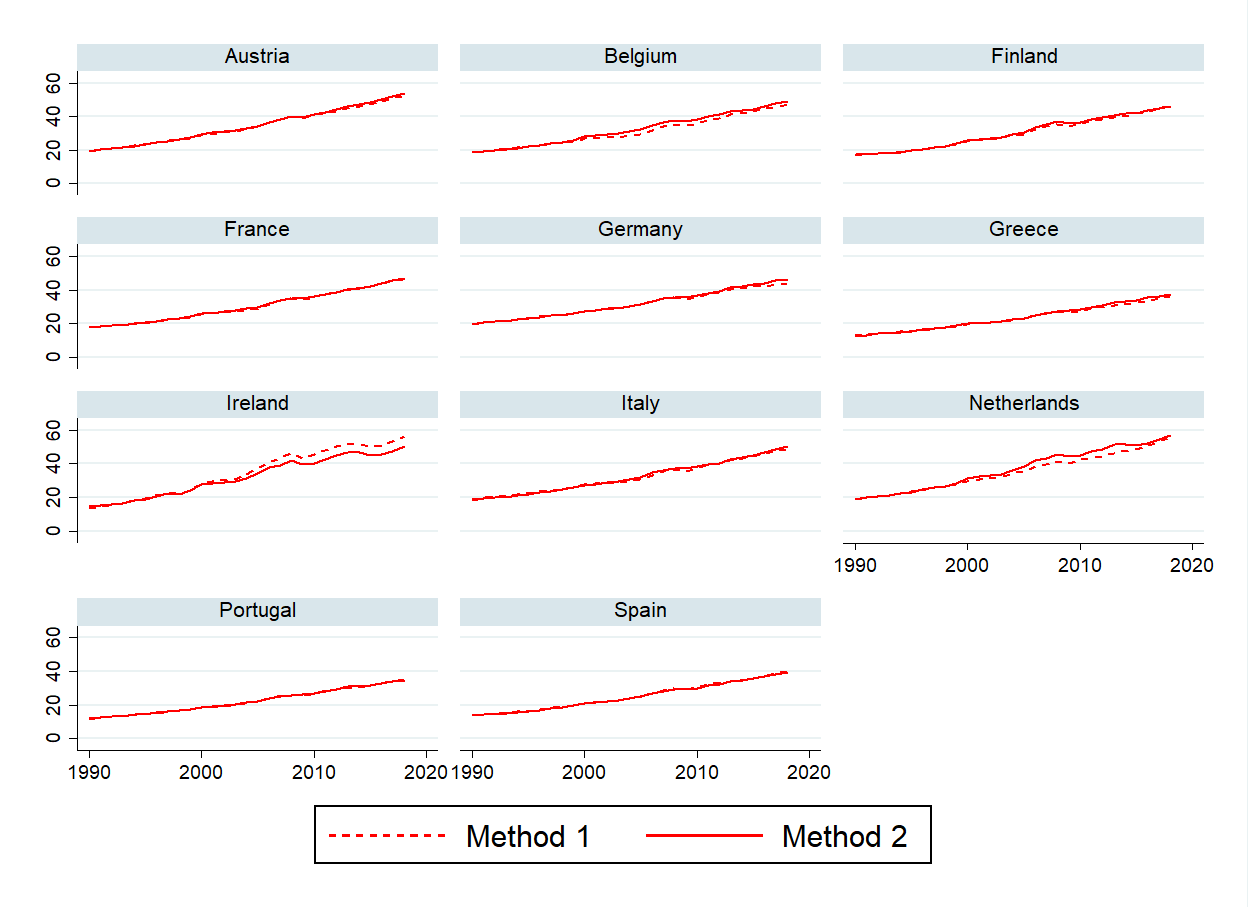}
    \label{identity}
\end{figure}
\footnotesize
\justify\emph{Note:} The figure shows synthetic GDP computed in two different ways. The first approach builds synthetic GDP by directly matching actual GDP. The second approach builds synthetic GDP by summing the synthetic version of GDP subcomponents. The goal is to show that the GDP accounting identity holds even if we have not imposed it ex ante. The series are in thousands of euros per capita.\normalsize

\begin{table}[H]
\caption{Country fixed weights for the synthetic variables: 1990-2018}
\centering
\scriptsize
\begin{tabular}{llccccc}
\toprule

          &&\multicolumn{1}{c}{Capital Markets}&\multicolumn{1}{c}{International Transfers}&\multicolumn{1}{c}{Public Savings}&\multicolumn{1}{c}{Private Savings}&\multicolumn{1}{c}{Unsmoothed}\\
          \midrule
          && \multicolumn{5}{c}{All countries}\\
\midrule
Pre euro&Synthetic &   -0.18\sym{***}&     0.00         &     0.48\sym{***}&    -0.11\sym{*}  &     0.81\sym{***}\\
        &  &  (-3.33)         &   (0.22)         &   (6.78)         &  (-1.98)         &  (14.49)         \\
[1em]
Post euro&Actual&     -0.07         &    -0.02         &     0.07         &    -0.30\sym{*}  &     0.33\sym{***}\\
      &    &  (-0.56)         &  (-0.96)         &   (0.87)         &  (-2.00)         &   (3.65)         \\

\midrule
$R^2$&   &     0.30         &     0.08         &     0.66         &     0.29         &     0.74         \\
\midrule
          && \multicolumn{5}{c}{Core countries}\\
    \midrule      
   Pre euro&Synthetic &       -0.27\sym{**} &     0.05         &     0.90\sym{***}&    -0.24         &     0.56\sym{***}\\
       &   &  (-2.56)         &   (1.16)         &   (7.52)         &  (-1.51)         &   (6.20)         \\
[1em]
Post euro&Actual&    -0.22         &    -0.00         &     0.49\sym{***}&    -0.33         &     0.06         \\
        &  &  (-1.24)         &  (-0.02)         &   (4.00)         &  (-1.27)         &   (0.52)         \\
\midrule
$R^2$&      &   0.38         &     0.16         &     0.70         &     0.44         &     0.82         \\
\midrule
          && \multicolumn{5}{c}{Periphery countries}\\
    \midrule      
   Pre euro&Synthetic &-0.22\sym{**} &    -0.01         &     0.45\sym{***}&    -0.10         &     0.89\sym{***}\\
        &  &  (-3.00)         &  (-0.73)         &   (4.75)         &  (-1.34)         &  (11.75)         \\
[1em]
Post euro&Actual&  -0.14         &    -0.02         &     0.05         &    -0.40\sym{**} &     0.51\sym{***}\\
        &  &  (-1.08)         &  (-0.98)         &   (0.91)         &  (-2.77)         &   (5.81)         \\
\midrule
$R^2$&  &      0.46         &     0.13         &     0.70         &     0.35         &     0.77         \\
\bottomrule
\end{tabular}
\label{fw1}
\end{table}
\footnotesize
\justify\emph{Note:} ***, **, and * denote significance at 1\%, 5\%,  10\% respectively. t-statistics are in parenthesis.
The table displays OLS estimates with clustered standard errors over the period 1990-2018 for the actual and the synthetic series before ({\sl Pre euro}) and after ({\sl Post euro}) the introduction of the euro. Differently from our baseline results, synthetic data are generated by using fixed weights for each country's variables as explained in section \ref{fixed_weights}. 
\normalsize

\begin{table}[H]
\caption{Country fixed weights for the synthetic variables: 1990-2007}
\centering
\scriptsize
\begin{tabular}{llccccc}
\toprule
          &&\multicolumn{1}{c}{Capital Markets}&\multicolumn{1}{c}{International Transfers}&\multicolumn{1}{c}{Public Savings}&\multicolumn{1}{c}{Private Savings}&\multicolumn{1}{c}{Unsmoothed}\\
          \midrule
          && \multicolumn{5}{c}{All countries}\\
\midrule
Pre euro&Synthetic &   -0.18\sym{***}&     0.00         &     0.48\sym{***}&    -0.11\sym{*}  &     0.81\sym{***}\\
         & &  (-3.33)         &   (0.22)         &   (6.78)         &  (-1.98)         &  (14.49)      \\
[1em]
Post euro&Actual&   -0.18         &    -0.04         &     0.15\sym{*}  &    -0.34\sym{**} &     0.41\sym{***}\\
      &    &  (-1.45)         &  (-1.61)         &   (1.81)         &  (-2.68)         &   (5.87)         \\

\midrule
$R^2$&  &     0.23         &     0.08         &     0.58         &     0.30         &     0.78         \\
\midrule
          && \multicolumn{5}{c}{Core countries}\\
    \midrule      
   Pre euro&Synthetic &        -0.27\sym{**} &     0.05         &     0.90\sym{***}&    -0.24         &     0.56\sym{***}\\
         & &  (-2.54)         &   (1.15)         &   (7.48)         &  (-1.50)         &   (6.16)         \\
        [1em]
Post euro&Actual&  -0.29         &    -0.02         &     0.53\sym{***}&    -0.36         &     0.15         \\
         & &  (-1.75)         &  (-0.56)         &   (4.18)         &  (-1.53)         &   (1.40)         \\
\midrule
$R^2$&      &   0.40         &     0.15         &     0.66         &     0.42         &     0.77         \\
\midrule
          && \multicolumn{5}{c}{Periphery countries}\\
    \midrule      
   Pre euro&Synthetic &   -0.22\sym{**} &    -0.01         &     0.45\sym{***}&    -0.10         &     0.89\sym{***}\\
       &   &  (-3.00)         &  (-0.73)         &   (4.75)         &  (-1.34)         &  (11.75)         \\
[1em]
Post euro&Actual&   -0.21         &    -0.04         &     0.15\sym{**} &    -0.45\sym{***}&     0.56\sym{***}\\
         & &  (-1.60)         &  (-1.65)         &   (3.02)         &  (-4.02)         &   (7.03)         \\
          \midrule
$R^2$&  &    0.42         &     0.15         &     0.60         &     0.43         &     0.82         \\
\bottomrule
\end{tabular}
\label{fw2}
\end{table}
\footnotesize
\justify\emph{Note:} ***, **, and * denote significance at 1\%, 5\%,  10\% respectively. t-statistics are in parenthesis.
The table displays OLS estimates with clustered standard errors over the period 1990-2007 for the actual and the synthetic series before ({\sl Pre euro}) and after ({\sl Post euro}) the introduction of the euro. Differently from our baseline results, synthetic data are generated by using fixed weights for each country's variables as explained in section \ref{fixed_weights}.  
\normalsize

\subsection{Placebo Studies}
\begin{table}[H]
\caption{Placebo studies}
\centering
\scriptsize
\begin{tabular}{llccccc}
\toprule
          &&\multicolumn{1}{c}{Capital Markets}&\multicolumn{1}{c}{International Transfers}&\multicolumn{1}{c}{Public Savings}&\multicolumn{1}{c}{Private Savings}&\multicolumn{1}{c}{Unsmoothed}\\
          \midrule
          && \multicolumn{5}{c}{1990-2018}\\
\midrule
Pre euro&Synthetic &    -0.09\sym{**} &     0.04\sym{**} &     0.00\sym{***}&    -0.01         &     1.06\sym{***}\\
       &   &  (-2.42)         &   (2.76)         &   (3.71)         &  (-0.14)         &  (11.29)         \\
[1em]
Post euro&Actual&     -0.12         &    -0.01         &     0.00         &    -0.05         &     0.18         \\
   &       &  (-0.91)         &  (-0.33)         &   (0.49)         &  (-0.26)         &   (1.48)         \\

\midrule
N    &&       728         &      728         &      728         &      728         &      728         \\
$R^2$&        &      0.20         &     0.06         &     0.49         &     0.38         &     0.70         \\
\midrule
          && \multicolumn{5}{c}{1990-2007}\\
    \midrule      
   Pre euro&Synthetic &      -0.09\sym{**} &     0.04\sym{**} &     0.00\sym{***}&    -0.01         &     1.06\sym{***}\\
       &   &  (-2.41)         &   (2.75)         &   (3.70)         &  (-0.14)         &  (11.26)         \\
[1em]
Post euro&Actual&    -0.01         &    -0.00         &     0.00         &    -0.11         &     0.12         \\
      &    &  (-0.09)         &  (-0.02)         &   (0.20)         &  (-0.66)         &   (0.73)         \\
\midrule
\(N\)    & &      442         &      442         &      442         &      442         &      442         \\
$R^2$&      &    0.15         &     0.06         &     0.42         &     0.32         &     0.69         \\
\bottomrule
\end{tabular}
\label{placebo1}
\end{table}
\footnotesize
\justify\emph{Note:} ***, **, and * denote significance at 1\%, 5\%,  10\% respectively. t-statistics are in parenthesis. The table displays OLS estimates over the period 1990-2011 for the actual and the synthetic series before ({\sl Pre euro}) and after ({\sl Post euro}) the introduction of the euro.  The analysis is run for placebo countries, which have never adopted the euro. Our difference-in-differences estimators are never significant, meaning that we find no effect of the adoption of the euro on our non euro area group.
\normalsize

\subsection{Anticipation Effects}

\begin{table}[H]
\caption{Anticipation effects: 1990-2018}
\centering
\scriptsize
\begin{tabular}{llccccc}
\toprule
          &&\multicolumn{1}{c}{Capital Markets}&\multicolumn{1}{c}{International Transfers}&\multicolumn{1}{c}{Public Savings}&\multicolumn{1}{c}{Private Savings}&\multicolumn{1}{c}{Unsmoothed}\\
          \midrule
          && \multicolumn{5}{c}{All countries}\\
\midrule
Pre euro&Synthetic &    0.02         &     0.05\sym{*}  &     0.13\sym{***}&     0.30\sym{***}&     0.50\sym{***}\\
    &      &   (0.47)         &   (1.90)         &   (6.55)         &   (9.29)         &  (10.68)         \\
[1em]
Post euro&Actual&      0.04         &     0.04         &    -0.03         &    -0.21         &     0.16\sym{*}  \\
         & &   (0.31)         &   (1.25)         &  (-0.56)         &  (-1.47)         &   (1.81)         \\

\midrule
$R^2$&    &     0.29         &     0.06         &     0.57         &     0.45         &     0.72         \\
\midrule
          && \multicolumn{5}{c}{Core countries}\\
    \midrule      
   Pre euro&Synthetic &       -0.07         &     0.11         &     0.18\sym{***}&     0.40\sym{***}&     0.39\sym{***}\\
     &     &  (-0.75)         &   (1.70)         &   (4.67)         &   (3.89)         &   (3.62)         \\
[1em]
Post euro&Actual&    -0.28\sym{**} &     0.06         &     0.02         &     0.16         &     0.04         \\
     &     &  (-2.24)         &   (0.79)         &   (0.47)         &   (0.94)         &   (0.29)         \\
\midrule
$R^2$&      &     0.45         &     0.16         &     0.68         &     0.53         &     0.79         \\\midrule
          && \multicolumn{5}{c}{Periphery countries}\\
    \midrule      
   Pre euro&Synthetic &  - 0.01         &     0.04         &     0.11\sym{***}&     0.28\sym{***}&     0.56\sym{***}\\
       &   &   (0.25)         &   (1.59)         &   (4.90)         &   (5.77)         &   (8.93)         \\
[1em]
Post euro&Actual&   0.07         &     0.05         &    -0.03         &    -0.43\sym{***}&     0.34\sym{***}\\
      &    &   (0.61)         &   (1.21)         &  (-0.59)         &  (-3.94)         &   (4.37)         \\
\midrule
$R^2$&  &     0.43         &     0.11         &     0.61         &     0.58         &     0.75         \\
\bottomrule
\end{tabular}
\label{anticipation1}
\end{table}
\footnotesize
\justify\emph{Note:} ***, **, and * denote significance at 1\%, 5\%,  10\% respectively. t-statistics are in parenthesis. The table displays OLS estimates with clustered standard errors over the period 1990-2018 for the actual and the synthetic series before ({\sl Pre euro}) and after ({\sl Post euro}) the introduction of the euro. The matching window is now 1990-1997 instead of 1990-1998. Our estimates are in line with our baseline results. With the adoption of the euro, we find an increase in the unsmoothed component. We still find that this result is driven by periphery countries. 
\normalsize

\begin{table}[H]
\caption{Anticipation effects: 1990-2007}
\centering
\scriptsize
\begin{tabular}{llccccc}
\toprule
          &&\multicolumn{1}{c}{Capital Markets}&\multicolumn{1}{c}{International Transfers}&\multicolumn{1}{c}{Public Savings}&\multicolumn{1}{c}{Private Savings}&\multicolumn{1}{c}{Unsmoothed}\\
          \midrule
          && \multicolumn{5}{c}{All countries}\\
\midrule
Pre euro&Synthetic &     0.02         &     0.05\sym{*}  &     0.13\sym{***}&     0.30\sym{***}&     0.50\sym{***}\\
     &     &   (0.41)         &   (1.90)         &   (6.50)         &   (9.29)         &  (11.00)         \\[1em]
          
Post euro&Actual&       -0.02         &     0.04         &    -0.07         &    -0.21         &     0.25\sym{***}\\
       &   &  (-0.12)         &   (0.86)         &  (-1.37)         &  (-1.54)         &   (3.27)         \\

\midrule
$R^2$&    &    0.15         &     0.06         &     0.44         &     0.39         &     0.75         \\

\midrule
          && \multicolumn{5}{c}{Core countries}\\
    \midrule      
   Pre euro&Synthetic &        -0.06         &     0.10         &     0.18\sym{***}&     0.39\sym{***}&     0.39\sym{***}\\
      &    &  (-0.72)         &   (1.72)         &   (4.47)         &   (3.93)         &   (3.84)         \\
[1em]
Post euro&Actual&   -0.20         &     0.09         &    -0.05         &     0.07         &     0.09         \\
      &    &  (-1.34)         &   (1.02)         &  (-0.98)         &   (0.35)         &   (0.68)         \\
\midrule
$R^2$&      &      0.38         &     0.19         &     0.59         &     0.47         &     0.76         \\
\midrule
          && \multicolumn{5}{c}{Periphery countries}\\
    \midrule      
   Pre euro&Synthetic &  0.01         &     0.04         &     0.11\sym{***}&     0.28\sym{***}&     0.56\sym{***}\\
   &       &   (0.25)         &   (1.58)         &   (4.86)         &   (5.73)         &   (8.87)         \\
[1em]
Post euro&Actual&     -0.00         &    -0.01         &    -0.06         &    -0.36\sym{**} &     0.43\sym{***}\\
&          &  (-0.04)         &  (-0.17)         &  (-1.43)         &  (-2.73)         &   (5.04)         \\
\midrule
$R^2$&  &   0.26         &     0.09         &     0.41         &     0.53         &     0.78         \\
\bottomrule
\end{tabular}
\label{anticipation2}
\end{table}
\footnotesize
\justify\emph{Note:} ***, **, and * denote significance at 1\%, 5\%,  10\% respectively. t-statistics are in parenthesis. The table displays OLS estimates with clustered standard errors over the period 1990-2007 for the actual and the synthetic series before ({\sl Pre euro}) and after ({\sl Post euro}) the introduction of the euro. The matching window is now 1990-1997 instead of 1990-1998. Our estimates are in line with our baseline results. With the adoption of the euro, we find an increase in the unsmoothed component. We still find that this result is driven by periphery countries, which reduce their shock absorption through private savings. \normalsize

\subsection{Exclusion of EU Members from non-Euro Area Group}

\begin{table}[H]
\caption{Exclusion of EU members from non euro area group of countries: 1990-2018}
\centering
\scriptsize
\begin{tabular}{llccccc}
\toprule
          &&\multicolumn{1}{c}{Capital Markets}&\multicolumn{1}{c}{International Transfers}&\multicolumn{1}{c}{Public Savings}&\multicolumn{1}{c}{Private Savings}&\multicolumn{1}{c}{Unsmoothed}\\
          \midrule
          && \multicolumn{5}{c}{All countries}\\
\midrule
Pre euro&Synthetic &   0.02         &     0.05\sym{***}&     0.12\sym{***}&     0.38\sym{***}&     0.43\sym{***}\\
       &   &   (0.16)         &   (3.01)         &   (5.96)         &   (7.93)         &   (5.90)         \\
[1em]
Post euro&Actual&      0.08         &    -0.02         &    -0.03         &    -0.12         &     0.08         \\
     &     &   (0.34)         &  (-0.35)         &  (-0.54)         &  (-0.63)         &   (0.83)         \\

\midrule
$R^2$&    &      0.24         &     0.12         &     0.53         &     0.39         &     0.70         \\
\midrule
          && \multicolumn{5}{c}{Core countries}\\
    \midrule      
   Pre euro&Synthetic &     -0.21         &     0.04         &     0.19\sym{***}&     0.56\sym{***}&     0.42\sym{***}\\
     &     &  (-1.26)         &   (0.88)         &   (5.70)         &   (3.84)         &   (3.12)         \\
[1em]
Post euro&Actual&     -0.30         &     0.06         &     0.01         &     0.36         &    -0.14         \\
      &    &  (-0.85)         &   (1.05)         &   (0.21)         &   (1.39)         &  (-0.84)         \\
\midrule
$R^2$&      &     0.29         &     0.15         &     0.65         &     0.47         &     0.76         \\\midrule
          && \multicolumn{5}{c}{Periphery countries}\\
    \midrule      
   Pre euro&Synthetic &  0.11\sym{**} &     0.05         &     0.10\sym{**} &     0.30\sym{***}&     0.45\sym{***}\\
      &    &   (2.38)         &   (1.66)         &   (3.15)         &   (4.51)         &   (4.82)         \\
          [1em]
Post euro&Actual&   0.17         &    -0.07         &    -0.01         &    -0.35\sym{**} &     0.26\sym{**} \\
      &    &   (0.90)         &  (-1.49)         &  (-0.24)         &  (-2.89)         &   (2.34)         \\
\midrule
$R^2$&  &     0.39         &     0.19         &     0.53         &     0.48         &     0.75         \\
\bottomrule
\end{tabular}
\label{noeu1}
\end{table}
\footnotesize
\justify\emph{Note:} ***, **, and * denote significance at 1\%, 5\%,  10\% respectively. t-statistics are in parenthesis. The table displays OLS estimates with clustered standard errors over the period 1990-2018 for the actual and the synthetic series before ({\sl Pre euro}) and after ({\sl Post euro}) the introduction of the euro. Only non-EU countries are included among the donors to generate synthetic variables. We confirm our main result that the adoption of the euro has decreased consumption smoothing of periphery countries, and this is due to a lower shock absorption through private savings.\normalsize

\begin{table}[H]
\caption{Exclusion of EU members from non euro area group of countries: 1990-2007}
\centering
\scriptsize
\begin{tabular}{llccccc}
\toprule
          &&\multicolumn{1}{c}{Capital Markets}&\multicolumn{1}{c}{International Transfers}&\multicolumn{1}{c}{Public Savings}&\multicolumn{1}{c}{Private Savings}&\multicolumn{1}{c}{Unsmoothed}\\
          \midrule
          && \multicolumn{5}{c}{All countries}\\
\midrule
Pre euro&Synthetic &   0.01         &     0.05\sym{***}&     0.12\sym{***}&     0.38\sym{***}&     0.43\sym{***}\\
      &    &   (0.14)         &   (3.01)         &   (5.88)         &   (7.87)         &   (5.93)         \\[1em]
Post euro&Actual&        0.05         &    -0.09         &    -0.06         &    -0.10         &     0.20\sym{*}  \\
     &     &   (0.23)         &  (-1.33)         &  (-1.24)         &  (-0.58)         &   (2.00)         \\
\midrule
%N    &&$     462         $&$     462         $&$     462         $&$     462         $&$     462        $\\
$R^2$&    &      0.24         &     0.12         &     0.53         &     0.39         &     0.70         \\
\midrule
          && \multicolumn{5}{c}{Core countries}\\
    \midrule      
   Pre euro&Synthetic &     -0.20         &     0.04         &     0.19\sym{***}&     0.55\sym{***}&     0.43\sym{***}\\
      &    &  (-1.17)         &   (0.80)         &   (5.40)         &   (3.86)         &   (3.25)         \\
[1em]
Post euro&Actual&     -0.21         &    -0.01         &    -0.04         &     0.39         &    -0.13         \\
      &    &  (-0.56)         &  (-0.06)         &  (-0.53)         &   (1.34)         &  (-0.70)         \\
\midrule
$R^2$&    &   0.25         &     0.12         &     0.58         &     0.45         &     0.70         \\\midrule
          && \multicolumn{5}{c}{Periphery countries}\\
    \midrule      
   Pre euro&Synthetic & 0.11\sym{**} &     0.05         &     0.10\sym{**} &     0.30\sym{***}&     0.45\sym{***}\\
      &    &   (2.36)         &   (1.65)         &   (3.13)         &   (4.48)         &   (4.78)         \\          [1em]
Post euro&Actual&    0.08         &    -0.13         &    -0.04         &    -0.32\sym{*}  &     0.41\sym{***}\\
         & &   (0.55)         &  (-1.51)         &  (-1.02)         &  (-2.21)         &   (3.57)         \\
          \midrule
$R^2$&  &     0.27         &     0.33         &     0.40         &     0.49         &     0.79         \\
\bottomrule
\end{tabular}
\label{noeu2}
\end{table}
\footnotesize
\justify\emph{Note:} ***, **, and * denote significance at 1\%, 5\%,  10\% respectively. t-statistics are in parenthesis. The table displays OLS estimates with clustered standard errors over the period 1990-2007 for the actual and the synthetic series before ({\sl Pre euro}) and after ({\sl Post euro}) the introduction of the euro. Only non-EU countries are included among the donors to generate synthetic variables. We confirm our main result that the adoption of the euro has decreased consumption smoothing of euro area countries. This  result is driven by periphery countries, which lowered their shock absorption through private savings.\normalsize

\subsection{Exclusion of EABCN recessionary periods}

\begin{table}[H]
\caption{Exclusion of EABCN recessionary periods}
\centering
\scriptsize
\begin{tabular}{llccccc}
\toprule
          &&\multicolumn{1}{c}{Capital Markets}&\multicolumn{1}{c}{International Transfers}&\multicolumn{1}{c}{Public Savings}&\multicolumn{1}{c}{Private Savings}&\multicolumn{1}{c}{Unsmoothed}\\
          \midrule
          && \multicolumn{5}{c}{All countries}\\
\midrule
Pre euro&Synthetic &  -0.01         &     0.02         &     0.13\sym{***}&     0.40\sym{***}&     0.46\sym{***}\\
       &    &  (-0.11)         &   (1.21)         &   (6.02)         &  (19.45)         &   (7.03)         \\
[1em]
Post euro&Actual&       0.04         &     0.02         &    -0.03         &    -0.24         &     0.21\sym{**} \\
     &      &   (0.21)         &   (0.47)         &  (-0.45)         &  (-1.46)         &   (2.52)         \\

\midrule
$R^2$&    &      0.25         &     0.07         &     0.54         &     0.44         &     0.71         \\
\midrule
          && \multicolumn{5}{c}{Core countries}\\
    \midrule      
   Pre euro&Synthetic &     -0.19\sym{**} &     0.07\sym{**} &     0.21\sym{***}&     0.43\sym{***}&     0.48\sym{***}\\
      &    &  (-2.23)         &   (2.49)         &   (4.75)         &   (3.71)         &   (3.84)         \\
[1em]
Post euro&Actual&    -0.32\sym{*}  &     0.02         &     0.05         &     0.12         &     0.13         \\
     &     &  (-2.19)         &   (0.46)         &   (1.12)         &   (0.54)         &   (0.90)         \\
\midrule
$R^2$&      &     0.31         &     0.14         &     0.65         &     0.46         &     0.77               \\\midrule
          && \multicolumn{5}{c}{Periphery countries}\\
    \midrule      
   Pre euro&Synthetic &     0.03         &     0.01         &     0.11\sym{***}&     0.42\sym{***}&     0.43\sym{***}\\
      &    &   (0.67)         &   (0.21)         &   (4.87)         &  (14.57)         &   (6.14)         \\
          [1em]
Post euro&Actual&     0.11         &     0.03         &    -0.02         &    -0.42\sym{***}&     0.30\sym{***}\\
    &      &   (0.63)         &   (0.66)         &  (-0.43)         &  (-3.27)         &   (3.72)         \\
\midrule
$R^2$&  &    0.41         &     0.12         &     0.57         &     0.61         &     0.74         \\
\bottomrule
\end{tabular}
\label{recessions}
\end{table}
\footnotesize
\justify\emph{Note:} ***, **, and * denote significance at 1\%, 5\%,  10\% respectively. t-statistics are in parenthesis. The table displays OLS estimates with clustered standard errors for the actual and the synthetic series before ({\sl Pre euro}) and after ({\sl Post euro}) the introduction of the euro. The regression is estimated over the period 1990-2018 excluding years 2008 and 2012, which are the years with at least three quarters of recessions according to the EABCN dating. We confirm our main result that the adoption of the euro has decreased consumption smoothing of periphery countries, and this is due to a lower shock absorption through private savings.\normalsize

\subsection{Differential Trends}

\begin{table}[H]
\caption{Allowing for different group-time-fixed effects}

\centering
\scriptsize
\begin{tabular}{llccccc}
\toprule
          &&\multicolumn{1}{c}{Capital Markets}&\multicolumn{1}{c}{International Transfers}&\multicolumn{1}{c}{Public Savings}&\multicolumn{1}{c}{Private Savings}&\multicolumn{1}{c}{Unsmoothed}\\

\midrule
Pre euro&Synthetic&     0.00         &     0.04\sym{*}  &     0.11\sym{***}&     0.40\sym{***}&     0.45\sym{***}\\
        &  &   (0.03)         &   (1.92)         &   (5.66)         &  (16.71)         &   (8.15)         \\
[1em]
&Actual&   -0.10         &     0.01         &     0.04         &     0.02         &     0.03         \\
      &    &  (-0.60)         &   (0.19)         &   (0.62)         &   (0.17)         &   (0.25)         \\
[1em]
Post euro&Synthetic&      0.06         &    -0.12\sym{***}&     0.03         &     0.33\sym{**} &    -0.29\sym{***}\\
      &    &   (0.29)         &  (-3.37)         &   (0.64)         &   (2.18)         &  (-6.34)         \\
[1em]
&Actual&     0.42\sym{*}  &     0.08\sym{*}  &    -0.16\sym{*}  &    -0.62\sym{***}&     0.28         \\
        &  &   (1.83)         &   (1.83)         &  (-1.97)         &  (-3.29)         &   (1.45)         \\
\midrule
$N$&    &      616         &      616         &      616         &      616         &      616 \\
$R^2$& &  0.36         &     0.16         &     0.60         &     0.53         &     0.73         \\
\bottomrule
\end{tabular}
\label{group_time_fe}
\end{table}
\footnotesize
\justify \emph{Note:}  ***, **, and * denote significance at 1\%, 5\%,  10\% respectively. t-statistics are in parenthesis. The table displays OLS estimates with clustered standard errors over the period 1990-2018 for the actual and the synthetic series before ({\sl Pre euro}) and after ({\sl Post euro}) the introduction of the euro. Separate time fixed effects for {\sl Synthetic} and {\sl Actual} are introduced to allow for differential trends across the two groups. 
  \normalsize

\subsection{Bias in the Difference-in-differences Estimation}\label{bias}
This section computes the bias that may be present in a difference-in-differences setup when the series used for estimation are estimated with some error.
Assume that there exists a true counterfactual for our experiment (the adoption of the euro). Denote by $X_i$ the unit of interest, by superscripts $T$ and $C$ the treatment and control group, and by $B$ and $A$ the before and after treatment period. We would like to estimate via the SCM $X_i^{CA}$, the behaviour of the control group after the treatment, and compare it to $X_i^{TA}$, the behaviour of the treatment group after the treatment.\\

We estimate the counterfactual based on the pre treatment matching and obtain $\tilde{X}_i^{CA}$ and $\tilde{X}_i^{CB}$, the series of interest for the control group after and before the treatment. The $\sim$ indicates that the series is estimated with some error $u$, that is $\tilde{X}_i^{CA}={X}_i^{CA}+u$.\\

The difference-in-differences in difference setup described in equation \ref{eq:DiD} provides an estimate of the treatment effect $\beta_4$ that is equivalent to 

\begin{align*}
\hat \beta_4= (\beta^{TA}-\tilde\beta^{CA})-(\beta^{TB}-\tilde\beta^{CB}).
\end{align*}
This estimate will be different from the true treatment effect 
\begin{align*}
\beta_4=(\beta^{TA}-\beta^{CA})-(\beta^{TB}-\beta^{CB}).
\end{align*}
The difference arises from the statistical error in the estimation of the coefficients of the counterfactual. The bias can be written as 
\begin{align*}
\hat \beta_4-\beta_4=(\beta^{TA}-\tilde\beta^{CA})-(\beta^{TB}-\tilde\beta^{CB})-(\beta^{TA}-\beta^{CA})+(\beta^{TB}-\beta^{CB}).
\end{align*}
Exploiting the standard result on additive measurement error on regressors we can denote the multiplicative bias by $\gamma$, meaning
\begin{align*}
\tilde \beta=\frac{\sigma^2_x+\lambda}{\sigma^2_x+\sigma^2_{ME}+2\lambda} \beta \equiv \gamma \beta, 
\end{align*}
where $\lambda$ is the covariance between the true variable and the measurement error, $\sigma^2_x$ is the variance of the true variable, and $\sigma^2_{ME}$ is the variance of the measurement error.

We now turn our attention to possible ways to estimate $\gamma$ in our setup in order to correct for or sign the bias.

\subsubsection{Solutions to Measurement Error Bias}
If we had a true counterfactual for the pre treatment period, we would get that the treated and the counterfactual group would behave the same way in the pre treatment period, that is $\beta^{TB}-\beta^{CB}= 0$.
As we do not have a true counterfactual, we have to generate it ourselves. We do so by using the SCM. The SCM minimises the distance between the actual and the synthetic series in the pre treatment period. Since the synthetic series are not the true counterfactual, but an estimated one, we do not compute exactly $\beta^{CB}$, but only $\tilde{\beta}^{CB}=\gamma\beta^{CB}$, where $\gamma$ is the bias. From this, we can compute $\gamma$ as follows: 
\begin{align*}
\tilde\beta^{CB}=\gamma\beta^{CB}=\gamma\beta^{TB}\Rightarrow\gamma=\frac{\tilde\beta^{CB}}{\beta^{TB}}.
\end{align*}
This bias $\gamma$ is derived by exploiting the difference between the actual and the synthetic series for the euro area countries in the pre treatment period.

\paragraph{Time Invariant Measurement Error}

If we assume that $\gamma$ is time invariant, and in particular that it does not change in the post treatment period, we can use it to correct for the bias as follows:

\begin{align*}
\hat \beta_4-\beta_4=&(\beta^{TA}-\tilde\beta^{CA})-(\beta^{TB}-\tilde\beta^{CB})-(\beta^{TA}-\beta^{CA})+(\beta^{TB}-\beta^{CB})\\
&=-\tilde{\beta}^{CA}+\tilde{\beta}^{CB}+\beta^{CA}-\beta^{CB}\\
&=(\beta^{CA}-\beta^{CB})(1-\gamma)
\end{align*}
Rewriting in terms of observables:
\begin{align*}	
\hat \beta_4-\beta_4&=\frac{1}{\gamma}(\tilde\beta^{CA}-\tilde\beta^{CB})\left(1-\frac{\tilde\beta^{CB}}{\beta^{TB}}\right)\\
&=(\tilde\beta^{CA}-\tilde\beta^{CB})\frac{\beta^{TB}-\tilde\beta^{CB}}{\tilde \beta^{CB}}.
\end{align*}
Hence, the true treatment effect is 
\begin{align*}
\beta_4&=\hat \beta_4-(\tilde\beta^{CA}-\tilde\beta^{CB})\frac{\beta^{TB}-\tilde\beta^{CB}}{\tilde \beta^{CB}}.
\end{align*}

\paragraph{Placebo Estimation of Bias}
In our robustness checks, we routinely estimate treatment effects for placebo countries. For these countries, we estimate the counterfactual and the treatment behaviour at the same time since, by definition, they coincide when no treatment occurs. We can then exploit this procedure to estimate the bias generated by statistical error. Note that for the placebo countries the estimated bias is available for both the pre treatment and the post treatment period.

More precisely, denoting $\hat\gamma^B$ the correction for the period before treatment and $\hat\gamma^A$ the one for the period after treatment, the corrected estimated treatment effect $\tilde{\hat{\beta}}_4$ is as follows: 
\begin{align*}
\tilde{\hat{\beta}}_4 = \beta^{AT}-\frac{\tilde\beta^{CA}}{\hat\gamma^A}-\beta^{BT}+\frac{\tilde\beta^{BC}}{\hat\gamma^B}.
\end{align*}

\paragraph{Neoclassical Measurement Error}
An alternative set of assumptions allows us to sign the bias. First note that the optimization problem of the SCM is effectively trying to minimize $\gamma^B$. Also, as shown before, $\gamma^B$ is  observable. In a difference-in-differences setting the second element of the vector $\beta$ is $1-1/\gamma^B$. Hence whenever this element is not significantly different from zero, which means that the generated counterfactual mimics the treated unit well, then there is no measurement error in the pre treatment period.\\

This implies that one can rewrite the bias as

\begin{align*}
\hat \beta_4-\beta_4=-\tilde{\beta}^{CA}\left(\frac{1}{\gamma^A}-1\right)
\end{align*}
As long as the econometrician is willing to assume that $\gamma^A<1$, which is a marginally weaker assumption than neoclassical measurement error,\footnote{Recall that the size of the bias $\gamma$ depends on the sign and magnitude of the covariance between the measurement error and X. Neoclassical measurement error assumes that that covariance is zero, generating attenuation bias. In general the following holds
\begin{align*}
\gamma<1\quad \text{if }-Cov(X,ME)<\sigma^2_{ME}\\
\gamma>1\quad \text{if }-Cov(X,ME)>\sigma^2_{ME}.
\end{align*}
}
then the sign of the bias follows from the sign of $\tilde{\beta}^{CA}$ since the term in the bracket is positive.\\

\subsubsection{Discussion}

There are a number of elements worth discussing  in these possible solutions. First, the pre treatment statistical error is what the SCM is trying to minimise, while the key unobserved parameter is the bias in the post treatment period. By this very token, we should think about the pre treatment bias as in sample and targeted, while the post treatment bias is out of sample.\\

In addition, in sample errors are relatively small if the matching procedure is successful. This can be tested by checking the second coefficient of the difference-in-differences outputs, that is the one that indicates the difference between the treated and the control group in the pre treatment period. Again this is not surprising given that the difference is minimised by the procedure.\\

Lastly, note that the entire set of solutions to the bias relies on the crucial fact that using the SCM, in the pre treatment period control and treatment group coincide. In more practical terms for our application this is just stating that Italy before the introduction of the euro is identical whether the euro will be introduced or not.\footnote{Recall that the lack of anticipation effects is one of the identifying assumptions of SCM. In our application we test this in the robustness checks by changing the treatment year to before the actual introduction of the common currency.} This is exactly what allows us to say that any observed difference between the treatment and the control group in the pre treatment period must be due to the statistical error, thereby providing an estimate of the bias itself. In more practical terms this implies that we can get a handle on the pre treatment bias both in the actual sample and in the placebo sample, whereas the post treatment bias is only available in the latter.

\bigskip

\begin{table}[H]
\caption{Bias-corrected estimates of the effect of the euro adoption on the risk sharing channels}
\centering
\scriptsize
\begin{tabular}{llcccc}
\toprule
&\multicolumn{1}{c}{Capital Markets}&\multicolumn{1}{c}{International Transfers}&\multicolumn{1}{c}{Public Savings}&\multicolumn{1}{c}{Private Savings}&\multicolumn{1}{c}{Unsmoothed}\\\midrule
  Placebo full sample &0.23 & -0.03 & 0.00 & -0.01 & 0.06 \\
 [1em]
  Placebo pre euro &0.22 & -0.01 & -0.02 & -0.21 & 0.07 \\
  [1em]
 Euro area pre euro  &0.24 & 0.04 & -0.06 & -0.12 & 0.08 \\  \bottomrule
\end{tabular}
\label{bias_correction}
\end{table}
\footnotesize
\justify\emph{Note:} The table reports the bias-corrected estimates of our baseline results. These estimates are obtained by correcting the fourth row of coefficients (Post euro Actual) in Table \ref{ols} for the bias. The row {\sl Placebo full sample} reports the bias-corrected estimates when the bias is computed from the placebo countries both pre and post adoption of the euro. The row {\sl Placebo pre euro} reports the bias-corrected estimates when the bias used for correction is computed only from the pre euro period using placebo countries. The row {\sl Euro area pre euro} reports the bias-corrected estimates when the bias used for correction is computed only from the pre euro period using euro area countries. Our main result that the adoption of the euro has increased the unsmoothed component of the shock is confirmed, even though this increase is quantitatively smaller than in our baseline estimation.
\normalsize

\end{document}